\documentclass[twocolumn]{openjournal}

\usepackage{natbib}

\usepackage[T1]{fontenc}
\usepackage{graphicx} 
\usepackage{graphicx}	
\usepackage{amsmath}	
\usepackage{amssymb}	
\usepackage{color}
\usepackage{xcolor}

\usepackage{hyperref}
\usepackage{orcidlink}
\newcommand{\orcid}[1]{\orcidlink{#1}}
\hypersetup{
	colorlinks=true,
	urlcolor=blue,    
	linkcolor=black,  
	citecolor=blue,   
}
\usepackage{orcidlink}

\usepackage{bm}
\usepackage{enumerate}
\usepackage{rotating}

\usepackage[normalem]{ulem}

\usepackage{longtable}
\usepackage{booktabs}
\usepackage{afterpage}  

\usepackage{array}
\usepackage{ragged2e}
\usepackage{xltabular}   

\newcolumntype{Y}{>{\RaggedRight\arraybackslash}X}
\newcolumntype{P}[1]{>{\RaggedRight\arraybackslash}p{#1}}

\renewcommand{\arraystretch}{1.1}

\newcommand{\mb}[1]{\textcolor{orange}{Matteo: #1}}

\newcommand{\mv}[1]{\textcolor{orange}{Marta: #1}}

\newcommand{\PC}[1]{\textcolor{cyan}{#1}}

\definecolor{mycol}{rgb}{0.0, 0.7, 0.0}

\definecolor{mygreen}{rgb}{0.0, 0.65, 0.20}

\newcommand{\soutPC}{\bgroup\markoverwith{\textcolor{cyan}{\rule[0.5ex]{2pt}{1pt}}}\ULon}

\newcommand{\arepo}{\textsc{Arepo}}
\newcommand{\illustris}{\textsc{Illustris}}
\newcommand{\tng}{\textsc{TNG}}
\newcommand{\tngl}{\textsc{TNG{50}}}
\newcommand{\tngm}{\textsc{TNG{100}}}
\newcommand{\tngh}{\textsc{TNG{300}}}
\newcommand{\E}[1]{\times\nobreak10^{#1}}

\newcommand*    \msun{{\,\rm{M}_{\odot}}}

\def \flares{\texttt{FLARES}\,}
\def \shark{\texttt{SHARK}\,}
\def \cat{\texttt{CAT}\,}
\def \newHorizon{\texttt{NewHorizon}\,}
\def \romulus{\texttt{Romulus25}\,}
\def \astrid{\texttt{Astrid}\,}
\def \horizonAGN{\texttt{Horizon-AGN}\,}
\def \obelisk{\texttt{Obelisk}\,}
\def \delphi{\texttt{Delphi}\,}
\def \ketju{\texttt{Ketju}\,}
\def \lgalaxies{\texttt{L-Galaxies}\,}
\def \massiveblack{\texttt{MassiveBlack-II}\,}
\def \illustris{\texttt{Illustris}\,}
\def \illustrisTNG{\texttt{TNG100}\,}
\def \barausse{\texttt{BACH}\,}
\def \renaissance{\texttt{Renaissance}\,}
\def \enzo{\texttt{Enzo}\,}
\def \simba{\texttt{Simba}\,}
\def \eagle{\texttt{Eagle}\,}
\def \rockstar{\texttt{Rockstar}\,}
\def \msun{\,\rm {M_\odot}}

\begin{document}


\title{The LISA Astrophysics MBHcatalogues Project: A comparison of predictions of simulated massive black hole binaries}

\author{David~Izquierdo-Villalba$^{1}$~\orcid{0000-0002-6143-1491}, Melanie~Habouzit$^{2}$~\orcid{0000-0003-4750-0187}, Matteo~Bonetti$^{3,4,5}$~\orcid{0000-0001-7889-6810}, Silvia~Bonoli$^{6,7}$~\orcid{0000-0002-6381-2052}, Alessia~Gualandris$^{8}$~\orcid{0000-0002-9420-2679}, Marta~Volonteri$^{9}$~\orcid{0000-0002-3216-1322}, Federico~Angeloni$^{10}$~\orcid{0009-0007-8669-6393}, Enrico~Barausse$^{11,12}$~\orcid{0000-0001-6499-6263}, Aklant.K.~Bhowmick$^{13,14}$~\orcid{0000-0002-7080-2864}, Laura~Blecha$^{15}$~\orcid{0000-0002-2183-1087}, Alexander~Bonilla$^{16,17}$~\orcid{0000-0002-7001-0728}, Elisa~Bortolas$^{18}$~\orcid{0000-0001-9458-821X}, Mesut~\c{C}al\i\c{s}kan$^{19}$~\orcid{0000-0002-4906-2670}, Pedro~R.~Capelo$^{20}$~\orcid{0000-0002-1786-963X}, Ana~Caramete$^{21}$~\orcid{0000-0002-8997-5730}, Laurentiu~Caramete$^{21}$~\orcid{0000-0002-3571-3145}, Nianyi~Chen$^{22,23}$~\orcid{0001-6627-2533}, Monica~Colpi$^{3,4}$~\orcid{0000-0002-3370-6152}, Thierry~Contini$^{24}$~\orcid{0000-0003-0275-938X}, Romeel~Dav{\'e}$^{25,26}$~\orcid{0000-0003-2842-9434}, Pratika~Dayal$^{27,28,29}$~\orcid{0000-0001-8460-1564}, Colin~DeGraf$^{30}$~\orcid{0009-0000-7729-3700}, Roger~Deane$^{31,32,33}$~\orcid{0000-0003-1027-5043}, Roberto~Decarli$^{34}$~\orcid{0000-0002-2662-8803}, R\'{e}mi~Delpech$^{24}$~\orcid{0009-0001-9154-5991}, Tiziana~Di~Matteo$^{35}$~\orcid{0000-0002-6462-5734}, 
Chi~An~Dong-P{\'a}ez$^{9}$~\orcid{0000-0002-8590-4409}, Alister~W.~Graham$^{36}$~\orcid{0000-0002-6496-9414}, Daryl~Haggard$^{37,38}$~\orcid{0000-0001-6803-2138}, Dimitrios~Irodotou$^{39}$~\orcid{0000-0003-2946-8080}, Peter~H.~Johansson$^{40}$~\orcid{0000-0001-8741-8263}, Atte~Keitaanranta$^{40}$~\orcid{0000-0001-7073-2977}, Luke~Kelley$^{41}$~\orcid{0000-0002-6625-6450}, Fezzel~Mahmood~Khan$^{42,43}$~\orcid{0000-0002-5707-4268}, Vivienne~Langen$^{44}$~\orcid{0000-0002-9523-3880}, Kunyang~Li$^{45,9}$~\orcid{0000-0002-0867-8946}, Shihong~Liao$^{46}$~\orcid{0000-0001-7075-6098}, Alberto~Mangiagli$^{47}$~\orcid{0000-0002-3689-1664}, Sylvain~Marsat$^{44}$~\orcid{0000-0001-9449-1071}, Joe~McCaffrey$^{48}$~\orcid{0000-0002-3921-2129}, Yueying~Ni$^{49}$~\orcid{0000-0001-7899-7195}, Coral~Pillay$^{33}$~\orcid{0000-0002-5756-4427}, Florentina-Crenguta ~Pislan$^{21,50}$~\orcid{0000-0002-0678-8148}, Alexander~Rawlings$^{40,22}$~\orcid{0000-0003-1807-6321}, John~A.~Regan$^{48}$~\orcid{0000-0001-9072-6427}, Basti\'{a}n~Reinoso$^{40}$~\orcid{0000-0002-1991-4465}, Jaelyn~S.~Roth$^{51}$~\orcid{0009-0009-5367-5603}, Milton~Ruiz$^{52}$~\orcid{0000-0002-7532-4144}, Olga~Sergijenko$^{53,54,55}$~\orcid{0000-0002-9212-7118}, Alberto~Sesana$^{3,4,5}$~\orcid{0000-0003-4961-1606}, Golam.M.~Shaifullah$^{3,4,56}$~\orcid{0000-0002-8452-4834}, Jasbir~Singh$^{5}$~\orcid{0000-0002-6260-1165}, Daniele~Spinoso$^{3,4}$~\orcid{0000-0002-9074-4833}, Alexandre~Toubiana$^{3,4}$~\orcid{0000-0002-2685-1538}, Michael~Tremmel$^{57}$~\orcid{0000-0002-4353-0306}, Alessandro~Trinca$^{58,59}$~\orcid{0000-0002-1899-4360}, Rosa~Valiante$^{59}$~\orcid{0000-0003-3050-1765}, Yihao~Zhou$^{35}$~\orcid{0000-0002-8828-8461}, Yohan~Dubois$^{9}$~\orcid{0000-0003-0225-6387}, Luca~Graziani$^{10,59,60}$~\orcid{0000-0002-9231-1505}, Christopher~C.~Lovell$^{61,62}$~\orcid{0000-0001-7964-5933}, Sebastien~Peirani$^{63,64}$~\orcid{0000-0001-6902-2898}, Will~Roper$^{65}$~\orcid{0000-0002-3257-8806}, Joop~Schaye$^{66}$~\orcid{0000-0002-0668-5560}, Raffaella~Schneider$^{10,59,60}$~\orcid{000-0001-9317-2888}, Maxime~Trebitsch$^{67}$~\orcid{0000-0002-6849-5375}, Aswin.~P.~Vijayan$^{65}$~\orcid{0000-0002-1905-4194}, Mark~Vogelsberger$^{68}$~\orcid{0000-0001-8593-7692}, Stephen.~M.~Wilkins$^{65}$~\orcid{0000-0003-3903-6935}, John~H.~Wise$^{69}$~\orcid{0000-0003-1173-8847}}\email{dizquierdo@ice.csic.es}\email{habouzit.astro@gmail.com}\email{matteo.bonetti@unimib.it}\email{silvia.bonoli@dipc.org}\email{a.gualandris@surrey.ac.uk}\email{martav@iap.fr}

\affiliation{
$^{1}$ Institute of Space Sciences (ICE, CSIC), Campus UAB, Carrer de Magrans, E-08193 Barcelona, Spain\\
$^{2}$ Department of Astronomy, University of Geneva, Chemin Pegasi 51, Versoix CH-1290, Switzerland\\
$^{3}$ Università degli Studi di Milano-Bicocca, Piazza della Scienza 3, I-20126 Milano, Italy\\
$^{4}$ INFN, Sezione di Milano-Bicocca, Piazza della Scienza 3, I-20126 Milano, Italy\\
$^{5}$ INAF - Osservatorio Astronomico di Brera, via Brera 20, I-20121 Milano, Italy\\
$^{6}$ Donostia International Physics Center (DIPC), Paseo Manuel de Lardizabal 4, 20018 Donostia-San Sebastian, Spain\\
$^{7}$ IKERBASQUE, Basque Foundation for Science, E-48013, Bilbao, Spain\\
$^{8}$ School of Mathematics and Physics, University of Surrey,  Guildford GU2 7XH, UK\\
$^{9}$ Institut d’Astrophysique de Paris, UMR 7095, CNRS and Sorbonne Universit\'e, 98 bis boulevard Arago, 75014 Paris, France\\
$^{10}$ Dipartimento di Fisica, “Sapienza” Università di Roma, Piazzale Aldo Moro 5, 00185 Roma, Italy\\
$^{11}$ SISSA, Via Bonomea 265, 34136 Trieste, Italy and INFN Sezione di Trieste\\
$^{12}$ IFPU - Institute for Fundamental Physics of the Universe, Via Beirut 2, 34014 Trieste, Italy\\
$^{13}$ Department of Astronomy, University of Virginia, 530 McCormick Road, Charlottesville, VA 22904\\
$^{14}$ Virginia Institute for Theoretical Astronomy, University of Virginia, Charlottesville, VA 22904, USA\\
$^{15}$ Department of Physics, University of Florida, Gainesville, FL, 32611-8440, USA\\
$^{16}$ Instituto de F\'{i}sica, Universidade Federal Fluminense, 24210-346 Niter\'{o}i, RJ, Brazil\\
$^{17}$ Facultad del Medio Ambiente, Universidad Distrital Francisco Jos\'e de Caldas, Carrera 5 Este \# 15-82 - Sede Vivero, Bogot\'a / Colombia\\
$^{18}$ INAF-Osservatorio Astronomico di Padova, Vicolo dell'Osservatorio 5, I-35122 Padova, PD, Italy\\
$^{19}$ William H.~Miller III Department of Physics and Astronomy, Johns Hopkins University, Baltimore, MD 21218, USA\\
$^{20}$ Department of Astrophysics, University of Zurich, Winterthurerstrasse 190, CH-8057 Z{\"u}rich, Switzerland\\
$^{21}$ Institute of Space Science - INFLPR Subsidiary, 409 Atomistilor St., Magurele, Romania\\
$^{22}$ Max-Planck-Institut f\"ur Astrophysik, Karl-Schwarzschild-Str 1, D-85748 Garching, Germany\\
$^{23}$ School of Natural Sciences, Institute for Advanced Study, Princeton, NJ 08540, USA\\
$^{24}$ Institut de Recherche en Astrophysique et Plan\'etologie (IRAP), Universit\'e de Toulouse, CNRS, UPS, CNES, Toulouse, France\\
$^{25}$ Institute for Astronomy, University of Edinburgh, Blackford Hill, Edinburgh EH8 9SQ, U.K.\\
$^{26}$ Department of Physics and Astronomy, University of the Western Cape, Robert Sobukwe Rd, Cape Town 7535, South Africa\\
$^{27}$ Canadian Institute for Theoretical Astrophysics, 60 St George St, University of Toronto, Toronto, ON M5S 3H8, Canada\\
$^{28}$ David A. Dunlap Department of Astronomy and Astrophysics, University of Toronto, 50 St George St, Toronto ON M5S 3H4, Canada\\
$^{29}$ Department of Physics, 60 St George St, University of Toronto, Toronto, ON M5S 3H8, Canada\\
$^{30}$ Truman State University, Kirksville, MO 63501, USA\\
$^{31}$ Inter-University Institute for Data Intensive Astronomy, Department of Astronomy, University of Cape Town, Cape Town, South Africa\\
$^{32}$ Department of Physics, University of Pretoria, Private Bag X20, Pretoria 0028, South Africa\\
$^{33}$ Wits Centre for Astrophysics, School of Physics, University of the Witwatersrand, 1 Jan Smuts Avenue, 2000, Johannesburg, South Africa\\
$^{34}$ INAF–Osservatorio di Astrofisica e Scienza dello Spazio, Via Gobetti 93/3, I-40129 Bologna, Italy\\
$^{35}$ McWilliams Center for Cosmology and Astrophysics, Carnegie Mellon University, 5000 Forbes Avenue, Pittsburgh, PA 15213, USA\\
$^{36}$ Centre for Astrophysics and Supercomputing, Swinburne University of Technology, Hawthorn, VIC 3122, Australia\\
$^{37}$ Department of Physics, McGill University, 3600 Rue University, Montr{\'e}al, Qu{\'e}bec, H3A 2T8, Canada\\
$^{38}$ Trottier Space Institute at McGill, 3550 Rue University, Montr{\'e}al, Qu{\'e}bec, H3A 2A7, Canada\\
$^{39}$ The Institute of Cancer Research, 123 Old Brompton Road, London SW7 3RP, UK\\
$^{40}$ Department of Physics, University of Helsinki, Gustaf H\"allstr\"omin katu 2, FI-00014 Helsinki, Finland\\
$^{41}$ Astrophysics Working Group, NANOGrav Collaboration, Berkeley, CA, USA\\
$^{42}$ New York University Abu Dhabi, PO Box 129188, Abu Dhabi, United Arab Emirates\\
$^{43}$ Center for Astrophysics and Space Science (CASS), New York University Abu Dhabi\\
$^{44}$ Université de Toulouse, CNRS/IN2P3, L2IT, Toulouse, France\\
$^{45}$ Center for Computational Astrophysics, Flatiron Institute, New York, NY 10010, USA\\
$^{46}$ Key Laboratory for Computational Astrophysics, National Astronomical Observatories, Chinese Academy of Sciences, Beijing 100101, China\\
$^{47}$ Max Planck Institute for Gravitational Physics (Albert Einstein Institute), Am M\"uhlenberg 1, DE-14476 Potsdam, Germany\\
$^{48}$ Centre for Astrophysics and Space Sciences Maynooth, Department of Physics, Maynooth University, Ireland\\
$^{49}$ Center for Astrophysics—Harvard \& Smithsonian, Cambridge, MA 02138, USA\\
$^{50}$ Doctoral School of Physics, Faculty of Physics, University of Bucharest, 405 Atomistilor, Magurele, Romania\\
$^{51}$ Department of Physics and Astronomy, Vanderbilt University, Nashville, TN 37235, USA\\
$^{52}$ Departamento de Astronomía y Astrofísica, Universitat de València, Dr. Moliner 50, 46100, Burjassot (València), Spain\\
$^{53}$ Faculty of Space Technologies, AGH University of Krakow, Aleja Mickiewicza 30, Kraków 30-059, Poland\\
$^{54}$ Main Astronomical Observatory of the National Academy of Sciences of Ukraine, Zabolotnoho str., 27, 03143, Kyiv, Ukraine\\
$^{55}$ Astronomical Observatory of Taras Shevchenko National University of Kyiv, 3 Observatorna Street, Kyiv, 04053, Ukraine\\
$^{56}$ INAF - Osservatorio Astronomico di Cagliari, via della Scienza 5, 09047 Selargius (CA), Italy.\\
$^{57}$ School of Physics, University College Cork, College Road, Cork, T12 K8AF, Ireland\\
$^{58}$ Institute for Astronomy, University of Edinburgh, Royal Observatory, Blackford Hill, Edinburgh EH9 3HJ, UK\\
$^{59}$ INAF/Osservatorio Astronomico di Roma, Via di Frascati 33, 00078 Monte Porzio Catone, Italy\\
$^{60}$ INFN, Sezione di Roma I, Piazzale Aldo Moro 2, 00185 Roma, Italy\\
$^{61}$ Kavli Institute for Cosmology, Madingley Road, Cambridge CB3 0HA, UK\\
$^{62}$ Institute of Astronomy, Madingley Road, Cambridge CB3 0HA, UK\\
$^{63}$ ILANCE, CNRS – University of Tokyo International Research Laboratory, 
Kashiwa, Chiba 277-8582, Japan\\
$^{64}$ Kavli IPMU (WPI), UTIAS, The University of Tokyo, Kashiwa, 
Chiba 277-8583, Japan\\
$^{65}$ Astronomy Centre, University of Sussex, Falmer, Brighton BN1 9QH, UK\\
$^{66}$ Leiden Observatory, Leiden University, PO Box 9513, 2300 RA Leiden, the Netherlands\\
$^{67}$ LUX, Observatoire de Paris, Universit\'e PSL, Sorbonne Universit\'e, CNRS, 75014 Paris, France\\
$^{68}$ Department of Physics and MIT Kavli Institute for Astrophysics and Space Research, 
77 Massachusetts Avenue, Cambridge, MA 02139, USA\\
$^{69}$ Center for Relativistic Astrophysics, School of Physics, Georgia Institute of Technology, Atlanta, GA 30332, USA\\
}

\begin{abstract}
In the hierarchical paradigm of galaxy formation, central massive black holes (MBHs) are expected to coalesce after the merger of their host galaxies. One of the main goals of the Laser Interferometer Space Antenna (LISA) is to constrain the origin and growth of MBHs through their merger rates and mass distribution.  Predicting MBH merger rates requires not only tracing their statistical population from large to small physical scales (kpc to sub-pc) but also modelling their formation, accretion, dynamics, mergers, and their galactic physical processes across cosmic time.
This project is the result of a large collaborative effort undertaken by the LISA Astrophysics Working Group, bringing together its collective expertise on MBH formation, evolution, and modelling, to build a comprehensive understanding of MBH merger rates across cosmic time. The project compares various theoretical predictions of MBH merger rates, quantifies the spread, and evaluates the global astrophysical uncertainties of the LISA event rates. To build a unique and complete view, our work is based on about 20 semi-analytical models and cosmological simulations from the literature, all employing distinct approaches to modelling MBH and galaxy physics. To compute the merger rates, we also incorporate delays arising from the dynamical phase of MBH hardening to coalescence. We present the expected LISA merger rates given current galaxy formation models and discuss how the merger rate depends on model assumptions, such as the seeding model and the resolution of cosmological simulations.
\end{abstract}

\maketitle

\section{Introduction}
\label{sec:intro}
ESA's flagship mission, LISA -- the Laser Interferometer Space Antenna --  will launch in the mid 2030s and aims to detect gravitational waves (GWs) in the low-frequency range from below $10^{-4}$ Hz to about $1$ Hz \citep{2017arXiv170200786A,2024arXiv240207571C}. This low-frequency window has known sources in the form of Galactic verification binaries  \citep{2023MNRAS.522.5358F,2024ApJ...963..100K}, and it is expected to be rich in a variety of other important sources. These are stellar compact binaries in the Milky Way \citep{2004MNRAS.349..181N,2022MNRAS.511.5936K}, inspiralling stellar-mass black hole binaries in nearby galaxies \citep{2016PhRvL.116w1102S,2022PhRvD.106j4034T,2025JCAP...01..084B}, extreme/intermediate mass-ratio inspirals at the centre of quiescent galactic nuclei \citep{2017PhRvD..95j3012B,2018LRR....21....4A}, and merging massive black holes (MBHs), which are the focus of this paper. A galactic foreground from unresolved compact object binaries in the Galaxy \citep{2003MNRAS.346.1197F,2024A&A...683A.139S,2024A&A...692A.165T} is present at $\sim$ mHz frequencies. In addition and over the entire LISA bandwidth, astrophysical backgrounds and cosmological backgrounds from the primordial Universe could be detected if they are sufficiently loud \citep{2023LRR....26....5A,2024PhRvD.109h3029P}.

Within the hierarchical paradigm of galaxy formation, central MBHs are expected to coalesce following the merger of their host galaxies. LISA will record their GW signal in the late inspiral, merger, and ringdown phases with enough sensitivity to allow detection of MBH binaries (MBHBs) with total masses in the largely unexplored range of $10^4 \msun \lesssim M_{\rm MBHB}\lesssim 10^7\msun,$ out to cosmological distances.  LISA's cosmic horizon is placed around redshift $z\sim 15$ (though it can in principle see MBHBs to $z>100$). LISA can thus discover early forming and growing MBHs,  observed across all cosmic epochs.

For each MBHB merger event, LISA provides the posterior distributions for the total mass,  mass ratio, MBH spins (when possible), luminosity distance, and sky localisation uncertainty map. Source-frame masses of MBHBs at $z\sim 8-15$, inferred with an accuracy at the 50--100\% level, bring unique insights into the earliest MBHs, their occupation fraction, and early growth from MBH seeds.  Moving to lower redshifts ($z \lesssim 8$), source-frame masses  are measured with increasing precision (down to per cent levels), and the spin of at least the primary, more massive, MBH is measured with an absolute error ranging between 0.1 to $10^{-4}$. This makes LISA a new and unique probe for understanding the origin, mass-growth, and spin evolution of MBHs \citep[for a review, see][]{2023LRR....26....2A}. 

In some cases, LISA's sources are expected to display an electromagnetic counterpart whose emission could span a wide range of wavelengths, from radio to gamma-rays, opening up concrete possibilities for multi-messenger astronomy \citep{2022PhRvD.106j3017M,2023MNRAS.521.2577P,2023MNRAS.519.5962L,2023A&A...677A.123I}. These counterparts will provide us with information on the environments in which MBHBs live and insight into accretion physics in a violently changing spacetime, a regime never probed before \citep{2022LRR....25....3B}, with far-reaching consequences for cosmography \citep{2023LRR....26....5A}.

Given LISA's cosmic reach, the mission will inform us, albeit indirectly, about the physical mechanisms driving MBH growth from seeds.  Seed models include: (i) black hole (BH) relics from the first generation of metal-free stars, the direct collapse of massive clouds, the collapse of super-massive stars formed in stellar runaway collisions, or a different process altogether, such as primordial black holes formed in the early Universe, before the epoch of galaxy formation \citep{2018MNRAS.478.3756C}. Multiple mechanisms, not mutually exclusive, introduce uncertainty when modelling the growth of MBHs, either single or in binaries, during galaxy evolution. 

Binaries of MBHs (MBHBs) form naturally in galaxy mergers and coalesce due to GW emission when the MBHs reach sub-mpc separations. In order to become LISA sources at a given redshift, they need to evolve and shrink from the kpc scales typical of the galactic merger to the pc scales where MBHs bind into a binary and further to the mpc scales of GW inspiral. Therefore, stellar and gas dynamical processes acting at different scales must be effective at hardening the binaries in less than a Hubble time \citep{1980Natur.287..307B}. Evolutionary timescales, referred to as {\it time delays} between the galaxy merger and the MBHB merger, are uncertain and depend on both MBH and galaxy properties, leading to significant uncertainties in merger rates and LISA detection rates \citep{2023LRR....26....2A}. Lastly, the MBH masses in a given merger are the result of, and keep memory of, the accretion (interrupted by feedback processes) and mergers experienced up to that redshift.

LISA's transformational impact in the study of MBHs and MBHBs relies on the ability of theoretical models to predict the distribution of GW events in redshift, mass, and mass-ratio, and to constrain how frequently these mergers occur. In light of the complexities cited above, the aim of this project is to compare various theoretical predictions for what LISA will observe, to identify key, global astrophysical uncertainties on the merger rates, and to quantitatively evaluate the {\it spread} in the predictive power of the current models of MBHB/MBH evolution. 

Estimates of LISA's merger rates for MBHBs come from a variety of models, broadly classified as either semi-analytical or numerical.  Semi-analytical models follow the evolution of MBHs and their host galaxies based on a range of prescriptions regarding the physical processes governing MBH dynamics and growth over cosmic time. Given their numerical efficiency, they can be used to explore a wide parameter space.  Cosmological hydrodynamic simulations, on the other hand, follow numerically the large scale assembly and growth of dark matter {DM} haloes together with their baryonic components from high redshifts, tracking galactic mergers and MBH evolution simultaneously, within a given spatial, temporal, and mass resolution. 

Over the last decade, both types of models have proven successful at reproducing observed properties of galaxies and black holes in either the quiescent or the active accretion phase \citep[e.g.,][]{2017MNRAS.470.1121T, 2018MNRAS.475..624N, 2018MNRAS.475..676S, 2019MNRAS.486.2336D, 2020ApJ...904...16B, 2021A&A...651A.109D, 2023MNRAS.519.2083I}. However, both approaches incur significant limitations, such as limited resolution and simplified prescriptions for physical processes. Predicted merger rates span a wide range from $\sim100\,\rm yr^{-1}$ to $\lesssim 1\,{\rm yr}^{-1}$ \citep[see Section 2.4 of][for a first compilation of predictions]{2023LRR....26....2A}. This large variation is due to the different techniques employed (semi-analytical models versus cosmological simulations) and their associated technical details (e.g., choice of resolution, softening) as well as varying assumptions regarding the physical processes included in the modelling of MBH physics (MBH seeding, MBH dynamics, gas accretion, feedback from active galactic nuclei (AGN) and galaxy physics (e.g., feedback from supernovae (SNae)).

In this paper, we present the first extensive collection of MBHBs catalogues derived from 5 semi-analytical models (\lgalaxies, \cat, \delphi, \shark, \barausse) and 15 numerical simulations (\astrid, \eagle, \flares,  \horizonAGN, \illustris, \tngl, \tngm, \tngh, \ketju, \massiveblack, \newHorizon, \obelisk, \renaissance, \romulus, \simba). The properties of the models are described in Section~\ref{sec:methods_models}. From these catalogues, we compute the merger rate of MBHBs as a function of black hole mass and galaxy properties, as well as redshift (Section~\ref{sec:methods_coalescence}). We present and discuss results in Section~\ref{sec:results} and \ref{sec:conclusions}.
We quantify the variations and assess the global astrophysical uncertainties in the LISA event rates. This comprehensive analysis provides a unique and complete view of the expected outcomes from LISA. We highlight the aspects where models differ the most, suggesting areas requiring attention from the community in the coming decade prior to LISA's launch.

\section{Description of the models}\label{sec:methods_models}

\subsection{Input physics and uncertainties}\label{sec:input}

In this section, we review the semi-analytical and hydrodynamical models available in the literature that we use to compute the merger rates of MBHs across cosmic time. We particularly highlight the characteristics of the models in terms of resolution, volume covered, range of MBH and galaxy mass that is captured, as well as the modelling of MBH seeding, accretion, dynamics, and criteria used for MBH coalescence. Finally, all of the models use a slightly different cosmological parameter setup, but these differences will have a very marginal effect on the results presented in this work.

Before entering into the details of each simulation, we briefly review aspects that are widely used in the literature and shared by most models.

\subsubsection{Seeding of massive black holes}

The semi-analytical models and hydrodynamical simulations analyzed in this paper mainly employ two primary theoretical channels of MBH formation: light seeds (with mass $M_{\rm BH} \leqslant 10^3\msun$) formed from the collapse of massive Population~III stars \citep[PopIII; e.g.,][]{2001ApJ...551L..27M, 2018MNRAS.480.3762S, 2021MNRAS.501.1413N, 2022MNRAS.511..616T}, and heavy seeds ($M_{\rm BH} \geqslant 10^3\msun$) formed possibly from runaway dynamical interactions in dense stellar environments \citep[e.g.,][]{2016MNRAS.455...35A, 2024MNRAS.531.377R}, from  the collapse of supermassive stars in the so-called atomic-cooling haloes  where metals/dust cooling is inefficient (with metallicity $Z \leq Z_{\rm cr}\sim 10^{-4} Z_\odot$) and molecular cooling is inhibited by an intense illuminating Lyman Werner flux \citep[e.g.,][]{2019PASA...36...27W}, or perhaps induced by galaxy interactions and mergers \citep[e.g.,][]{2019Natur.566...85W,  2019RPPh...82a6901M, 2024ApJ...961...76M}.

Semi-analytical models often implement these models directly, sometimes even exploring the coexistence of these channels, where MBHs form through both light and heavy seed formation mechanisms (see Section~\ref{sec:SAMs}), closely following the theoretical prescriptions of the mechanisms (e.g., local gas density, metallicity, intensity of photo-dissociating Lyman-Werner radiation). Depending on their resolution, cosmological hydrodynamical simulations (see Section~\ref{sec:SIMU}) may not be able to apply the same flexibility and in most cases cannot easily explore a large parameter space. In particular, until relatively recently simplistic approaches were employed in hydrodynamical simulations for the location of the seeds (e.g., MBHs that form at the centre of all resolved haloes or galaxies) and their initial mass (e.g., often employing a fixed seed mass). However, more recent explorations have begun to employ more sophisticated (subgrid) dynamical locations and a mass distribution for seeding (albeit without a strong physical underpinning, as the initial mass function of MBHs is largely unconstrained). Fig.~\ref{fig:SeedMass} provides an overview of the seed masses employed by the models, both semi-analytical and hydrodynamic, and the wide range of simulated volumes that they cover.

\subsubsection{Growth of massive black holes}
Being unable to resolve scales close to the accretion disc, all models need to make simplified assumptions regarding the growth phase. In the hydrodynamical simulations covered here, accretion is often modelled using the Bondi-Hoyle-Lyttleton formalism \citep[BHL;][]
{1944MNRAS.104..273B,1952MNRAS.112..195B}: 
\begin{eqnarray}
\dot{M}_{\rm BHL}=\alpha \, 4\pi\,  G^{2}\, M_{\rm BH}^{2}\,  \frac{\bar{\rho}}{(\bar{c}_{\rm s}^{2}+\bar{v}^{2})^{3/2}}~,
\label{eq:BHL}
\end{eqnarray}
\noindent where $\bar{\rho}$ and $\bar{c}_{\rm s}$ are the kernel weighted density and ambient sound speed, $\bar{v}$ is the average velocity of the gas relative to the BH, $G$ is the gravitational constant, and $\alpha$ is a boost factor (which may or may not be applied depending on the simulation setup), taken as a free parameter or as a function of the local environment properties. $\alpha$ is applied to compensate for the unresolved interstellar medium multiphase structure, which can artificially lower the accretion rates compared to what it ``should'' be. Accretion rates can additionally be capped at the Eddington limit
\begin{equation}
\dot{M}_{\rm Edd} = \frac{4\pi G M_{\rm BH}\, m_{\rm p}}{\epsilon_{\rm r}\sigma_{\rm T}c}~,
\label{eq:Eddington}
\end{equation}
\noindent with $m_{\rm p}$ the proton mass, $\epsilon_{\rm r}$ the radiative efficiency, $\sigma_{\rm T}$ the Thomson cross-section, and $c$ the speed of light in vacuum. Unless otherwise stated, the value of $\epsilon_{\rm r}$ adopted in Eq.~\eqref{eq:Eddington} is 0.1, and we additionally define the Eddington ratio as $f_{\rm Edd} = \dot{M}_{\rm BH}/\dot{M}_{\rm Edd}$.

In semi-analytical models, growth generally is linked to events that disturb the host galaxy and can lead to the infall of gas towards the nuclear regions, such as galaxy mergers and disc instabilities. How accretion proceeds varies significantly across semi-analytical models: some assume BHL accretion, others funnel a fraction of the available gas towards the MBH, modulated by dynamical processes such as mergers or dynamical instabilities, others differentiate between accretion from discs and from the hot gas component. Additionally, some models allow for super-Eddington accretion.

\subsubsection{Feedback from massive black holes}

The injection of energy and momentum from MBHs is an important physical process in regulating the gas content and star formation in galaxies \citep{2005MNRAS.361..776S}. In hydrodynamical simulations, feedback is modelled with either an injection of thermal energy, of kinetic energy or both (dual-mode feedback). The latter usually accounts for the different properties of MBHs at high and low Eddington ratio. MBHs accreting at a substantial fraction of the Eddington rate are radiatively efficient and produce copious photons in optical/UV as well as winds, while MBHs accreting at very sub-Eddington rates are characterized by suppressed production of optical/UV photons, but often exhibit jets \citep[e.g.,][]{2008MNRAS.388.1011M}. Besides the technical differences in the implementations (spherical, collimated, region of the injection) the main parameter is the feedback efficiency:

\begin{equation}
E_{\rm f}=\epsilon_f E_{\rm AGN},
\label{eq:fedd_eff}
\end{equation}

where $E_{AGN}$ is the accretion energy from the MBH. Depending on the type of feedback $E_{\rm AGN}$ is defined as $\dot{M}_{\rm BH} c^2$ multiplied by a model-defined timescale (coarse or fine timestep, or even multiple timesteps) and a radiative efficiency (or not, for the case of jets).

Similarly, semi-analytical models account for the feedback of active black holes both during phases of high and low Eddington ratios.  Feedback taking place when MBHs are accreting at low rates is generally  assumed to be taking place in high-mass haloes: a fraction of the energy emitted by the accreting MBH is assumed to couple to the medium and heat part of the cooling gas within the galaxy, leading to star formation quenching  \citep[the so-called ``radio-mode'' feedback, e.g.,][]{2006MNRAS.365...11C}. ``Quasar-mode'' feedback, instead, takes place during episodes of high accretion rates, often associated to major merger events, and can also lead to removal of cold gas and the regulation of MBH growth itself.

\subsubsection{Dynamics of massive black holes}

\begin{figure*}
\centering
\includegraphics[width=2.\columnwidth]{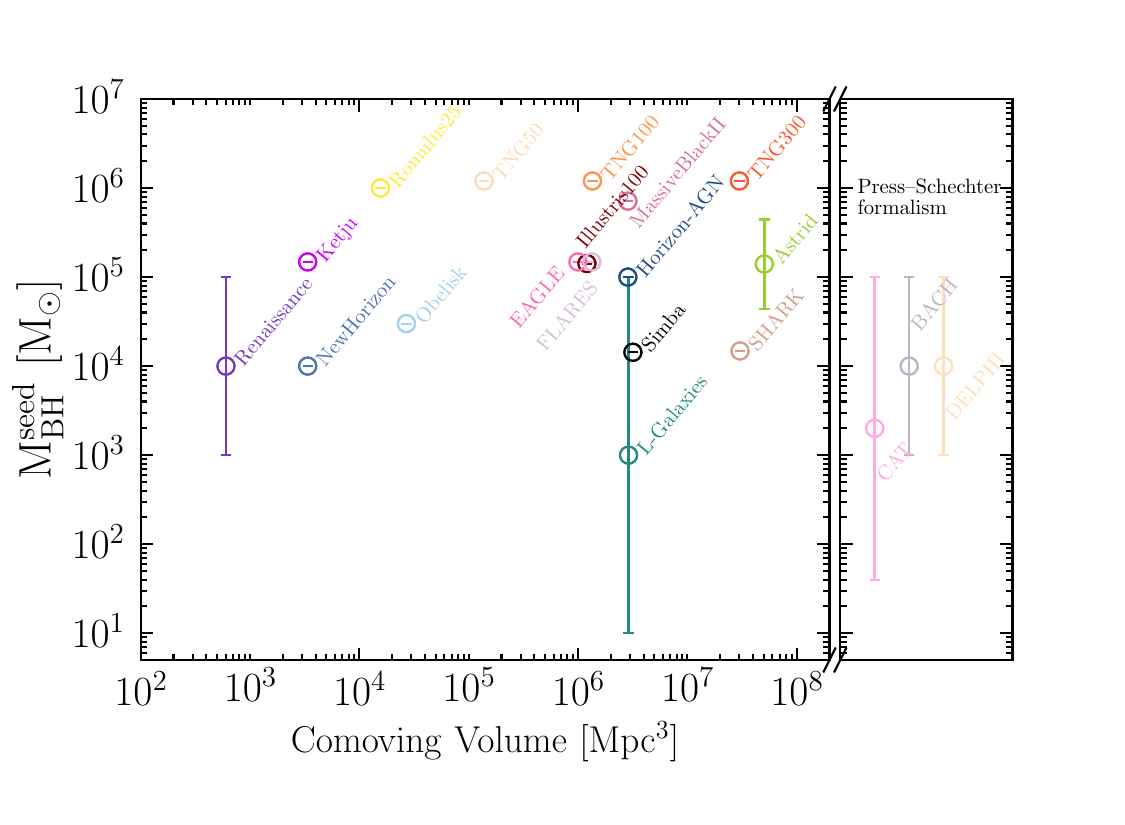}
\caption{Comoving volumes and seed mass ranges covered by the different simulations and semi-analytical models analysed in this work. For models with a distribution of seed masses, the line joins the minimum and maximum seed mass, and the symbol represents logarithmic midpoint.
High-resolution zoom-in regions, with small simulated volumes, are localised at the extreme left of the figure (e.g., \renaissance, \ketju, \newHorizon, \obelisk). \flares is a suite of 40 spherical regions, each with a radius of 20.66 cMpc, functioning together through a weighting scheme. The volume referenced here is the total of the volumes of these 40 regions. Semi-analytical models based on the Press-Schechter formalism (\cat, \barausse, and \delphi), by not tracking the spatial information of haloes, do not have a well defined associated volume, thus we show them at the extreme right of the figure in a separated panel. Large-volume hydrodynamical simulations or semi-analytical models based on $N$-body simulations cluster in the middle-right of the plot. Focusing on the seed BH mass, we can clearly infer that the masses span a quite large range among different models. Hydrodynamical simulations generally input a single and quite high value ($\gtrsim 10^4\msun$) for the seed BH mass (with the exception of \astrid not being limited to a unique injected initial mass). On the contrary, semi-analytical models generally explore different seed formation mechanisms, yielding a wide spectrum of possible seed masses going from $\sim 10^5\msun$ down to $\sim 10\msun$ for e.g., \lgalaxies and \cat.}
\label{fig:SeedMass}
\end{figure*}

As illustrated in Fig.~\ref{fig:SeedMass}, the models cover different volumes and resolutions, which significantly impact the modelling of MBH dynamics. In hydrodynamical simulations, resolution impacts which physical processes can be followed directly or have to be implemented as sub-grid recipes. In some semi-analytical models, the lack of spatial information reflects in simplified approaches that typically assume spherical or disc-like configurations. In the description of each model, we note the typical separation at which binaries are numerically merged. This information is required to calculate likely delay times in post-processing. 

As MBHs move, they generate an overdensity of material (gas, stars, and DM) in their wake, which in turn drags and decelerates them, ultimately helping them sink to the centre of their host galaxies. Some simulations use a drag force to simulate the dynamical friction from one or more components of the medium (i.e., gas, DM, stars). Other models do not employ any dynamical frictional modelling and rely on the gravitational force accuracy of their solver (and the resolution of their setup). At the other end of the spectrum, implementations may also `reposition' MBHs to the spatial location with the lowest gravitational potential. Only one model in our sample follows MBHBs  to sub-pc separations via direct N-body integration (\ketju, Section~\ref{sec:Ketju}). The other models, whether they include dynamical friction or not, are unable to track the later stages of the binaries' coalescence (i.e., binary hardening phase, circumbinary disc, GW regime). 

Semi-analytical models also cannot follow MBH dynamics directly, and therefore adopt different approaches to merge MBHs or model their dynamical evolution. A first approach is to merge MBHs when halos or galaxies do, without accounting for the actual MBH dynamical evolution causing a delay between halo/galaxy merger and MBH merger. In this case, a mass ratio threshold is often employed to distinguish major mergers leading to MBH mergers from minor mergers where binary formation is unwarranted. More refined models either calculate timescales for the various physical processes (dynamical friction, hardening, migration in circumbinary discs) or integrate the equations of motion for the same processes.

\begin{figure*}
\centering
\includegraphics[width=2.0\columnwidth]{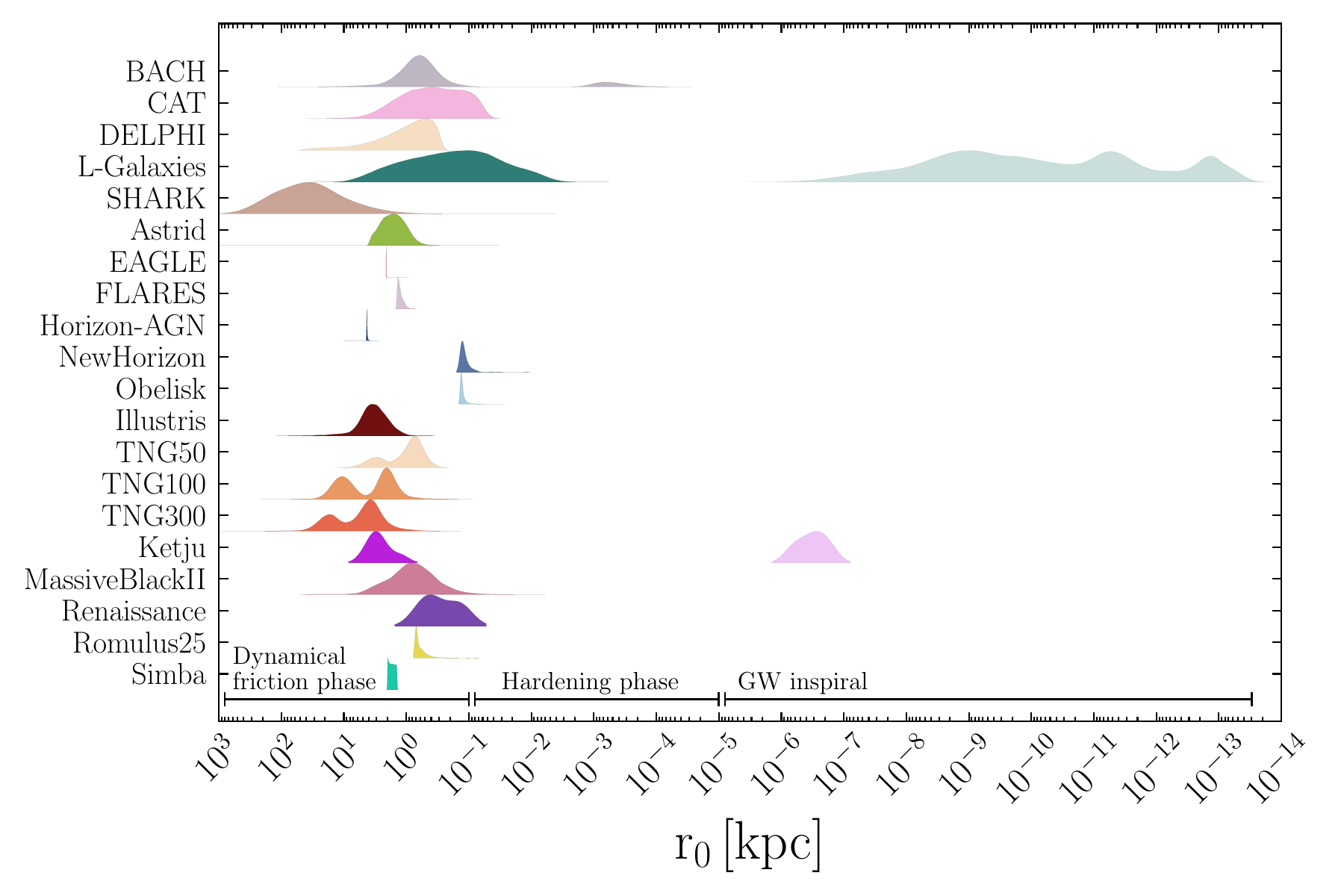}
\caption{
Distribution of MBHB separations across different models (in proper kpc), stacking all redshifts. To guide the reader, we have added some indicative distances corresponding to phases where the MBHs are expected to be in the dynamical friction stage, hardening phase or the gravitational wave-driven stage. The distances presented for each model are defined differently: (i) the galaxy half-mass radius (\lgalaxies, \barausse and \ketju) (ii) a few softening lengths (\horizonAGN, \newHorizon, \obelisk, \obelisk, \tng, \eagle, \flares, \astrid and \simba) (iii) the distance between the two MBH positions in the snapshot immediately before the host merger (\shark and \renaissance), and (iv) a fraction of the host galaxy virial radius (\cat and  \delphi). We refer the reader to the main text for further information.  High resolution cosmological simulation-based models resolve separations down to $\gtrsim 10^{-2}\, \rm kpc$, with those run in $100\, \rm cMpc$ boxes typically peaking at $\sim1\, \rm kpc$. We notice that \lgalaxies and \ketju display an additional (light shaded) distance distribution at $r_0 < 10^{-5}\,\rm kpc$. This occurs because these two models are the only ones that track the evolution of the MBHB semi-major axis down to the point of final coalescence. To illustrate this, we show the distribution of merged MBHs measured at a separation of $6GM_{\rm tot}/c^2$, where $M_{\rm tot}$ is the total mass of the binary system. The bimodal distributions for the \tng\ simulations are an artifact of the AGN feedback prescription, as discussed in Section~\ref{sec:tgn_model}.}
\label{fig:Separation}
\end{figure*}

Depending on their MBH seeding and dynamics prescriptions, models can either permit a single MBH per galaxy or allow multiple MBHs to co-exist and evolve within galaxies. This is crucial for interpreting the MBH merger rates predicted by the models. In the following subsections, we clarify this aspect for each model and define the binary separation employed in this paper. Distributions of binary separations are shown in Fig.~\ref{fig:Separation}.

\subsection{Semi-analytical models}\label{sec:SAMs}

Semi-analytical models follow the evolution of the baryonic component of the Universe via a set of analytical prescriptions that are linked to the cosmological evolution of the host DM haloes, expressed in the form of merger trees. DM merger trees can either be constructed analytically via the extended Press-Schechter formalism  ~\citep{1974ApJ...187..425P} or from the outputs of large $N$-body simulations. In what follows, we summarize the most relevant features of the semi-analytical models used in this work. 

\subsubsection{The Cosmic Archaeology Tool (CAT)}

\cat is a semi-analytical model that predicts the formation and evolution of the first MBHs and galaxies through cosmic epochs while being consistent with the general population of AGN and galaxies observed at $4\leq z\leq 7$~\citep{2022MNRAS.511..616T,2016MNRAS.457.3356V}. \cat relies on the galaxy formation model GALFORM~\citep{2008MNRAS.383..557P}, which is based on the extended Press-Schechter formalism, to reconstruct a large sample of DM hierarchical merger histories representative of the evolution of the entire galaxy population. For the model used here, the merger trees run from $z=4$ to $z= 24$ with a time resolution of $\Delta t\sim 0.5 \,(z=24) - 4 \, (z=4)\rm \,Myr$.  The DM halo mass resolution is set to correspond to a halo virial temperature of $T_{\rm vir}=1200$ K and ranges from $\sim 10^6 \, \rm M_\odot$ at $z=24$  to $\sim 10^7 \, \rm M_\odot$ at $z=4$. 

In \cat, MBH seeds ranging from a few tens to a few hundreds of  $\msun$ (light seeds) and up to $10^5\msun$ (heavy seeds) co-exist. The light seeds form as remnants of $[40-140]\msun$  and $[260-300]\msun$ PopIII stars, depending on the galaxy star formation efficiency while the most massive seeds ($10^5\msun$) form in atomic-cooling haloes \citep[see][for details on the model implementation]{2016MNRAS.457.3356V,2022MNRAS.511..616T}. This model does not include the intermediate-mass seeds that are modeled in the newest version of \cat (Davari et al. in prep.).

After their formation, MBH seeds grow through gas accretion and mergers with other MBHs, as detailed in~\citet{2022MNRAS.511..616T,2001ApJ...551L..27M,2005ApJ...620...59S}. MBH accretion is described using two alternative prescriptions. In the Eddington-limited model, the accretion proceeds at the BHL rate (assuming a boost factor $\alpha = 90$) and cannot exceed the Eddington limit. In the super-Eddington model, short phases of enhanced accretion that exceed the Eddington limit are triggered by major halo mergers \citep{2016MNRAS.458.3047P,2017MNRAS.471..589P}. Super-Eddington growth is generally sustained for $\lesssim 2-5\rm \, Myr$, consistent with predictions from high-resolution hydrodynamical simulations \citep{2020OJAp....3E...9R,2023MNRAS.519.1837S,2023A&A...669A.143M,2024A&A...686A.256L}. The AGN radiative efficiency is set to the standard value of $\epsilon_r = 0.1$ for sub-Eddington accretion rates, while for super-Eddington rates is calculated relying on the fitting formulas proposed in \citet{2014ApJ...784L..38M}.

The dynamical evolution of MBHBs is not accounted for in the present model: it assumes that two MBHs coalesce instantaneously during  mergers wherein the mass ratio of the host DM haloes is $> 0.1$~\citep{2011MNRAS.416.1916V}. In the present project, the \cat{} catalogs consider a single MBH per galaxy. After a galaxy major merger (of mass ratio $> 0.1$), the corresponding catalog contains a remnant MBH whose mass is the sum of the two progenitor MBH masses. After a minor merger (of mass ratio $< 0.1$) only the most massive MBH is kept and no MBH merger is reported in the catalog. The separation of MBHs before their coalescence is set to $10\%$ of the merger-remnant galaxy virial radius. 

So far, \cat has been used to probe observational imprints of light and heavy seeds and their growth mode on the redshift evolution of the MBH mass/luminosity function \citep{2022MNRAS.511..616T}. It has also been applied to study the nature of the UV-bright galaxies detected by JWST at $z \ge 10$ \citep{2024MNRAS.529.3563T}, the detectability of MBHs within ongoing JWST surveys at $5\leq z\leq 15$ \citep{2023MNRAS.519.4753T}, and the origin of JWST-detected AGN and their host galaxies at $z\geq 9$ \citep{2023MNRAS.526.3250S, 2024arXiv241214248T}. Additionally, \cite{2023MNRAS.520.3609V} used \cat to investigate the expected imprints of the first stars and accreting MBHs on the global 21cm signal at cosmic dawn.\\

\subsubsection{L-Galaxies}

\lgalaxies\footnote{The \lgalaxies{} public releases and catalogs are available online \href{https://lgalaxiespublicrelease.github.io/}{L-Galaxies}.} is a semi-analytical model which tracks the evolution of the baryonic component of the Universe using DM merger trees extracted from $N$-body simulations, such as the \texttt{Millennium} suite \citep{2005Natur.435..629S,2009MNRAS.398.1150B}. This allows \lgalaxies to model self-consistently the evolution of galaxies, MBHs and MBHBs within large-volume, cosmological environments. The \lgalaxies results presented in the present paper are based on the merger trees of the \texttt{Millennium-II} simulation, which follows the evolution of DM in a periodic box of side $\rm L_{box}=96\,cMpc\,\mathit{h}^{-1}$, with a halo mass resolution of $\rm M_{halo}=3.84\times10^7\msun\,\mathit{h}^{-1}$. We note that these values are referred to the re-scaling of \texttt{Millennium-II} \citep{2010MNRAS.405..143A} to cosmological parameters close to those of the \texttt{Planck 2015} cosmology \citep[][]{2016A&A...594A..13P}. The simulation produces 63 snapshots between $z=56$ and $z=0$, with \lgalaxies then evolving galaxy properties over substeps (of $\sim 1$ to $\sim20$ Myr) between two consecutive snapshots.

\lgalaxies was recently updated with detailed models for the formation and evolution of MBHs and MBHBs \citep{2020MNRAS.495.4681I,2022MNRAS.509.3488I,2023MNRAS.518.4672S}. Specifically, MBH seeds form in the mass-range $\rm10\lesssim M_{seed}/M_\odot\lesssim10^5$ according to four different scenarios: light seeds ($\rm M_{seed}\lesssim500M_\odot$), intermediate-mass seeds ($\rm 10^3\lesssim M_{seed}/M_\odot\lesssim10^4$), heavy direct-collapse BHs (DCBHs, $\rm M_{seed}=10^5M_\odot$) and heavy merger-induced BHs (miDCBHs, $\rm M_{seed}\sim10^5M_\odot$). These scenarios are activated depending on the spatial variations of chemical enrichment and Lyman-Werner flux, which are tracked self-consistently across the entire \texttt{Millennium-II} box \citep{2023MNRAS.518.4672S}. Due to resolution limits, light-seeds are accounted for as the evolved counterparts of PopIII remnants inherited by the GAMETE-QSO/dust model outputs \citep[][]{2016MNRAS.457.3356V}. On the other hand,  miDCBHs are modelled as in \cite{2014MNRAS.437.1576B}.

MBH seeds form at the centre of their hosts, so that they can readily grow mass as it becomes available by galaxy mergers or disc instabilities. During these phases of efficient gas accretion, growth proceeds at high rates (capped at the Eddington limit), followed by a quiescent phase characterized by a time-declining accretion rate  \citep[see][]{2009MNRAS.396..423B, 2020MNRAS.495.4681I}. During the initial high-rates accretion phase, a thin-disk accretion model is assumed with radiative efficiency $\epsilon=\eta(a)$, with $\eta(a)$ being the BH-spin dependent mass-accretion efficiency. On the other hand, during the time-declining accretion rate phase an advection-dominated accretion flow (ADAF) mode is considered when the Eddington rate ($f_{\rm Edd}$) falls below a critical threshold ($f_{\rm Edd}^{\rm crit}$), assuming a radiative efficiency $\epsilon=\eta(a)\times \frac{f_{\rm Edd}}{f_{\rm Edd}^{\rm crit}}$ \citep[see][for details]{2020MNRAS.495.4681I}. This ADAF mode is also responsible for AGN feedback and galaxy quenching  \citep[see][]{2006MNRAS.365...11C}.  The model version used in this work includes the tracking of MBH spin evolution, where the spin-up/-down is regulated by the stellar bulge and gas environments in which MBHs evolve \citep[see][]{2014ApJ...794..104S,2020MNRAS.495.4681I}. 

The model accounts for the formation and  evolution of MBHBs from galactic scales down to their final coalescence \citep{2022MNRAS.509.3488I}. In particular, it follows the MBHBs evolution through the different phases of pairing, hardening, and final inspiral, via large-scale dynamical friction, small-scale interactions with single stars or a circumbinary disc, and finally GW emission \citep[see][]{2022MNRAS.509.3488I,2024arXiv240611985F}. Furthermore, \lgalaxies describes three-body interactions, when MBHBs hosted in galactic centres are perturbed by the presence of a third MBH, deposited by a recent galaxy merger \citep[as in][]{2018MNRAS.477.2599B}. Details of the MBH dynamics, their inclusion and effect are discussed and analyzed in Section~\ref{ModelsWithDelays}. 

On top of being able to describe a wide range of processes that drive  galaxy formation and evolution, such as galaxy stripping and tidal disruptions, and to successfully predict key galaxy properties at $z<4$ across a wide dynamic range \citep[$\rm10^7<M_*/M_\odot<10^{12}$; e.g.,][]{2015MNRAS.451.2663H,2020MNRAS.491.5795H,2021MNRAS.503.4474Y}, the inclusion of MBHB formation and evolution has also allowed to draw predictions about the properties of galaxies hosting MBHBs and LISA events as well as to interpret Pulsar Timing Array (PTA) data \citep{2023MNRAS.519.5962L, 2023A&A...677A.123I,  2023MNRAS.519.2083I, 2024arXiv240110983I}. 

Although \lgalaxies{} can track several MBHs in each galaxy, namely two ``centrals'' and a fixed number of ``pairing'' ones, only the most massive among the central ones is used to build the catalogs for this project. Furthermore, after a galaxy merger, the separation between their central MBHs used to compute the dynamical friction time-scale is set to the galaxy-radius of the merger remnant, defined as the mass-weighted average between the bulge and disc radii \cite[as in][]{2011MNRAS.413..101G}. In Section ~\ref{ModelsWithDelays}, we also analyze a catalogue of \lgalaxies{} in which delays between galaxy-galaxy mergers and MBH-MBH mergers are tracked self-consistently.

\subsubsection{Black holes Across Cosmic History (BACH)} 

BACH is a semi-analytical model for the formation and co-evolution of galaxies and MBHs \citep{2012MNRAS.423.2533B,2014ApJ...794..104S, 2015ApJ...812...72A,2018MNRAS.477.2599B,2019MNRAS.486.4044B,2020ApJ...904...16B,2023PhRvD.108j3034B}.  For most of the sections of the present work, we make use of the `HS-nod-SN-high-accr (B+20)' version of the model \citep{2023PhRvD.108j3034B}, which is calibrated to PTA data \citep{2024A&A...685A..94E} by enhancing MBH accretion and switching off delays between galaxy and MBHB mergers.\footnote{See \url{https://grams-815673.wixsite.com/barausse/publications-miscellaneous}.} However, in Section ~\ref{ModelsWithDelays}, we study a version of the same model wherein delays are included on the fly in the semi-analytical calculation.

BACH adopts an extended Press-Schechter formalism (augmented to match results from $N$-body simulations) to model the cosmic history of DM haloes~\citep{1974ApJ...187..425P,2008MNRAS.383..557P}. 
The results presented here were produced with merger trees extending to $z = 20$, with adaptive mass resolution set to a fixed fraction ($3\times 10^{-3}$) of the maximum halo mass at given redshift. Baryonic components (chemically unprocessed hot gas, stellar and cold-gas discs and bulges, nuclear star clusters and black holes) and chemical composition are also evolved along the DM merger trees. Rather than assuming a constant time-step, the equations for the baryonic evolution are integrated with a fourth-order Runge-Kutta integrator with adaptive stepsize and Richardson extrapolation, with (fractional) tolerance set to $10^{-2}$. 

The model allows for choosing either a light or heavy seed scenario. The former assume that PopIII remnant light seeds form in the most massive haloes ($>3.5\sigma$ peaks of the primordial density field) at $z>15$, with the mass of the original stars drawn from a log-normal distribution centred at $300\msun$ and standard deviation of 0.2 dex, with an exclusion region between 140 and $260\msun$ (i.e., the mass gap of pair instability SNae). The mass of the resulting seed MBH is 2/3 of that of the original star. The heavy-seed scenario implemented in BACH is currently that of~\cite{2008MNRAS.383.1079V}, where heavy ($\sim10^3$--$10^5\msun$) seeds originate from bar instabilities of proto-galactic discs, c.f. the prescription given in~\cite{2008MNRAS.383.1079V} for the exact relation between galaxy properties and seed mass/occupation fraction. For the purpose of this work, we only consider the heavy-seed scenario.

MBHs accrete cold gas in quasar mode on a timescale set to the gas viscous timescale evaluated at the MBH influence radius (defined as the  radius at which the enclosed bulge stellar mass is twice the binary mass) and from the hot gas component at the BHL accretion rate. Mass accretion is capped at the Eddington rate for heavy seeds, and at moderately super-Eddington rates ($\approx 2$--3) for light seeds. However, the bolometric luminosity is modelled through Eq.~40 of \citet{2012MNRAS.423.2533B}, which accounts for the fact that accretion is expected to become radiatively inefficient at super-Eddington rates. Radiative and jet-driven AGN feedback on baryonic structures is also included, as well as the possible quenching effect of SN winds on MBH accretion in small haloes~\citep{2017MNRAS.468.3935H}.

The dynamics of MBHBs is included in BACH, although these features can be switched off -- like for model  `HS-nod-SN-high-accr (B+20)' presented here.  Delays are described, included and analyzed in Section~\ref{ModelsWithDelays}. The catalogues used to derive the mass and luminosity functions contain all MBHs present in central and satellite galaxies, with, for example, up to 4 MBHs per central galaxy. In practice, the contribution from MBHs in satellites does not significantly impact the mass and luminosity functions. In catalogues of merging MBHs, the MBHB separation is set to the half-light radius of their host as defined in \cite{2012MNRAS.423.2533B}. If the host bulge-mass is null, the MBHB separation is set to $r_{\rm inf} = G(\rm M_{BH_1}+M_{BH_2})/\sigma^2$ where $\sigma$ is defined by the scaling relation as in Eq.~(9) of \cite{2016MNRAS.463L...6S}.

\subsubsection{SHARK}
SHARK\footnote{SHARK is hosted at: \url{https://github.com/ICRAR/shark}} is a semi-analytical model of galaxy formation specially designed to be flexible and to allow easy extensions and modifications of physical models, solving the associated suite of differential equations with an adaptive stepsize \citep{2018MNRAS.481.3573L}. The model adopted here is described in \cite{2018MNRAS.481.3573L} and its DM halo assembly history is based on a {\small GADGET-2} $N$-body simulation. This run covers a box with size L$_{\rm box}=210$~cMpc~$h^{-1}$ and produces 200 snapshots in evenly spaced logarithmic intervals in the scale factor $a = 1/(1+z)$ from $z=24$ to $z=0$. The snapshot cadence ranges between $\approx 6$--80~Myr \citep{2018MNRAS.475.5338E}. This simulation resolves DM haloes down to $\approx 10^9\msun$. 

In \shark, MBHs are seeded once a DM halo surpasses a minimum mass of $10^{10}\msun{}~ \mathit{h}^{-1}$, with an initial MBH mass of $10^4\msun~\mathit{h}^{-1}$. MBHs can grow through mergers with other MBHs that occur instantaneously following the merger of their host galaxies.\footnote{We note that \citet{2024MNRAS.528.1053C} presented an extension to \shark including the dynamics of MBHs. However, this is not yet available on the public branch of \shark.} MBHs can also grow by gas accretion following central starbursts and gas inflow from instabilities generated in the hot haloes surrounding galaxies. Central starbursts are triggered by accretion during galaxy mergers and disc instabilities that drive gas to the bulge. The mass accretion during starbursts follows the phenomenological description of \cite{2000MNRAS.311..576K} with a free parameter calibrated to reproduce the MBH-bulge mass relation at $z=0$. The hot halo accretion mode assumes an unboosted BHL accretion rate and approximating the gas density as in \cite{2006MNRAS.365...11C}. Only the radio mode of AGN feedback is included, via a radiative efficiency of 0.1 that acts by modifying the cooling timescale of the gas in the galaxy. The accretion rate is not limited to the Eddington accretion rate but only a small fraction of the MBHs surpass this value and only at high redshift \citep{2019MNRAS.485.2694A}. The bolometric AGN luminosity is calculated by assuming a radiative efficiency of 0.1 and with the total MBH accretion rate, which is a contribution from accretion due to starbursts and gas inflow from instabilities, as well as accretion in the hot halo mode.

Since MBHs are assumed to merge instantaneously following the merger of their host galaxies, the SHARK MBH catalogue includes one MBH per galaxy, obtained from those hosted in the merging galaxies. The separation of MBHs before their coalescence is set equal to the host-galaxies separation in the snapshot prior to their numerical merger. 

\shark has been successful in reproducing several observed relations, such as the optical colour bimodality of galaxies and its dependence on stellar mass \citep{2020MNRAS.497.3026B}; the atomic hydrogen (H I)–halo mass relation \citep{2021MNRAS.506.4893C}; the panchromatic emission of galaxies from the far-ultraviolet (FUV) to the far-infrared (FIR) across cosmic time \citep{2020MNRAS.497.3026B,2023MNRAS.518.1378C}; and the redshift distribution of bright FIR galaxies and the redshift evolution of their number density \citep{2019MNRAS.489.4196L,2024ApJ...970...68L}.

\subsubsection{DELPHI}

{\sc Delphi} ({\bf D}ark Matter and the {\bf e}mergence of ga{\bf l}axies in the e{\bf p}oc{\bf h} of re{\bf i}onization) is a semi-analytical model that uses a binary merger tree approach to jointly track the build-up of DM haloes, their baryonic components (gas, stellar, dust, and metal masses) and MBHs at $z \geqslant 4.5$ \citep{2014MNRAS.445.2545D,2019MNRAS.486.2336D, 2022MNRAS.510.5661P, 2024arXiv240111242D}.

In brief, we follow the assembly of DM haloes between $\log_{10}(M_{\rm h}/\rm M_\odot)=8$--14 from $z \sim 40$ down to $z = 4.5$ with a mass resolution of $10^8\msun$. At any time-step, the available gas mass (from both mergers and accretion) can form stars with an ``effective'' star formation efficiency  - this is the minimum between the efficiency that produces enough type II SN (SNII) energy to eject the remainder of the gas and an upper maximum threshold ($f_*$); each SNII is assumed to produce $10^{51}$ erg of energy. The two free parameters concerning star formation are $f_* \sim 10\%$ and the fraction of SNII energy that can couple to the gas ($f_{\rm w} \sim 7.5\%$) - these are obtained by matching to the faint and bright ends of the star-forming Lyman Break Galaxy (LBG) UV luminosity function and stellar mass function at $z \sim 5-9$, respectively \citep[e.g.,][]{2022MNRAS.512..989D,2023MNRAS.526.2196M}. While the details of MBH physics - growth from both Eddington-limited accretion and instantaneous mergers - and the associated feedback, with a radiative efficiency of $\epsilon_r = 10\%$) remain the same as in our previous works \citep{2019MNRAS.486.2336D}, we have included a number of key new ingredients required to match the AGN populations being observed by the JWST: we seed the starting haloes of any merger tree at $z \geq 13$ with heavy seeds of masses randomly sampled in the range $10^{3-5}\msun$. Our model also includes a ``critical'' halo mass for efficient MBH accretion with a value that evolves with redshift as $M_{\rm bh}^{\rm crit}(z) = 10^{11.25}[\Omega_{\rm m}(1+z)^3 +\Omega_\Lambda ]^{-0.125}$ on which we include a scatter of 0.5 dex, motivated by the results of cosmological simulations \citep[e.g.,][]{2017MNRAS.465...32B}. In order to explain the number density of JWST-detected AGN, at any time-step black holes are allowed to accrete a gas mass of $M_{\rm bh}^{\rm crit}(z)/{\rm M}_{\odot} = \min[\epsilon_r f_{\rm bh}^{\rm ac}  M_{\rm g}^{\rm sf}, f_{\rm Edd}~ m_{\rm Edd} ]$, where $M_{\rm g}^{\rm sf}$ is the gas mass left after star formation and its associated SNII feedback, 
and $f_{\rm Edd}~ m_{\rm Edd}$ is the black hole mass accreted in a given time-step. Allowing very weak AGN feedback ($0.01\%$ of MBH luminosity being coupled to the gas), we require values of $f_{\rm bh}^{\rm ac} = 0.1 ~ (5 \times 10^{-4})$ and $f_{\rm Edd} = 1.0 ~ (10^{-4})$ for haloes above (below) the critical mass (we allow 0.5 dex of scatter on all of these quantities). This results in MBHs in high-mass (low-mass) haloes accreting the minimum between 10\% ($0.05\%$) of the available gas mass and $100\% ~(0.01\%)$ of the Eddington fraction. We include the impact of dust attenuation to calculate the UV luminosity from both the star-forming and AGN components that are calibrated by matching to observations. 

In the {\it fiducial} model presented here, MBHs merge as soon as their host haloes merge -- i.e., any halo hosts a single BH at any given time. The separation of MBHs before their coalescence is set to the gas radius ($\sim 0.18$ of the virial radius) of the merged halo in the snapshot under consideration. Finally, the time-step of \delphi{} is fixed to 30~Myr. 

This model has been used to study the GW event rates from the high-redshift Universe \citep{2019MNRAS.486.2336D}, the mass assembly of early MBHs \citep{2021MNRAS.500.2146P} and their impact on the total UV luminosity of early systems \citep{2022MNRAS.510.5661P}, the contribution of MBHs to the reionization process \citep{2020MNRAS.495.3065D,2024arXiv240111242D} and been pushed to its extreme limits to explain the enormous MBH-to-stellar mass ratios being observed with the JWST \citep{2024Natur.628...57F}.

\subsection{Hydrodynamical simulations}\label{sec:SIMU}

In the following subsections, we review each of the cosmological hydrodynamical simulations used in the catalogue. Each setup inevitably has its own specifications, including volume size, mass and spatial resolution, cosmology, as well as differing subgrid prescriptions for MBH seeding, accretion, and feedback. In the following subsections we briefly describe each of the models while also outlining how the MBHs from each of the simulation suites are extracted and used in this analysis.

\subsubsection{Horizon-AGN and NewHorizon}\label{Horizon}
The Horizon suite\footnote{https://www.horizon-simulation.org/} \citep{2015MNRAS.452.1502D} provides two hydrodynamical cosmological simulations: the \horizonAGN \citep{2014MNRAS.440.2333D} and the \newHorizon\footnote{https://new.horizon-simulation.org/} \citep[][]{2021A&A...651A.109D}. 

\horizonAGN \citep{2016MNRAS.463.3948D} simulates a comoving volume of $(100/h)^3$~cMpc$^3$, whereas in \newHorizon \citep{2021A&A...651A.109D} a smaller volume (a sphere of radius of approximately 6.7 cMpc/$h$) is simulated with very high resolution. These simulations employ the adaptive mesh refinement (AMR) code RAMSES \citep{2002A&A...385..337T}, wherein mesh refinement is triggered when the cell mass exceeds 8 times the initial mass resolution. Additionally, for the \newHorizon simulation, another refinement step is triggered when the cell length falls below a Jeans length and the gas number density is greater than 5 H cm$^{-3}$.

\horizonAGN has a mass resolution of $8 \times 10^7\msun$ for DM particles and $2\times 10^6\msun$ for star particles. Starting from a coarse grid of 1024$^3$ cells, the spatial resolution is refined using quasi-Lagrangian triggering to a proper value of 1 kpc physical. Gas cooling is modelled down to 10$^4$~K using curves from \citet{1993ApJS...88..253S} and a uniform UV background heating is invoked after redshift $z_{\rm reion} = 10$ \citep{1996ApJ...461...20H}. The gas obeys the ideal monoatomic gas equation of state with an adiabatic index of $\gamma_{\rm ad}$ = 5/3. For this simulation, star formation follows the Schmidt relation, with a constant efficiency of $\epsilon_{\ast} = 0.02$ where the gas hydrogen density threshold is greater than  0.1~H\,cm$^{-3}$ following a Poisson random process. The simulation also includes feedback from stellar winds and Type Ia and Type II SNae, assuming a Salpeter initial mass function with cutoffs at 0.1 and 100 $\msun$. 

\newHorizon simulates a smaller ``average'' spherical volume of 6.7 cMpc/$h$ comoving chosen within the box of \horizonAGN. This smaller volume allows for sampling multiple haloes with a finer grid of 4096$^3$ cells. This simulation reaches a maximum resolution of 34 proper pc for the densest regions. The mass resolution is $1.2 \times 10^6\msun$ for DM particles and $10^4\msun$ for star particles.

Both simulations include a model for the growth of MBHs and their feedback on the surrounding gas. The model for MBH seeding is based on local properties as follows. MBHs of $10^{5}\msun$ (\horizonAGN) or $10^{4}\msun$ (\newHorizon)  are created in cells where the gas density is larger than the density threshold for star formation \citep{2014MNRAS.440.2333D, 2016MNRAS.463.3948D}, and where the gas velocity dispersion is larger than $100\, \rm km\,s^{-1}$ in Horizon-AGN and $20\,\rm km\,s^{-1}$ in \newHorizon; an exclusion radius of 50 comoving ckpc is imposed to avoid formation of multiple MBHs in the same galaxy. MBH formation in \horizonAGN is stopped at $z = 1.5$, while no limit has been imposed \newHorizon. The model for MBH accretion is based on the BHL formalism, boosted à la \citet{2009MNRAS.398...53B} in Horizon-AGN to account for the low resolution and unboosted (i.e., $\alpha = 1$) in New-Horizon. The rate is capped at the Eddington luminosity with a radiative efficiency of 10\%  in \horizonAGN, while the radiative efficiency is spin-dependent in \newHorizon, since the simulation evolves MBH spin on the fly.

AGN feedback is implemented as a dual-mode feedback with isotropic thermal energy injection at high Eddington ratios and bipolar kinetic energy injection at low Eddington ratios, with the threshold between the two set at $f_{\rm Edd} = 0.01$. In Horizon-AGN, the feedback efficiencies are set at 15\% and 100\% for the two modes, while in \newHorizon the jet efficiency is spin-dependent, mimicking the Blandford-Znajek process. The model for MBH dynamics includes an explicit drag force from gas, and the criterion for MBH mergers is based on the distance between two BHs being less than 4 resolution elements.

The catalogues generated from the \horizonAGN and \newHorizon datasets are taken from the simulation outputs. Full snapshots in \newHorizon are outputted every $\sim 15$ Myr, while in \horizonAGN full snapshots occur every $\sim 150$ Myr. MBH information is outputted more frequently - every coarse time-step and so (i.e., approximately 0.5~Myr). We summarise some catalogue-specific details below but direct the reader to \citet{2022MNRAS.514..640V, 2016MNRAS.463.3948D, 2021A&A...651A.109D} for more details.

For the full MBH population, catalogues include a single MBH per galaxy, with the most massive MBH selected being the MBH that is within $2 \times R_{\rm eff}$ of the host galaxy, where $R_{\rm eff}$ is the half-mass 2D radius. MBH and galaxy properties are obtained from snapshots where both galaxy and MBH properties are available at the same time and it is this information that is used in the MBH catalogues for \horizonAGN and \newHorizon.

For the merging population, MBH mergers are identified at coarse time-steps and the host galaxy is defined as the galaxy hosting the remnant at the first available snapshot after the MBH merger. The separation of merging MBHs is calculated at the (coarse) time-step previous to the merger and it is almost always 4 resolution elements (see Figure \ref{fig:Separation}, where the separation distance is very close to 4 resolution elements with a narrow dispersion in physical space). Additional wandering MBHs are excluded, since they are generally too faint to be detected. MBH mergers include all merging MBHs within $2 \times R_{\rm eff}$ of the host galaxy. This removes ``spurious'' off-centre mergers occurring because of the limited resolution. Such off-centre mergers in low-density regions would not lead to an actual MBH merger. Both \horizonAGN and \newHorizon also provide information on MBH merger delays, described and analyzed in Section~\ref{ModelsWithDelays}. 

\subsubsection{\textsc{Obelisk}}\label{subsubsec:Obelisk}

\textsc{Obelisk}\footnote{https://obelisk-simulation.github.io/} \citep{2021A&A...653A.154T} is a zoom-in cosmological simulation, which follows the evolution down to $z\sim3.5$ of an overdense region of the \horizonAGN simulation (see Section~\ref{Horizon}). The initial conditions of the zoom are taken from all particles within $4R_\mathrm{vir}$ of the most massive halo of \horizonAGN at $z\sim2$, a halo that  reaches a mass of $M_\mathrm{vir}\simeq2.5\times10^{13}\,M_\odot$ at $z=0$. The convex hull enclosing all these particles in the initial conditions is defined as the high-resolution region, with a volume of approximately $10^4$ cMpc$^3$. This region is re-sampled with a DM mass resolution of $1.2\times10^6\,M_\odot$. The rest of the $100\,h^{-1}\,\mathrm{cMpc}$ \horizonAGN box is maintained at a lower resolution.

\textsc{Obelisk} was run with the radiative transfer hydrodynamical code \textsc{Ramses-RT} \citep{2013MNRAS.436.2188R,2015MNRAS.449.4380R}, based on the AMR \textsc{Ramses} code \citep{2002A&A...385..337T}. If the mass of a cell exceeds $8$ times the mass resolution, the cell is refined, up to a maximum resolution of $\Delta x\simeq35\,\mathrm{pc}$ physical. Stars are modelled as $10^4\, \rm M_\odot$ particles representing a stellar population following a \citet{2001MNRAS.322..231K} initial mass function. SN explosions take place $5\,\mathrm{Myr}$ after the birth of a star particle and release $10^{51}\,\mathrm{erg}$ according to the implementation by \citet{2014ApJ...788..121K}.

The MBHs are seeded with an initial mass of $3\times10^4\msun$ if for a given cell both the gas and stellar density exceed $100\,\mathrm{H}\,\mathrm{cm}^{-3}$, the gas is Jeans unstable, and there are no pre-existing MBH at a distance smaller than $50\,\mathrm{ckpc}$.

The MBHs can grow by accreting gas. This is modelled as unboosted BHL accretion. The accretion rate is capped at the Eddington rate.  Radiative efficiency is spin-dependent, and further if $f_\mathrm{Edd}<f_{\mathrm{Edd,crit}} = 0.01$, the accretion is modelled as radiatively inefficient, and the radiative efficiency is reduced further by a factor $f_\mathrm{Edd}/f_{\mathrm{Edd,crit}}$.

The energy accreted by the MBH is released in part according to a dual-mode AGN feedback model. The `radio mode' is assumed at low accretion rates $f_\mathrm{Edd}<f_{\mathrm{Edd,crit}}$. In this regime, a fraction $\varepsilon_\mathrm{MCAF}$ of the rest-mass accreted energy is released as kinetic energy in jets in the direction of the MBH angular momentum. $\varepsilon_\mathrm{MCAF}$ is a fit to the simulations of Magnetically Choked Accretion Flows by \citet{2012MNRAS.423.3083M}. The `quasar mode' is assumed in the high-accretion rate regime $f_\mathrm{Edd}\geq f_{\mathrm{Edd,crit}}$. In this case, $15\%$ of the accretion luminosity is injected isotropically as thermal energy.

The MBHs can also grow by MBH-MBH mergers. Two MBHs are numerically merged in the simulation if their separation is smaller than 4 resolution elements ($4\Delta x$), which is approximately 140 (physical) pc. As a result, the distribution of separations shown in Fig.~\ref{fig:Separation} has a narrow dispersion around 140 pc. To account for the unresolved dynamical evolution of the MBH pairs, the simulation includes sub-grid models to account for dynamical friction from both collisionless particles and gas \citep{2013MNRAS.428.2885D,2019MNRAS.486..101P}. The population of MBH mergers was studied in previous works \citep{2023A&A...673A.120D,2023A&A...676A...2D}.  These studies consider both the population of MBH mergers occurring numerically in the simulation and the population of `delayed' MBH mergers obtained after applying a semi-analytical model for sub-grid delays in post-processing \citep{2020MNRAS.498.2219V}

MBH spin is additionally tracked in the simulation. The spin-up by gas accretion at $f_\mathrm{Edd}<f_{\mathrm{Edd,crit}}$ is modelled with the \citet{1970Natur.226...64B} formula, assuming that the gas angular momentum is conserved from the resolved scales down to the innermost stable circular orbit. At a lower accretion rate, the MBHs are spun down, since the spin rotational energy is assumed to power the jets, following fits to \citet{2012MNRAS.423.3083M}. Following a merger, the spin of the remnant MBH is updated according to the fitting formulae from \citet{2008PhRvD..78d4002R}.

The catalogues generated from the \textsc{Obelisk} datasets are taken from the simulation outputs. The information for the full MBH population is obtained from snapshots where both galaxy and MBH properties are recorded at the same time (every 15~Myr). Masses and separations of merging MBHs are obtained from MBH outputs which occur at each coarse time-step (approximately 0.1~Myr apart). Although galaxies in \textsc{Obelisk} can contain multiple MBHs, the catalogue of the full MBH and AGN populations that we use in this paper includes one MBH per galaxy, which is the most massive MBH within the 3D half-mass radius of a galaxy. MBH mergers that occur at larger distances from galaxy centers are considered spurious. The catalogues are built with the same procedure as described for \horizonAGN and \newHorizon. 

\indent Since \obelisk simulates an overdensity, the number density of each MBH/binary has been rescaled to mimic a ``normal'' region. To obtain the cosmic average, we have rescaled the \obelisk{} galaxy mass function to that of the \textsc{NewHorizon} simulation, which simulates an average region of the Universe. Delays are  also included for \textsc{Obelisk}, described and analyzed in Section~\ref{ModelsWithDelays}.

\subsubsection{Romulus}\label{subsubsec:Romulus}

Romulus25 \citep[][]{2017MNRAS.470.1121T} is a uniform-volume cosmological $N$-body, smoothed particle hydrodynamics (SPH) simulation run using the \textsc{ChaNGa} code \citep{2015ComAC...2....1M}. \textsc{ChaNGa} is a Tree-SPH code built using physics modules previously used in the \textsc{Gasoline} code \citep{2004NewA....9..137W}. The simulations used in this catalogue were run in a volume of (25~cMpc)$^3$ down to $z = 0$.  The initial number of particles is $1152^3$ for DM and $768^3$ for gas (note that the ratio of DM to gas particles is larger than unity to decrease numerical noise), yielding a mass resolution of $m_{\rm DM} = 3.39 \times 10^5$~M$_{\odot}$ and $m_{\rm gas} = 2.12 \times 10^5$~M$_{\odot}$ and a minimum resolvable halo (galaxy) mass of $3.39 \times 10^7$~M$_{\odot}$ ($2.12 \times 10^7$~M$_{\odot}$), assuming a minimum of 100 particles per system. The spatial resolution for all particles (gas, stars, and DM) is given, for gravity, by a spline force softening of 350~pc (physical) [i.e., a Plummer equivalent force softening of 250~pc (physical)] for $z < 8$ and $3.15/(1+z)$~kpc (physical) for $z \geq 8$ and, for hydrodynamics, by a minimum smoothing length that is 20 per cent of the spline force softening value. Star formation and stellar feedback are modelled following the recipes of \citet{2006MNRAS.373.1074S}.

MBHs are seeded, with a mass of $10^6$~M$_{\odot}$, by converting gas particles that have already been selected to form a star and that additionally have very low metallicity ($Z_{\rm gas} < 3 \times 10^{-4}$), a density at least 15 times the threshold for star formation ($\rho_{\rm gas} > 3 \, m_{\rm H}$~cm$^{-3}$), and have a temperature $9500 < T_{\rm gas}/{\rm K} < 10000$.

MBHs can grow via mergers and gas accretion. Gas accretion is modelled with an Eddington-capped BHL recipe, which has been modified with a density-dependent boost \citep[][]{2009MNRAS.398...53B} and with a formalism that accounts for the angular momentum of the gas. Part of the rest energy of the accreting mass is then released and coupled isotropically to the thermal energy of the nearest gas particles, assuming a radiative efficiency of 10 per cent and a coupling efficiency of 2 per cent.

Two MBHs are allowed to merge if $|\Delta {\bf r}|$ is smaller than two softening lengths and if $\Delta {\bf v}/2 < \Delta {\bf a} \cdot \Delta {\bf r}$, where $\Delta {\bf r}$, $\Delta {\bf v}$, and $\Delta {\bf a}$ are the relative distance, velocity, and acceleration, respectively, between the two MBHs, i.e., if they are close enough and if they are gravitationally bound to each other \citep[][]{2011ApJ...742...13B}. 

The dynamics of MBHs is modelled with an additional subgrid model that mimics the effects of unresolved dynamical friction from DM and stars, following the recipe of \citet{2015MNRAS.451.1868T}. Not every galaxy merger results in a MBH merger and many galaxy mergers result in MBH pairs that take Gyrs to merge, or fail to merge entirely and create a substantial population of wandering MBHs \citep{2018MNRAS.475.4967T, 2018ApJ...857L..22T}. The most massive haloes in the simulation, therefore, have several hundreds of MBHs within them \citep{2021MNRAS.503.6098R}. \textsc{Romulus} outputs snapshot information every 10--30~Myr at high redshift and every 100--300~Myr at low redshift, resulting in an average of about 100 Myr between outputs (there are 124 snapshots saved in total between $z = 0$ and 20).

The catalogues for \textsc{Romulus} are built as follows: the host galaxies of the MBHs are determined based on outputs from the amiga halo finder \citep[AHF,][]{2009ApJS..182..608K}, which determines the halo that each particle (including MBHs) is bound to. For subhaloes, the lowest level host halo is used, meaning that an MBH would be considered a member of a satellite galaxy, rather than that satellite's host halo. The current mass, accretion rate, and position of every MBH is saved at 1.5 Myr intervals, including at the time of each saved snapshot of the simulation. Merger events are collected in their own file as they occur in the simulation, which includes the properties of each MBH at the time of the (numerical) merger. The final separation of merging MBHs is taken at the last saved step prior to the numerical merger, with a maximum of 0.7 kpc (physical), the maximum allowed separation for a merger to take place. Host haloes of mergers are determined based on the remnant MBH at the next available snapshot following the numerical merger event.

Given that minor galaxy mergers can result in wandering MBHs, massive galaxies can have dozens or even hundreds of MBHs. For the full MBH population, the catalogues include only the most luminous MBH in each halo. Because MBHs are allowed to form based on local gas properties, spurious merger events can occur between new MBHs forming very close together. To avoid counting these in the catalogue of MBH mergers, only mergers between MBHs that have existed for 10 Myr prior to the merger event and that start at an initial separation greater than 1.4 kpc (or 4 times the physical gravitational softening length) are included. The merger separations shown in Figure \ref{fig:Separation} show mergers occuring close to but less than 0.7 (physical) kpc. We also exclude any MBH which failed to attain its target initial mass of $10^6\msun$.  No limits are placed on the location of the mergers. The host halo/galaxy is identified as the host post merger.

\subsubsection{Illustris}\label{subsubsec:Illustris}

The \illustris cosmological simulation suite \citep{2014Natur.509..177V, 2014MNRAS.444.1518V, 2014MNRAS.445..175G, 2015MNRAS.452..575S, 2015A&C....13...12N} uses the moving-mesh hydrodynamics code \arepo{} \citep{2010MNRAS.401..791S, 2011MNRAS.418.1392P, 2016MNRAS.455.1134P}. The primary simulation, as used in this catalogue, covers a volume of 106.5 cMpc on the side with $(1820)^3$ gas cells (initially) and $(1820)^3$ DM particles. \illustris has a baryonic and DM mass resolution of $1.26\times10^6\, {\rm M_{\odot}}$ and $6.26\times10^6\, {\rm M_{\odot}}$, respectively.

The gravitational and collisionless baryonic particle softening is 1.4 ckpc for $z\geqslant 1$; at lower redshifts, the gravitational softening remains the same and for all particle types other than DM the softening is fixed to the $z = 1$ value, such that their softening length is 0.7~kpc at $z = 0-1$. Sub-grid models are included for galaxy formation processes including star formation, kinetic stellar feedback, gas cooling, chemical enrichment, and a two-phase interstellar medium, as described in \citet{2013MNRAS.436.3031V}. 

MBH seeds form with a mass of $M_{\rm seed}=1.42 \times 10^{5}\, \rm M_{\odot}$ in all haloes reaching a mass of $M_{\rm h}= 7.1 \times 10^{10}\, \rm M_{\odot}$ that do not already host an MBH. MBH accretion is modelled with a boosted BHL formalism ($\alpha=100$), and accretion is capped at the Eddington limit. AGN luminosities have been computed with a radiative efficiency of 0.2. AGN impact their surroundings via thermal (``quasar-mode"), kinetic (``radio-mode"), and radiative feedback. MBHs switch between thermal feedback at high Eddington ratios and kinetic feedback at low Eddington ratios, using the formalism of \citet{2007MNRAS.380..877S}. 

To prevent spurious motion of MBHs driven by numerical noise, MBHs are repositioned at every time-step at the potential minimum of their host halo. Two MBHs are considered to have merged when their separation falls within the MBH smoothing length, which is typically $\sim 1$--10~ckpc; no criterion is imposed on the MBH velocity. Full details of the Illustris sub-grid MBH models are given in \citet{2013MNRAS.436.3031V} and \citet{2015MNRAS.452..575S}. The snapshot output times in \illustris are pre-defined, such that \illustris has 133 snapshots between $z = 127$ and $z = 0$. 

The MBH catalogues for \illustris are primarily based on the MBH merger files \citep{2017MNRAS.464.3131K,2016MNRAS.456..961B}, rather than the galaxy merger trees\footnote{In the \illustris MBH merger files built with the ``illustris blackholes'' code, an incorrect mass is recorded for one of the MBHs in each merger. Here we have reconstructed the masses based on the ``details'' files, which record the masses and properties of local gas at each time-step in which each MBH particle is updated.  The details-files entries are identified immediately following the merger, and the remnant mass is used to reconstruct the missing MBH mass.}. Therefore, in principle, they correspond to every MBH merger in the simulation, which are tracked at each integration step ($\lesssim \ \textrm{Myr}$) instead of each snapshot. The separation at the time of numerical merger is set to be the largest smoothing length of the two constituent MBH particles.  It is possible to have multiple MBHs in a galaxy, though the dynamics scheme to reposition MBHs on the potential minimum tends to lead to efficient mergers once MBHs enter the same galaxy. The repositioning scheme can also result in rather large ``merger separations'' since the last recorded separation can be relatively large and up to more than 10 physical kpc in some case (see Fig.~\ref{fig:Separation}). All MBHs within a galaxy are included in the catalogue. The host halo/galaxy is identified in the snapshot immediately following the MBH merger. 

A small fraction of mergers in the MBH merger files appear to be spurious either because both MBH particles continue to be seen in the simulation, and/or because the mass of the `remnant' is inconsistent with the sum of component masses.  At other times, one of the MBH particles is no longer seen but the remnant mass is inconsistent. This situation is categorized as a type of `missing' merger.  This is also the case at times when MBH particles no longer appear in the simulation, while a merger event is not recorded.  The underlying cause, of both the spurious and missing mergers, is unclear but often seems to be due to acute errors in the output files, often coinciding with simulation restarts. To build the \illustris merger catalogue, every merger event is cross-checked with the snapshot data, host galaxy catalogues, and MBH details files to validate the merger events and identify missing events.

In the case of missing mergers, the galaxy merger trees are cross-referenced, and each MBH in the same galaxy and any merging galaxies are examined.  If a unique merger-companion candidate can be identified, it is assigned as the merger companion and recorded in the catalogue.  Candidates are identified based on instantaneous growths in mass, of an amount consistent with the missing MBH and at a time coincident with the missing merger.  When a companion (and remnant) MBH cannot be uniquely identified for a missing merger, the merger is not included in the catalogue.  Similarly, mergers that are inconsistent with snapshot or `details' data are also denoted as spurious and removed from the catalogue.

\subsubsection{IllustrisTNG}\label{IllustrisTNG}
\label{sec:tgn_model}

Like \illustris{}, the TNG suite was generated using the \arepo{} moving-mesh code, but with the addition of magnetic fields \citep{2011MNRAS.418.1392P, 2016MNRAS.455.1134P} and updated physics modules \citep{2017MNRAS.465.3291W, 2018MNRAS.473.4077P}. \tng{} includes 18 simulations in total that differ in the  size of the computational domain, the mass resolution, and some differences in the included physics \citep{2018MNRAS.477.1206N,2018MNRAS.475..624N,2018MNRAS.480.5113M,2018MNRAS.475..648P,2018MNRAS.475..676S}. There are three simulation volumes: $(51.7\,{\rm cMpc})^3$, $(110.7\,{\rm cMpc})^3$, and $(302.6\,{\rm cMpc})^3$, which are referred to as TNG50, TNG100, and TNG300, respectively. The highest-resolution TNG50 volume is made up of $2160^3$ gas elements (initially) and $2160^3$ DM particles, providing a resolution of $8.5 \times 10^4\msun$  for the baryonic component and $4.5 \times 10^5\msun$ for the DM component. 
The gravitational softening length for collisionless (DM and star) particles is 0.575 ckpc at $z\geq1$ in TNG50, and it is fixed to its $z=1$ proper value of 0.288 kpc at $z<1$. The minimum value of the adaptive gas gravitational softening length is 0.072 ckpc. The mass resolution for \tngm{} is $1.4\E{6} \msun{}$ and $7.5\E{6} \msun{}$ for baryons and DM respectively. The TNG100 gravitational softening length for collisionless particles is 1.48 ckpc at $z\geq1$ and 0.738 proper kpc at $z<1$, and the minimum gas softening is 0.19 ckpc. 
Finally, for \tngh{} \citep{2019ComAC...6....2N,2019MNRAS.490.3234N,2019MNRAS.490.3196P}, the masses are $1.1\E{7}\msun$ and $5.9\E{7} \msun$, the collisionless gravitational softening is 2.96 ckpc ($z\geq1$) and 1.48 kpc ($z<1$), and the minimum gas softening is 0.37 ckpc. Each \tng{} simulation volume records 100 complete snapshots equally spaced over their evolution. 

\tng{} forms MBHs with a seed mass of $1.2\E{6} \msun$ in any DM halo exceeding a mass of $7.4\E{10} \msun$ that does not already contain an MBH.  
Merging of MBHs in \tng{} operates using the same principle as \illustris: MBHs are merged when their separation falls within the MBH smoothing length. As in \illustris, this can lead to relatively large merger separations due to the repositioning scheme, as shown in Fig.~\ref{fig:Separation}. Note that the bimodality seen in the \tng{} separations is due to a different implementation of AGN feedback in \tng{}, as described below.

Accretion onto MBHs and feedback from MBHs are treated differently in \tng{} compared to \illustris \citep[see][for details]{2018MNRAS.475..648P}. In brief, the MBH seed masses are eight times higher, the accretion is unboosted (i.e., $\alpha = 1$), not computed on the same physical scale (i.e., Illustris's model computes the accretion from the MBH gas parent cell, while the TNG model uses a kernel-weighted average over neighboring cells), and the AGN feedback model is different (i.e., different energy injection rates in the radiative mode, for the low-accretion mode Illustris uses thermal hot bubbles displaced from the MBH position and the TNG model injects kinetic energy in a number of cells around the MBH in a random direction at each feedback event). Radiative efficiency is 0.2, as in \illustris. The new kinetic feedback mode of \tng{} is responsible for the bimodality in the MBH separations (Fig.~\ref{fig:Separation}). When a MBH enters this feedback mode (i.e., mostly when MBHs reach a mass threshold of $10^8\msun$), the decrease in central gas density causes an increase in the size of the region containing ``nearest-neighbor" gas cells around the MBH. The size of this region is used not only in calculating the 3BH accretion and feedback rates, but also for the MBH merger criterion, causing an abrupt increase in the pre-merger physical separation of merging MBHs when the primary MBH is in the kinetic AGN feedback mode.

As described for \illustris{} above, MBH mergers are extracted from the \textit{blackhole mergers and details} supplementary data catalogues, including the same data validation and correction.

\subsubsection{EAGLE}\label{sec:EAGLE}

The Evolution and Assembly of GaLaxies and their Environments (\eagle) simulations \citep{2015MNRAS.446..521S,2015MNRAS.450.1937C, 2016A&C....15...72M} is a suite of hydrodynamical, cosmological simulations that use the \textsc{Anarchy} SPH code \citep{2015MNRAS.454.2277S}, which is a modified versions of \textsc{Gadget}--3 \cite[last described in][]{2005MNRAS.364.1105S}. In this work, we use the (100 cMpc)$^3$ volume of the \eagle suite (RefL0100N1504), which has DM resolution of $m_\mathrm{DM} = 9.7 \times 10^6\msun$  and baryonic resolution of $m_\mathrm{baryons} = 1.8 \times 10^6\msun$. The Plummer--equivalent gravitational softening length is 2.66 ckpc for $z\geqslant 2.8$ and is limited to a maximum of 0.70 pkpc at lower redshifts.

In \eagle, MBH particles of $1.48 \times 10^5\msun$ are seeded in all haloes more massive than $1.48 \times 10^{10}\msun$ that do not already contain an MBH. Accretion of mass onto MBHs is modelled following an Eddington--limited and modified but unboosted (i.e., $\alpha = 1$) BHL prescription \citep{2015MNRAS.454.1038R} given by

\begin{flalign}\label{dotM_EAGLE}
\dot{M}_\mathrm{BH} = \mathrm{min} \left( \dot{M}_\mathrm{BHL} \times \mathrm{min} \left( (c_\mathrm{s}/ V_\phi)^3 / C_\mathrm{visc}, 1\right), \dot{M}_\mathrm{Edd} \right),
\end{flalign}

\noindent where $c_\mathrm{s}$ is the local sound speed, $V_\phi$ is the SPH-averaged circular velocity of the gas surrounding the black hole, and $C_\mathrm{visc}$ is a free parameter controlling the sub--grid viscosity, which and is set equal to $2\pi$.

The above accretion of mass onto the MBH results in AGN feedback, which in \eagle is modelled as thermal energy \citep{2009MNRAS.398...53B}. Gas particles surrounding the MBH are stochastically heated by increasing their temperature by $10^{8.5}$ K, every time enough energy to do so has been accumulated in the MBH's energy ``reservoir''. Energy is added to this reservoir at a rate given by $\epsilon_\mathrm{r}\epsilon_\mathrm{f}\dot{M}_\mathrm{BH}c^2$, where $\epsilon_\mathrm{r}=0.1$ and $\epsilon_\mathrm{f}=0.15$ describe, respectively, the radiative efficiency and the fraction of the AGN luminosity coupled to the surrounding gas.

Finally, since in \eagle dynamical friction is poorly captured for ``low-mass'' MBH particles, MBHs whose mass is less than $100 \times m_{\mathrm{baryons}}$ are moved to their local potential minimum if (i) their velocity with respect to their neighbouring particles is less than $0.25 \times c_\mathrm{s}$; and (ii) their relative distance is less than three gravitational softening lengths. In \eagle, MBHs merge together if (i) their relative separation is smaller than both one smoothing length and three gravitational softening lengths; and (ii) their relative velocity is smaller than the circular velocity evaluated using the most massive (of the pair) MBH's mass and a distance of one smoothing length. Fig.~\ref{fig:Separation} shows that merging in \eagle occurs at a well defined physical radius of approximately 2.1~pkpc (corresponding to about 3 gravitational smoothing lengths at low redshift).

For \eagle, the full MBH/AGN catalogues are retrieved from the SUBFIND galaxy catalogues, which are identified using the SUBFIND algorithm \citep{2001MNRAS.328..726S,2009MNRAS.399..497D} from the output snapshots. The snapshots in \eagle are unevenly spaced both in terms of redshift and the age of the universe. The output intervals are finer at higher redshifts ($\sim$100~Myr at $z \ga 4$). At lower redshifts ($z \la 4$), the output interval gradually increases from $\sim$200~Myr to $\sim$1~Gyr.

The MBH mass and AGN luminosity are the sum of the masses and luminosities of all MBHs hosted in a given galaxy.  Galaxies with DM fractions with $f_{\rm dm} \equiv M_{\rm dm} / M_{\rm tot} \leq 0.2$ are excluded from the catalogue, as such galaxies are either tidally stripping galaxies or artifacts of the substructure finder. Galaxy masses are given as stellar mass within twice the half-mass radius.

The MBHB information is obtained from the MBH detailed files saved during the simulation. Note that these MBH detailed files, saved at every time-step when an MBH is active (i.e., they have moved since the last output), have a much higher time resolution than the previously mentioned snapshot and galaxy catalogues. Specifically, when two MBHs merge in the simulation, the scale factor of the merger event, along with the IDs and masses of the merging MBHs, are recorded in these files. However, the coordinates of the merging MBHs are not saved in the detailed files. Therefore, we estimate the final separation of a MBHB in this study by using three gravitational softening lengths in physical units, which is one of the merger separation criteria adopted in \eagle.

The host galaxies of the MBHBs are identified from the closest snapshot after the MBH merger. This approach is influenced by the snapshot output interval, which ranges from approximately 200 Myr to 1 Gyr. In some instances, no associated host galaxies can be identified for a MBHB, in which case the host galaxies are labeled as -1 in the catalogue. 

\subsubsection{FLARES}

The First Light And Reionization Epoch Simulations \citep[\flares,][]{2021MNRAS.500.2127L,2021MNRAS.501.3289V} are a suite of 40 zoom re-simulations targeting galaxy formation at the Epoch of Reionisation ($z > 5$). Each of the 40 spherical regions has a radius of 20.66~cMpc and has been selected from the (3.2 cGpc)$^3$ DM-only simulation of the \textsc{C--Eagle} project \citep{2017MNRAS.471.1088B}. 

These regions, when re-simulated with the \eagle astrophysical model, provide galaxies residing in a range of overdensities and collectively capture both the first massive galaxies in the Universe and the environmental impact on galaxy formation. Hence, \flares uses identical simulation parameters (e.g., SPH implementation, gravity solver parameters) to the \eagle simulations described in Section~\ref{sec:EAGLE}. \textsc{Flares} uses an identical resolution to the fiducial \eagle simulation, with gas particle mass $m_{\rm gas} = 1.8 \times 10^6\msun$  and a softening length of 2.66 ckpc. 

The only difference between \flares and \eagle is that in \flares the AGNdT9 variant of the \textsc{Eagle} model is used \citep{2015MNRAS.446..521S,2015MNRAS.450.1937C}, which affects the MBH physics implementation. In this variant, the $C_\mathrm{visc}$ factor in Eq.~\eqref{dotM_EAGLE} is set to $2\pi \times 10^2$ (i.e., much slower MBH growth). In addition, the temperature given to the surrounding gas particles -- reflecting the AGN feedback -- is increased to 10$^9$ K (i.e., feedback episodes are more intense but less frequent). Apart from these two differences, in \flares MBHs are seeded, repositioned, and merged as described for \eagle in Section~\ref{sec:EAGLE}.

The catalogues for this project have been created using the same procedure as for the \eagle catalogues, except that the number density of each MBH/binary has taken the weight of each \flares zoom region, $f_i$, into account. This weight is required to reproduce the correct distribution of those overdensities averaged over the whole universe \citep[see][for details]{2021MNRAS.500.2127L}. Thus, the `number density' used to weight the data in our catalogues is computed as $f_i \times 1 / V_{\rm zoom}$ for an MBHB in the $i$-th region. Otherwise the procedure for computing the MBH catalogue information is identical to that of \eagle. One additional characteristic of \flares is that, since \flares is a high-redshift ($z \ge 5$) zoom simulation, mergers occur at smaller physical separations compared to \eagle (see Fig.~\ref{fig:Separation}), since the criteria for merging are based on comoving separations which are smaller physically at high redshift. Hence, in Fig.~\ref{fig:Separation}, there is an offset between the \eagle simulation and the \flares simulation results.  

\subsubsection{MassiveBlack-II}

The \textsc{MassiveBlack-II} simulation \citep{2015MNRAS.450.1349K, 2015MNRAS.454..913D, 2014MNRAS.441..470T} is a large-volume simulation evolved to $z=0.06$, which was run as a followup to the \textsc{MassiveBlack} simulation \citep{2012ApJ...745L..29D, 2012MNRAS.423.2397K, 2012MNRAS.424.1892D}.  The simulation was run using \textsc{P-GADGET}, an upgraded version of \textsc{Gadget}--3 \citep{2005MNRAS.364.1105S}.  The simulation covers a comoving volume of (100~$h^{-1}$ cMpc$)^3$, with $2 \times 1792^3$ particles, with a DM mass resolution of $m_{\rm{DM}} = 10^7 h^{-1}\msun$, gas mass resolution of $m_{\rm gas} = 2.2 \times 10^6 h^{-1}\msun$, and a comoving gravitational softening length of $\epsilon = 1.85 h^{-1}$ ckpc. 

MBHs are seeded in \textsc{MassiveBlack-II} with a mass of ${\rm M_{seed}}=5 \times 10^5 h^{-1}\msun$, in any halo with mass above $5 \times 10^{10} h^{-1}\msun$  that does not already contain an MBH.  These MBHs grow by gas accretion and merging with other MBHs.  Gas accretion is modelled by an BHL formalism without a Bondi boost, with an imposed cap of two times the Eddington limit to allow for mildly super-Eddington accretion.  While accreting, the MBHs radiate with a fixed radiative efficiency of 0.1 (i.e., $L = 0.1 \, \dot{M}_{\rm{BH}} c^2$).  Feedback is modelled by thermally coupling five per cent of the radiated energy to the surrounding gas, accomplished by isotropically depositing energy on the gas particles which lie within the MBH kernel.

In \textsc{MassiveBlack-II}, MBHs are re-positioned at the local gravitational minimum at each time-step to prevent spurious dynamical movement due to unresolved frictional forces. MBHs merge together when one MBH is within the kernel of another MBH, as long as their relative velocity is below the sound speed of the local gas (to prevent spurious mergers of flyby MBHs). \textsc{MassiveBlack-II} had a total of 85 snapshots down to $z = 0.06$.

The catalogues for \textsc{MassiveBlack-II} include all MBHs from the simulation. A single galaxy is assigned all the MBHs it contains, each MBH representing an individual entry. For mergers, MBH position data is saved at every time-step and outputted to an MBH file from which the catalogue information is obtained. The separation for any numerical mergers is defined as the distance between the two black holes at the last time-step prior to the merger event itself. Typically, this will be of order the 2--3 ckpc (see Fig.~\ref{fig:Separation}, where the separations for \massiveblack show a spread around 1 (physical) kpc reflecting the relatively low redshift of mergers in \massiveblack). The post-merger galaxy is assigned to be the one where the remnant MBH is found in the snapshot immediately after the merger.

\subsubsection{Astrid}

\texttt{Astrid} \citep{2022MNRAS.512.3703B} is a large-volume cosmological hydrodynamical simulation performed using a new version of the SPH code \textsc{MP-Gadget} \citep{2018zndo...1451799F}. The simulation evolves a cube of 250~$h^{-1}$~cMpc per side with $2\times5500^3$ initial tracer particles comprising DM and baryons using the Planck20 cosmology. \texttt{Astrid} has a DM particle mass resolution of $M_{\rm DM} = 6.7 \times 10^6~h^{-1}\msun$ and $M_{\rm gas} = 1.3 \times 10^6~h^{-1}\msun$ in the initial conditions. The gravitational softening length is $\epsilon_{\rm g} = 1.5 {\rm \ ckpc/{\it h}}$ for both DM and gas particles. The simulation includes a full-physics sub-grid treatment for modelling galaxy formation, MBHs and their associated AGN feedback, as well as inhomogeneous hydrogen and helium reionization. We refer the reader\soutPC{s} to the introductory papers \cite{2022MNRAS.512.3703B, 2022MNRAS.513..67N} for detailed descriptions of physical models deployed in the simulation, and to \cite{2022MNRAS.514.2220C} for detailed descriptions of the MBH merger catalogue.

MBHs are seeded in haloes with $M_{\rm halo} > 5 \times 10^9 h^{-1}\msun$ and $M_{\rm \star} > 2 \times 10^6~h^{-1}\msun$ with seed masses stochastically drawn between $3\times10^{4}~h^{-1}\msun$ and $3\times10^{5} h^{-1}\msun$. The gas accretion rate onto the MBH is estimated via a BHL-like prescription without additional boosting factor. We allow for short periods of super-Eddington accretion in the simulation but limit the accretion rate to two times the Eddington accretion rate. The MBH radiates with a bolometric luminosity $L_{\rm bol}$ proportional to the MBH accretion rate, with a mass-to-energy conversion efficiency of $0.1$ in an accretion disc according to \cite{1973A&A....24..337S}. 5\% of the radiated energy is coupled to the surrounding gas as AGN feedback. The dynamics of the MBH is modelled with a sub-grid dynamical friction model \citep{2015MNRAS.451.1868T,2022MNRAS.514.2220C} to replace the original implementation that directly repositioned MBHs to the local minimum of the potential. This gives well-defined MBH trajectories and velocities. Per this implementation, two MBHs can merge if their separation is within two times the gravitational softening length $2\epsilon_g$, once their kinetic energy is dissipated by dynamical friction and they are gravitationally bound to the local gravitational potential. \astrid outputs snapshot information at a frequency of $\Delta z = 0.1$.

Additional to the full Astrid snapshots, MBH-specific information is outputted at every active MBH time-step (typically of order every 1 Myr). Following a merger, host galaxies are identified with subfind. Finally, typical merger separations in \astrid are of the order of 1--2~kpc (physical) -- see also Fig.~\ref{fig:Separation}. The catalogues provided for \astrid include all MBHs in the simulation; multiple MBHs in a given galaxy are given as separate entries.  The simulation included a kinetic energy check for mergers, and separation for merging MBHs is calculated at the last time-step before the merger. 

\subsubsection{KETJU}\label{sec:Ketju}

The \ketju code \citep{2017ApJ...840...53R,2023MNRAS.524.4062M} extends the widely used \textsc{GADGET}--3 and \textsc{GADGET}--4 codes \citep{2005MNRAS.364.1105S,2021MNRAS.506.2871S} by introducing spherical regions around MBHs in which the integration of the MBHs and their surrounding stellar particles is performed with the algorithmically regularised integrator MSTAR \citep{2020MNRAS.492.4131R}. The code has been used to simulate both disc and early-type galaxy mergers \citep{2023MNRAS.520.4463L,2024MNRAS.528.5080L,2024MNRAS.530.4058L} and galaxies forming in cosmological zoom-in simulations \citep{2021ApJ...912L..20M,2022ApJ...929..167M}.

The simulation presented in \citet{2022ApJ...929..167M} 
includes a high-resolution region with a volume of (10~cMpc$/h)^3$ in a (100~cMpc$/h)^3$ box centred on a group-sized DM halo with a mass of $\rm{M_{200}}\sim 2.5 \times 10^{13}\msun$. The high-resolution region contains $2\times 410^{3}$ particles, evenly split between gas particles with initial masses of $m_\mathrm{gas}=3\times10^5\msun$ and DM particles with masses of $m_\mathrm{DM}=1.6\times 10^6\msun$.      

The simulation is initially evolved with the standard \textsc{GADGET}-3 code including MBH seeding (see below for details) and MBH repositioning until a redshift of $z = 0.815$, at which time the simulation volume contains tens of MBHs with masses in excess of $>7.5\times10^7\msun$. At this point, we switch on \ketju, as the MBH-to-stellar particle mass ratio is now sufficiently high to allow for accurate binary dynamics \citep{2021ApJ...912L..20M} and the simulation is further evolved until $z = 0.19$. In the \ketju simulation, the gravitational softening length is $\epsilon_\mathrm{DM}=93\ \mathrm{pc/h}$ for DM, $\epsilon_\mathrm{gas}=40\ \mathrm{pc/}h$ for gas particles and $\epsilon_\mathrm{star}=20\ \mathrm{pc/}h$ for stellar particles, resulting in a regularised \ketju region with a radius of $r_{\rm KETJU}=60\ \mathrm{pc/}h$ (all in physical units). The dynamics between MBHs and between MBHs and stellar particles is unsoftened, however stellar-stellar interactions are softened also inside the \ketju region. 

MBHs with masses of $10^5\msun$/$h$ are seeded into galaxies when the 
DM halo mass reaches $M_\mathrm{DM}=10^{10}\msun$/$h$, and then grow through gas accretion and merging. The MBHs accrete gas using the standard BHL model including $\alpha=25$ as a dimensionless multiplier to account for the limited spatial resolution \citep{2009ApJ...690..802J}. The accretion rate is capped by the Eddington limit and the radiative efficiency is set to $\epsilon_r=0.1$. A total of 0.5\% of the accreted rest mass energy is coupled to the surrounding gas as thermal energy \citep{2005MNRAS.361..776S}. The hydrodynamics is modelled using the SPHGal SPH implementation of \citet{2014MNRAS.443.1173H}. The model includes metal-dependent cooling tracks of 11 individual elements and stellar feedback from type Ia and type II SNae and the slow winds from asymptotic giant branch (AGB) stars \citep{2013MNRAS.434.3142A,2017MNRAS.468..751E}. Stars are stochastically formed above a critical hydrogen number density threshold of $n_{\rm H}=0.1 \ \rm cm^{-3}$. 

The \textsc{GADGET}-3 simulation uses repositioning for MBHs, fixing them to the local minimum of the gravitational potential at every timestep and the MBHs merge using the standard merger criterion, in which two MBHs merge when they are within the SPH kernel from each other and their relative velocity is smaller than half of the local sound speed. In the \ketju simulation the dynamics of MBHBs including three-body scattering and GW emission can be modelled to sub-pc separations. PN corrections are included in the interactions between MBHs to model relativistic effects, with terms up to 3.5PN included for MBHB systems and the leading order 1PN corrections included for general $N$-body systems. The final orbital evolution of the MBHB is driven by GW emission and the MBHB will be merged at a separation of $12G(M_{1}+M_{2})/c^{2}$, where $M_{1,2}$ are the individual MBH masses. \textsc{GADGET}-snapshots are generated at time intervals of $\sim 100\ \mathrm{Myr}$, while the \ketju output for MBHs is generated every $\sim 10\ \mathrm{kyr}$.

The \ketju catalogues include the MBHs above the \ketju integration mass limit ($>7.5\times10^7\msun$). For the MBH population catalogue, multiple MBHs per galaxy are allowed: if two MBHs above the \ketju mass limit reside in the same galaxy, both MBHs are included separately in the catalogue. For MBH mergers, we give the final separation saved in the \ketju output for MBHs, which typically is of the order of $\sim 10^{-3}\ \mathrm{pc}$. As \ketju resolves the dynamics of MBHBs down to sub-pc separations, these catalogues should be considered to already include delays. Therefore, in analysis that apply delays, the starting radius of the binaries is set to the stellar half-mass radius of the galaxy merger remnant (shown in Fig.~\ref{fig:Separation}) instead of the actual final separation, which is at sub-pc scales for \ketju. The properties of the post-merger galaxies are obtained from the first snapshot produced after the merger of the MBHs. 

\subsubsection{Renaissance}
\renaissance \citep{2015ApJ...807L..12O,2014ApJ...795..144C,2013ApJ...773...83X,2018MNRAS.480.3762S,2014ApJ...791..110X,2016ApJ...823..140X} is a simulation suite which utilised the \enzo \citep{2014ApJS..211...19B, 2019JOSS....4.1636B} code to perform high resolution simulation of early galaxies formation. \enzo is an structured AMR code with grids selectively refined based on user-inputted criteria (typically based on the Jeans criteria or particle/baryon overdensity criteria).
In particular, the \renaissance suite consists of three separate zoom-in regions each of approximately 6 cMpc $h^{-1}$ on the side. The zoom-in regions consist of the Rarepeak (RP) region, which is run to $z = 15$ and has dimensions of $(3.8 \times 5.4 \times 6.6)$ cMpc$^3$ $h^{-3}$; the Normal region, run to $z = 11.6$, and the Void region, run to $z = 8$. Both the Normal and Void regions have dimensions $(6.0 \times 6.0 \times 6.125)$ cMpc$^3$ $h^{-3}$. Each zoom-in region was extracted from a parent volume of 40 cMpc $h^{-1}$ on the side \citep{2015ApJ...807L..12O}. 
Only the Normal region is used in the analysis presented here. 

For the \renaissance suite, the DM particle resolution inside the zoom-in regions is $2.4 \times 10^4\msun$  and for the spatial resolution, the AMR capabilities of \enzo allowed a maximum spatial resolution of 19~cpc comoving (so approximately 1.5 physical pc at the times of interest). DM haloes of approximately $10^6\msun$ are resolved within \renaissance with a minimum of 32 particles required to define a halo as is standard in the literature. 

A subgrid model for modelling MBHs is not part of \renaissance. For this catalogue, the results from \cite{2024arXiv240916413M}, wherein the MBHs are modelled in post-processing, are used. In post-processing, the empirical scaling relation of \cite{2015ApJ...813...82R} and \cite{2023ApJ...957L...3P} are used to seed the MBHs. This results in MBH seed masses between $10^{3-5}\msun$.

Accretion onto the MBHs is modelled using a boosted BHL prescription for unresolved black hole masses (see below for details). A radiative efficiency of 0.1 is assumed. To model accretion, the MBH seeds are `positioned' at two distinct reference points in post-processing. In the simplest case, the MBHs are placed at the highest density point in the halo (in a methodology similar to that of MBH repositioning in large-scale cosmological simulations). MBH seeds are additionally assigned to the position of the nearest DM particle at the time of seeding. This allows for the calculation of the accretion rate onto a dynamically active (tracer) particle within the simulation. This latter methodology mimics wandering MBHs in (early) galactic centres.

Many of the black holes seeded are relatively low mass ($10^3 - 10^4$ M$_{\odot}$) and their BHL radius is not resolved. In order to account for this, accretion rates are boosted when the BHL radius is not resolved using the methodology set out in \cite{2009MNRAS.398...53B}. MBHs then grow by both accretion and mergers (see below) with other haloes' MBHs. Accretion rates are not capped but rarely exceed the Eddington rate.

No modelling of AGN feedback is taken into account in this analysis. Mergers between MBHs occur when the DM host haloes \citep[as determined using \rockstar,][]{2013ApJ...762..109B} 
merge. The actual timescale for mergers is determined in one of two ways. In the most simplistic case, the MBH mergers occur instantly at the same time the DM host halo merger occurs, whereas in the other case an explicit treatment of dynamical friction is considered \citep{2024arXiv240916413M}.

The \renaissance catalogues for both the full MBH population and the MBH mergers are built by post-processing the \textsc{Renaissance} datasets as described above. Full snapshot outputs in Renaissance occur with $\Delta z = 0.1$. No MBH specific output files are available. The \textsc{Renaissance} catalogue of MBHs and MBHBs include one MBH per galaxy. In the case of mergers, the initial separation between the two MBHs is defined as the minimum of (a) the virial radius of the resulting halo or (b) the distance between the positions of the two MBHs in the snapshot just before host merger. This relatively large separation distance is reflected in the spread seen in Fig.~\ref{fig:Separation}, where we see MBHs merging in \renaissance between approximately 2 (physical) kpc and 0.05 (physical) kpc.

For the catalogues, the MBHs are assumed to sit at the highest density point in the galaxy at all times, and  scaling relations from \cite{2023ApJ...957L...3P} are used to compute the initial MBH masses. Both of these assumptions lead to an optimistic growth and seeding prescription - see \citet{2024arXiv240916413M} for an alternative method, wherein the MBHs are allowed to dynamically wander in the halo and the scaling relations of \cite{2015ApJ...813...82R} are used. Finally, only information from the \textit{Normal} region of the \renaissance suite is provided for the catalogues. This is done so as to ensure that a non-biased region is used. The host galaxy of the merged MBHB is identified from the first snapshot post-merger. 

\subsubsection{SIMBA}

The \simba\footnote{There is a new available version of the \simba simulation, \texttt{Simba-C}, with improved chemical evolution and improved tracking of low-mass MBHs \citep{2023MNRAS.525.1061H}.} simulation suite~\citep{2019MNRAS.486.2827D} is run with the {\sc Gizmo} cosmological solver using its meshless finite mass (MFM) hydrodynamics scheme~\citep{2015MNRAS.450...53H}. 

\simba is run in a $(100 \, h^{-1} \, \rm cMpc)^3$ volume containing \(1024^3\) gas elements and \(1024^3\) DM particles, evolved from \(z = 249\) to \(z = 0\). It includes primordial and metal cooling; star formation using an $H_2$-based subgrid prescription; stellar evolution tracking of Type II+Ia SNae and AGB stars; and decoupled kinetic winds following prescriptions based on the {\sc FIRE} zoom simulations \citep{2015MNRAS.454.2691M, 2017MNRAS.470.4698A}. The minimum gravitational softening length is $0.5 \, h^{-1} \, \rm cMpc$, with a baryonic mass resolution (initial) of $1.81 \times 10^7 \, \msun$  and a DM mass resolution of $9.6 \times 10^7 \, \msun$.

In \simba, accretion is modeled via a two-phase approach, using gravitational torque-limited accretion~\citep{2011MNRAS.415.1027H,2017MNRAS.464.2840A} for cool gas ($T<10^5 \rm \, K$), and BHL accretion with $\alpha=0.1$ for hot gas ($T>10^5 \rm \, K$), with the corresponding accretion rates capped at $3 \times f_{\rm Edd}$ and $f_{\rm Edd}$, respectively. Most of the growth of MBHs occurs in the torque-limited mode, whereas the BHL mode becomes important for MBHs at late times. MBHs are seeded at $10^4 \, h^{-1} \, \msun$ in galaxies with stellar mass $M_{\star} \gtrsim 10^{9.5} \, \msun$, and merge instantaneously when they are within one smoothing length, which encloses 256 neighboring particles, of each other. MBHs are repositioned at every time-step to the location of the potential minimum within the FOF host group.

\indent MBH feedback occurs in three modes, and is mostly kinetic and bi-directional. At high Eddington ratios ($f_{\rm Edd}>0.2$), the radiative wind mode is assumed to have a wind speed of $\sim 1000 \, \rm km \, s^{-1}$ that rises slowly with the MBH mass following observations of $H\alpha$ outflows. At low Eddington ratios ($f_{\rm Edd}<0.02$), the jet mode has wind speeds that are capped at $\sim 10^4$~km~s$^{-1}$, with a smooth transition between the two regimes. The mass ejection rate is computed by assuming that the momentum output is always $20 \, L/c$, where $L=0.1 \, \dot{M}_{\rm BH}c^2$ is the luminosity with a radiative efficiency of $10\%$. These two modes of feedback are always outputted along the angular momentum axis, where the angular momentum is computed from the particles within the MBH kernel; this direction is typically stable over hundreds of Myr. Finally, there is a third mode, X-ray feedback following \citet{2012ApJ...754..125C}, which is applied spherically to particles within the MBH kernel for galaxies with low cold gas content, producing a small outwards kick or modest heating. This mode is only applied for galaxies with full velocity jets and with a gas fraction of $f_{\text{gas}} < 0.2$, where $f_{\text{gas}} = M_{\text{gas}} / M_{\star}$.

Generally, the jet mode is responsible for galaxy quenching and halo evacuation, the X-ray mode is responsible for clearing out the last remaining cold gas in massive galaxies, and the radiative winds have a minimal impact. For more information and physical justifications on all these modules, see \citet{2019MNRAS.486.2827D,2023MNRAS.525.1061H}.

\simba outputs include snapshots and {\sc CAESAR}\footnote{CAESAR is a python \texttt{yt}-extension to read simulations' outputs and produce HDF5 catalogs: \url{https://caesar.readthedocs.io/}.} catalogs, with the latter providing pre-computed galaxy properties derived from the snapshots at specific redshifts. Snapshots are generated at time intervals of $\sim 0.01 \rm \, Gyr$ at high redshifts to $0.2\, \rm Gyr$ at low redshifts. While these snapshots include particle lists that detail individual MBH particle properties, this information is also stored at every active MBH time-step in separate output files.

The full MBH population is characterized using galaxy properties from the {\sc CAESAR} catalogs and MBH properties from the \simba snapshots. This is achieved by cross-matching particle IDs of both central and satellite MBHs in the catalogs with the IDs of ‘Type 5’ or ‘boundary’ particles in the snapshots. Central MBHs are the most massive MBHs if multiple exist within a single galaxy. \simba does not contain MBHs prior to \(z \sim 8.7\).

The MBHB population in \simba is identified using the ‘bhMRG.hdf5’ file from the simulation's flagship run repository. This file contains detailed information on the IDs, masses, positions, and redshifts of MBH particles involved in a merger. This occurs when two MBH particles come within each other’s smoothing length, or within a maximum radius of $2 \, h^{-1} \, \rm ckpc$ for the MBH accretion kernel, and if their relative velocity is lower than three times their mutual escape velocity. Host galaxies of the post-merger remnant are determined by cross-matching MBH IDs from the merger file with those in the snapshot at the nearest redshift following the merger. Binary separations are calculated as the Euclidean distance between each merging MBH pair. Figure \ref{fig:Separation} shows the separation between MBHs in \simba. Separations peak at approximately $1.5 \rm \, ckpc$ with a small spread. 

\section{Methods}\label{sec:methods_coalescence}

\subsection{Binary formation and evolution}

MBHBs naturally form via galaxy mergers~\citep{1980Natur.287..307B}. Although the pairing and final coalescence spans 12 orders of magnitude in spatial scale, the physics governing the process can be broken into three characteristic phases  \citep{2005LRR.....8....8M, 2006PhRvD..73f4030B, 2013GWN.....6....4A,2023LRR....26....2A}. In the earliest phase, MBHs sink to the centre of the newly-merged galaxy by {\it dynamical friction} furnished by background DM, stars, and gas. Soon after forming a binary, an MBHB enters the {\it hardening phase}, in which energy and angular momentum loss in the orbit is driven by close few-body encounters with stars \citep[e.g.,][]{1996NewA....1...35Q,2001ApJ...563...34M,2012ApJ...744...74G,2012ApJ...756...30K,2022MNRAS.511.4753G} and, in the presence of gas, through interactions with a circumbinary disc \citep[e.g.,][]{2009MNRAS.393.1423C,2013ApJ...777L..14F,2014MNRAS.439.3476R,2014ApJ...780...84D,2015ApJ...811...59D,2015MNRAS.449..494R,2018MNRAS.480..439D,2020A&A...641A..64H,2020ApJ...901...25D,2021ApJ...914L..21D,2022ApJ...929L..13F}. The timescale of the hardening phase depends strongly on the detailed properties of the host galaxy, such as gas content, gas cooling, AGN feedback, galaxy shape, structure, and kinematics, as well as properties of the MBHB orbit, such as eccentricity and inclination between the initial galaxy orbits. This means that delay times can vary from Myrs to Gyrs from hardening alone~ \citep[e.g.,][]{2011ApJ...732...89K,2015ApJ...810...49V,2017MNRAS.464.2301G,2025ApJ...995L..32H}. Once the MBHB separation is of the order of a mpc, GWs drive the inspiral and final coalescence.

In this work, for simplicity, we neglect the hardening and GW phases when determining the timescale of a MBHB merger. This is motivated by the large limitations of both semi-analytical models and hydrodynamical simulations in resolving and evolving the detailed structure of the stars and gas surrounding the MBHB and in accurately modelling the accretion rate onto the MBHs. In addition, given the relatively small mass of the black holes explored here, the dynamical friction phase may arguably account for the largest portion of the MBHB coalescence timescale.

\subsection{Modelling of merger delays}\label{ModellingDelays}

Due to the resolution limit of the models included in this project, the MBHs are usually numerically ``merged'' at a scale determined by the spatial resolution of the simulation, typically of the order of a kpc for a cosmological simulation. In reality, the MBHs at this separation are not yet bound and would still need to navigate the three evolutionary phases described above before becoming a LISA source. 

In semi-analytical models, the MBH evolution from the merger of the host haloes or galaxies has been implemented using analytical timescales for the dynamical decay and hardening of the binary \citep{2003ApJ...582..559V,2012MNRAS.423.2533B,2020ApJ...904...16B,2020MNRAS.495.4681I}. In cosmological simulations, post-processed analytical models are often used to estimate the delay time between the numerical merger, flagged by the simulation, and the expected MBHB coalescence \citep{2017MNRAS.464.3131K,2020MNRAS.498.2219V,2022MNRAS.514.2220C,2023MNRAS.523..758C}. There have been recent attempts to model MBHB evolution by stellar hardening down to coalescence using fast integrators \citep{2023MNRAS.524.4062M}, or using sub-grid models on the fly below the resolution limit \citep{2024arXiv241007856L}. However, since MBHB evolution depends strongly on the unresolved details near the MBH influence radius, accurately estimating the binary's lifetime in this phase remains challenging. In this paper, we focus exclusively on modelling the delay caused by dynamical friction of the lightest MBH of each pair (i.e., the secondary MBH), as it will have a longer sinking time than the more massive primary MBH. Our model, detailed below, incorporates several assumptions, the uncertainties of which are further discussed in Appendix~\ref{sec:uncertainties}.

We model the dynamical friction timescale according to  the canonical Chandrasekhar prescription \citep{1943ApJ....97..255C}, assuming that the host galaxy that once surrounded the incoming MBH has been stripped away, leaving the MBH embedded in a spherical isothermal galaxy remnant. In this case, the time required for the secondary MBH to sink to the centre of the remnant galaxy, starting from an initial radius $r_0$, is given by \citet{2008gady.book.....B}:
\begin{equation}
T_{\mathrm{dyn}}^{\mathrm{BH}}=19 \left(\frac{r_0}{5 \,\mathrm{kpc}}\right)^2\left(\frac{\sigma}{200 \mathrm{~km}\, \mathrm{s}^{-1}}\right)\left(\frac{10^8 \,\mathrm{M}_{\odot}}{\mathrm{M}_{\mathrm{BH}}}\right) \frac{1}{\Lambda}[\mathrm{Gyr}]~,
\label{Eq_df_old}
\end{equation}
\noindent where $\sigma$ is the velocity dispersion of the remnant host galaxy. Here $\Lambda$ refers to the Coulomb logarithm, computed according to $\Lambda=\ln\left(1+M_{\star}/M_{\rm BH}\right)$, with $M_{\star}$ being the total stellar mass of the host galaxy. 

Owing to the uncertainty regarding the evolving mass of the stellar nucleus surrounding the MBH, which will depend on both the galactic environment and the merger dynamics, we use only the MBH mass in the dynamical friction calculation. In this sense, the resulting dynamical friction timescales obtained here can be considered upper limits. In cases wherein the secondary galaxy undergoes many orbits about the primary, tidal stripping may efficiently remove most of the stellar mass around the MBH before it reaches the kpc-scale phase. If instead a kpc-scale dual nucleus forms after just a few pericentric passages (a scenario that likely describes most major mergers), the effective mass contributing to the dynamical friction deceleration could be significantly larger than the MBH mass. 

In this paper, we compute $T_{\mathrm{dyn}}^{\mathrm{BH}}$ from an initial separation $r_0$ according to two different procedures.
For procedure (i), we take $r_0$ to be the separation between the merging MBHs, as identified in each model and reported in Fig.~\ref{fig:Separation}. This corresponds to the resolution limit for most hydrodynamical simulations and to a merger remnant scale radius for the SAMs (if no further modeling of MBH dynamics is included). Taking the separation between MBHs is more straightforward to compute consistently across the various models considered in this work, compared to calculating the distance between the secondary MBH and the center of the remnant galaxy.
However, as $r_0$ remains model dependent we use a second procedure (ii) to homogenize the inclusion of delays by placing the secondary MBH at the effective radius of the merger remnant according to the scaling relations:

\begin{equation}
R_{\rm eff} = \gamma \left(\frac{M_{\star}}{M_{\rm 0}}\right)^{\alpha} \left(1+\frac{M_{\star}}{M_{\rm 0}}\right)^{\beta - \alpha} \, \text{[kpc]}~. 
\label{eq:Reff_scale}
\end{equation}

\noindent where we assume that all simulated galaxies, independently of their true nature, obey either the early-type relation from \citet{2015MNRAS.447.2603L} derived from the GAMA survey ($0.01<z<0.1$), with $\alpha=0.11, \, \beta=0.76, \gamma = 0.11,$ and $M_{\rm 0}=2.01\times 10^{10}\, \rm M_{\odot}$. In Appendix~\ref{sec:uncertainties}, we test the 
the late-type relation derived from SDSS galaxies \citep[$z<0.3$][]{2003MNRAS.343..978S}, with $\alpha=0.14, \, \beta=0.39, \gamma = 0.10$, and $M_{\rm 0}=3.98\times 10^{10}\msun$. We also account for the redshift evolution $R_{\text{eff}} \propto (1 + z)^{\delta}$ from \citet{2014ApJ...788...28V}, which results in a higher degree of galaxy compactness with increasing redshift, assuming $\delta=-1.48$ for early-type galaxies and $\delta=-0.75$ for late-type galaxies. Our approach with these scaling relations is motivated by our choice to use the separation between MBHs as $r_0$ and the fact that this separation is defined differently across the various catalogues and depends on several factors including subgrid modelling and resolution.

Finally, we define the stellar velocity dispersion $\sigma$ in two different ways. First, when using the initial separation as given by each model we employ the simple log-linear scaling relation presented in \citet{2013ARA&A..51..511K} \citep[see also][]{2013ApJ...764..151G, 2013ApJ...764..184M}:

\begin{equation}
\sigma = 200 \, \left( \frac{M_\mathrm{BH}}{10^{9}\, \mathrm{M}_{\odot}} \right)^{1/4.38} \,\text{[km\, s}^{-1}]~.
\label{eq:sigma_1}
\end{equation}

\noindent Second, when using the galaxy effective radius of Eq.~\eqref{eq:Reff_scale} as initial separation, we employ a virial relation: 
\begin{equation}
\sigma=\left(\dfrac{0.25 G M_{\star}}{R_{\rm eff}} \right)^{1/2} \,\text{[km\, s}^{-1}]~,
\label{eq:sigma_2}
\end{equation}
\noindent where $M_{\star}$ is the total stellar mass of the galaxy and $R_{\rm eff}$ the effective radius of the galaxy taken from Eq.~\eqref{eq:Reff_scale}. 

\section{Results}
\label{sec:results}
\subsection{Massive black hole population}

This section shows the predictions of the different models regarding fundamental demographics of the MBH populations they produce, such as the $M_{\rm BH}$ mass function, the $M_{\rm BH}$--$M_{\rm *}$ relation, and the AGN luminosity function. Data from different models cover different redshift intervals, depending on the model setup (as an example: the \flares{} model focuses on the epoch of reionization, hence it only produced outputs at $z>5$). Overall, predictions from hydrodynamical simulations and semi-analytical models are shown, respectively, as solid lines and dashed/dotted lines. Most of the results shown in these sections are adapted from the reference papers of each model, as reported in the descriptions provided in Section~\ref{sec:methods_models}, although the catalogues used for this project, in some cases, differ from those used in previous papers. For instance, for the results presented in Sections~\ref{BHMF},~\ref{MBHMstar},~\ref{AGNLF} some models have opted to include multiple MBHs hosted in a given galaxy as separate entries (\simba, \illustris, \massiveblack, \barausse, the {\sc TNG} suite), others have opted for the sum of the masses and luminosities of all MBHs hosted in a given galaxy (e.g., \eagle, \flares), and others to select only the main MBHs (e.g., \romulus, \horizonAGN, \newHorizon, \obelisk, \cat, \lgalaxies). Finally, some models allow for only one MBH per galaxy (\shark, \delphi).  

\subsubsection{Black hole mass function}\label{BHMF}


\begin{figure}
\centering
\includegraphics[width=1.\columnwidth]{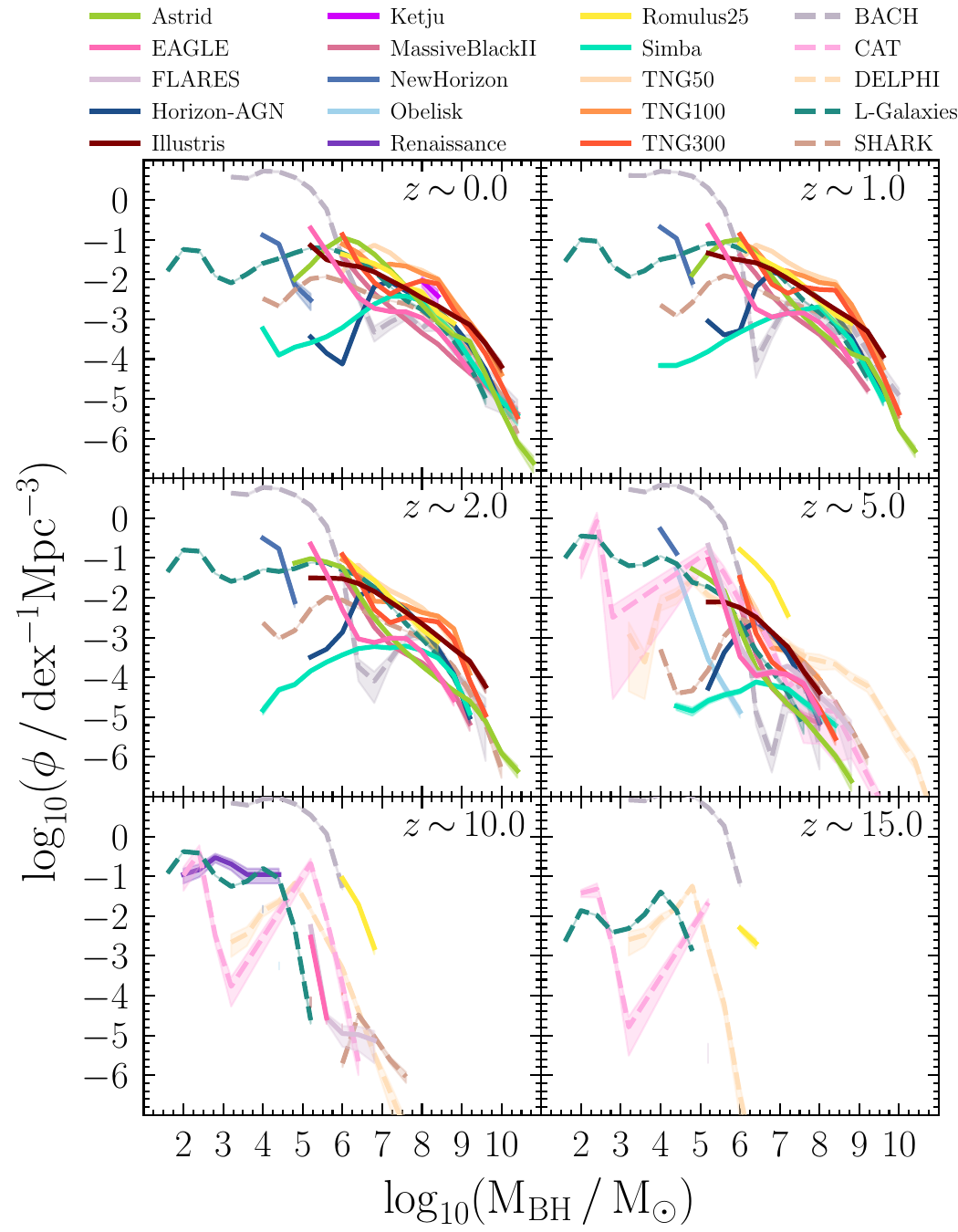}
\caption{MBH mass function obtained from the semi-analytical models and cosmological simulations. Full (dashed) lines show the predictions for cosmological simulations (semi-analytical models), and shaded areas indicate the 1$\sigma$ region accounting for Poisson errors. For each model, we show only bins with a minimum of 5 counts. All MBHs produced by the different models are included, although in some simulations (\tng, \illustris, \massiveblack, \simba, \eagle) we retain only the most massive MBHs per galaxy, a choice that primarily affects the most massive galaxies, which can host a population of non-evolved MBHs. This selection has no effect on the MBH mass function. No threshold in halo DM mass or galaxy stellar mass is applied, except for \simba where we remove MBHs located in galaxies with $M_{\star}< 10^{9.4}\, \rm M_{\odot}$, i.e. below the seeding threshold. 
While the massive end of the MBH mass functions ($M_{\rm BH}\geqslant 10^{7}\msun$) tend to converge at $z\leqslant 2$, differences in modeling lead to large discrepancies across the models at higher redshifts. Models always diverge significantly for the low-mass end of the MBH mass function.}  
\label{fig:FullMBHpopulation_BHmassfunction}
\end{figure}


Fig.~\ref{fig:FullMBHpopulation_BHmassfunction} presents the evolution of the MBH mass function for all the models. All MBHs provided by a given model and located in resolved galaxies are included in the figure. No other threshold on halo DM mass or galaxy total stellar mass is used, besides those used to define resolved systems based on the definitions of each of the models. With the exception of \simba, for which we remove MBHs located in galaxies with $M_{\star}< 10^{9.4}\, \rm M_{\odot}$, i.e. below the seeding threshold. 

The largest discrepancies are seen at $z \geqslant 5$, given the wide number of assumptions made by different models regarding the genesis of the first MBHs. The highest redshift bin ($z \sim 15$) is mainly populated by semi-analytical models (\barausse{}, \cat{}, \lgalaxies{}, and \delphi{}) or high-resolution simulations such as \romulus and \flares (very light symbol in the panel), which also have seeds at this redshift. Among these models, only \cat{} and \lgalaxies{} can shed light on the formation of MBHs with masses smaller than $\rm 10^3\, M_{\odot}$. Indeed, while the Press-Schechter formalism of \cat{} and the high-resolution 
of \renaissance{} allow to accurately trace the so-called mini-haloes ($M_{\rm h}\rm {\sim}\,10^6\, M_{\odot}$) where light seeds form, \lgalaxies{} improves the resolution of its $N$-body-based merger trees by including evolved counterparts of light MBH seeds provided by GAMETE/QSOdust outputs \citep{2016MNRAS.457.3356V}, while tracking self-consistently the formation of $M_{\rm BH}\gtrsim10^3\, \rm M_\odot$ MBH seeds.

The two peaks in the MBH mass functions of \lgalaxies{} and \cat{} represent the contributions by light seeds and more heavy ones\footnote{The normalization of the high-mass peak produced in \lgalaxies at $z\sim 15$ is higher than that of the low-mass peak, which may seem counterintuitive, as we would typically expect more light seeds than heavy ones. However, although heavy seeds do contribute to the peak, the dominant contribution comes from ``grafting'' already-evolved MBHs (with a light-seed origin) from GAMETE/QSOdust into Millennium-II haloes with typical masses of $\rm\sim10^4\,M_\odot$ at these high redshifts. At $z>10$, the dynamical range of Millennium-II primarily overlaps with the most massive haloes in GAMETE/QSOdust, which host the most evolved light-seed remnants in terms of BH mass.}, respectively.
In particular, the lack of MBHs in the mass range $\sim$$10^4$--$10^5$ M$_\odot$ in the \cat{} mass function is a consequence of the adopted Eddington-limited gas accretion model in which the growth of light seeds is inefficient (at all redshifts). \delphi produces an SN-feedback-regulated MBH mass function extending between $10^{3}$ and $10^{6}\msun$, with the number density falling off at masses larger than $10^5\msun$; this is driven by the imposed Eddington-limited accretion. The peak at $10^5\msun$ is driven by a combination of Eddington (sub-Eddington) accretion in sources with stellar mass $\sim 10^{6-7.5} ~(10^{7.5-9})\msun$. \barausse employs initial masses in the range $\sim 10^3-10^{5}\, \rm M_{\odot}$ (at $z>15$), and produces the MBH mass function with the highest normalization at both high and low redshift for $\rm M_{BH}\,\leqslant\,10^6\msun$. \barausse number density values range between 1--$10 \, \rm Mpc^{-3} \, dex^{-1}$, about 1 to 2 dex higher than the other models. This is because the realization of the \barausse model that we use here enhances accretion rates throughout cosmic history to reproduce PTA data at high masses and low redshift \citep{2024A&A...685A..94E,2023PhRvD.108j3034B}; note that the results still agree with the local MBH luminosity function, as a result of accretion becoming radiatively inefficient at high rates \citep[see the implementation in][]{2012MNRAS.423.2533B}. Finally, all models that feature low-mass seeds (i.e., \cat, \lgalaxies, \barausse, and \renaissance) produce a low-mass tail in their MBH mass function below $\rm 10^4 \, M_{\odot}$, down to lower redshift, as a consequence of the presence of ungrown seeds. 

At $z\,{\sim}\,10$, cosmological hydrodynamical simulations such as \obelisk and \eagle, or the post-processed model used on \renaissance start forming MBHs (some models like \obelisk have a very light symbol in the panel). These are typically more massive than those included in semi-analytical models, with seed masses of $\geqslant \rm 10^{5}\, \rm M_{\odot}$. \renaissance forms MBHs with a lower seed mass, more similar to the lower seed masses that are used in the \cat, \lgalaxies, and \delphi semi-analytical models. As expected, \eagle and \flares (zoom-in simulations using the \eagle subgrid model) provide similarly shaped mass functions. They both fall several orders of magnitude below the predictions of \romulus{}, in terms of normalization. \shark starts forming the first massive MBHs at these redshifts as well. Unlike other semi-analytical models, it does not include any sophisticated model of seeding and all the haloes of the DM-based merger trees are seeded with an MBH of $\rm 10^4\, M_{\odot}$. This relatively high seed mass, coupled with an efficient MBH growth, causes \shark to produce MBHs with $M_{\rm BH}>10^7\, \rm M_\odot$ already at $z\,{\sim}\,10$. 

Finally, the high normalization of the $z\geq 10$ MBH mass function produced by a few models (namely: \romulus{} and \barausse{}) is due to the high occupation fraction of MBHs in $z>10$ haloes, consequential to the BH-seeding prescriptions they employ. In addition, more efficient mergers in simulations like \eagle{} will further decrease the number of MBHs relative to simulations like \astrid{} and \romulus{}, which can have significant delays between galaxy and (numerical) MBH merger.

At $2\leqslant z\leqslant5$, all the models show a certain degree of convergence at $M_{\rm BH}\geqslant10^8\, \rm M_{\odot}$. This trend is driven by both the hierarchical assembly of MBHs and their growth via gas accretion, which tends to smooth out the initial differences produced by the diverse BH-seeding prescriptions. At $z \sim 5$, \romulus predicts a significantly higher normalization to the MBH mass function, but, as shown below, the luminosity function predicted by this model at the same redshift is less of an outlier compared to the remaining simulations. Despite the general agreement at $M_{\rm BH}\geqslant 10^8 \rm M_\odot$, the MBH mass functions of different models show significant discrepancies at $\rm M_{BH}\,{<}\,10^8 \, \rm M_{\odot}$. Much of these discrepancies are due to seeding models. However, \barausse in particular displays the largest number densities for $\rm M_{BH}\,\leqslant\,10^6\msun$, up one order of magnitude larger than the results produced by hydrodynamical simulations such as \astrid, \newHorizon, and  \eagle, as well as by the \lgalaxies, \cat, and \delphi semi-analytical models. This is due to the high redshift seeding model~\cite{2008MNRAS.383.1079V} employed by \barausse.

Finally, at $z\,{<}\,1$, the convergence between different models becomes more evident in terms of predicted number densities and extends down to $M_{\rm BH}\geqslant10^7\, \rm M_{\odot}$. As discussed above, this can be ascribed to the cosmological growth of MBHs, which generally tends to erase the initial differences due to BH-seed-mass distributions. Furthermore, the agreement amongst different results on the low-redshift MBH mass function is favored by the latter being one of the key observational constraints against which different numerical models and simulations which track MBHs evolution are calibrated.

\subsubsection{$M_{\rm BH}$--$M_{\star}$ relation}\label{MBHMstar}
\label{sec:MBH-stellar relation}

\begin{figure}
\centering
\includegraphics[width=1.\columnwidth]{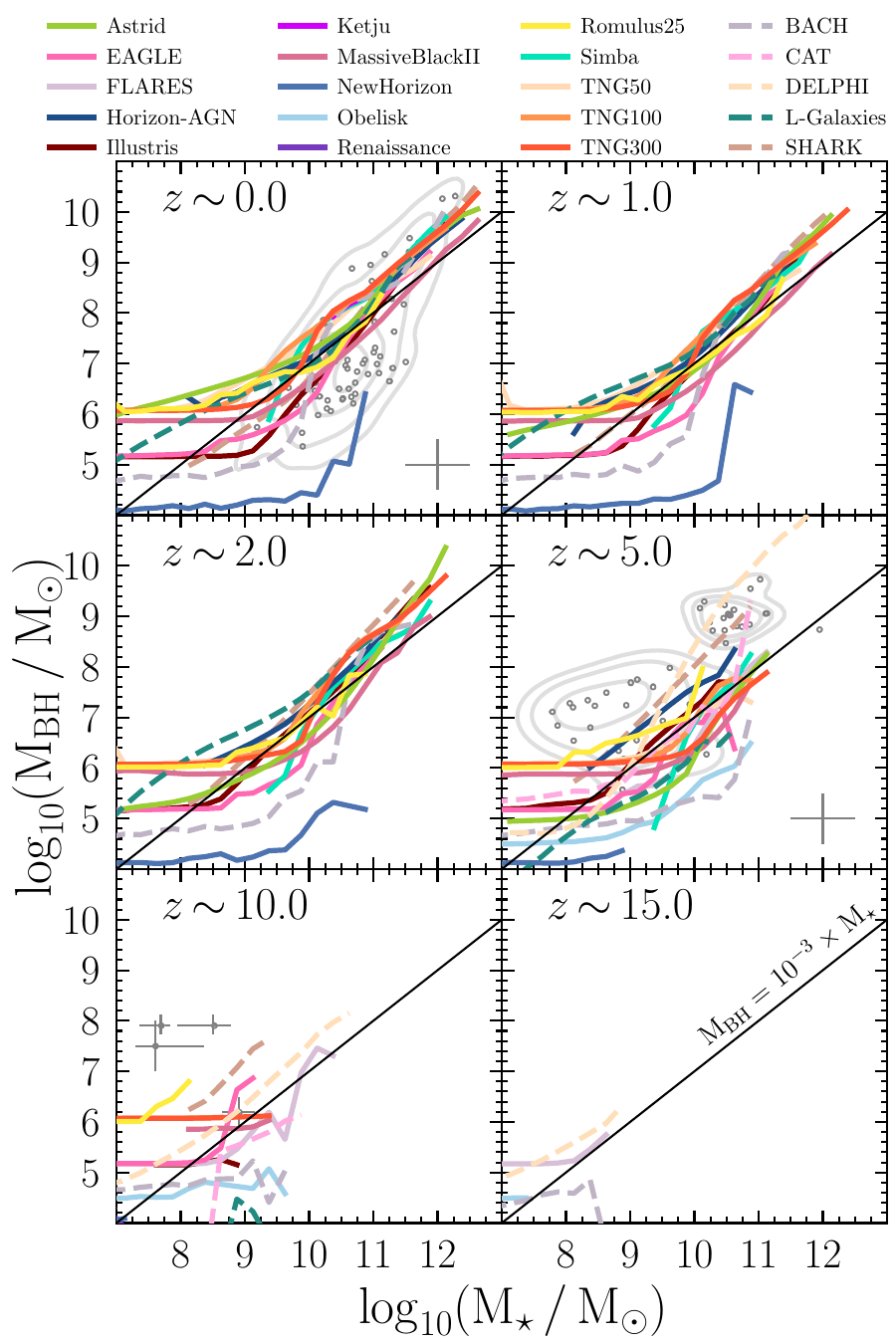}
\caption{MBH-galaxy stellar mass relations at different redshifts. For the models, we show the MBH mass median in bins of stellar mass. Only bins containing at least five galaxies are shown for $M_{\star}\geqslant 10^{11}\, \rm M_{\odot}$. In the \illustris, \massiveblack, \simba and \eagle simulations, the median MBH mass in these massive galaxies can be dominated by un-evolved MBHs. In such cases, we only consider the most massive MBH per galaxy for these models.
For \simba, we also display only bins with $M_{\star}\geqslant 10^{9.4}\, \rm M_{\odot}$, corresponding to the mass threshold at which seeding occurs. To guide the eyes, we add the $M_{\rm BH}=10^{-3}\times M_{\star}$ relation as a black line in each panel. For $z \sim 0$, we show the observational samples of \citet{2015ApJ...813...82R,2018MNRAS.478.2576M,2023MNRAS.518.2177G} and associated kernel density estimate (KDE) contours. At $z\geqslant 5$, we show the sample of high-redshift quasars compiled in \citet{2020ApJ...905...51S} (upper contours) and some of the AGN candidates revealed by JWST \citep[][lower contours]{2023arXiv230801230M,2023arXiv230311946H,2023arXiv230308918L}. Note that MBHs at $z \sim 10$ have uncertainties provided from \cite{2024Natur.627...59M,2024ApJ...965L..21K,2024NatAs...8..126B}, while for $z<10$ we provide a sample uncertainty of $\pm 0.5$ dex in the lower-right corner.}
\label{fig:MBH_Mstellar_Comparison_Observations}
\end{figure}

We show the $M_{\rm BH}$--$M_{\star}$ relation in Fig.~\ref{fig:MBH_Mstellar_Comparison_Observations},  with the median relations produced by all the models. 
We show the $M_{\rm BH}$--$M_{\star}$ relation in Fig.~\ref{fig:MBH_Mstellar_Comparison_Observations},  with the median relations produced by all the models. We notice that in the \illustris, \massiveblack, \simba and \eagle simulations, the median MBH mass in massive galaxies of $M_{\star}\geqslant 10^{11}\, \rm M_{\odot}$ can be dominated by un-evolved wandering MBHs, which results in a decrease at the massive end. In such cases, we only consider the most massive MBH per galaxy for these models. For the low redshifts ($z \sim 0$--2), we compare the results to observations in the local Universe from \citet{ 2015ApJ...813...82R, 2018MNRAS.478.2576M} and \citet{2023MNRAS.518.2177G}. At $z\geqslant 5$, we show the sample of high-redshift quasars compiled in \citet[][$z\geqslant6$]{2020ApJ...905...51S} with $L_{\rm bol}\geqslant 10^{45}\, \rm erg$~s$^{-1}$ and MBH MgII-based mass estimates. For this sample, we assume the galaxy stellar mass to be the dynamical mass of the host system. We also add some of the AGN candidates revealed by JWST \citep{2023arXiv230801230M,2023arXiv230311946H,2023arXiv230308918L,2023ApJ...957L...7K,2023arXiv230308918L,2024Natur.627...59M,2024ApJ...965L..21K,2024NatAs...8..126B}. To guide the reader, we add a line at $M_{\rm BH}/M_{\rm *}=10^{-3}$ in all panels. Finally, we note that some of the dispersion seen in the scaling relations may reflect physical, morphology-dependent evolutionary tracks rather than just intrinsic scatter. Recent empirical work demonstrates that galaxies follow distinct relations based on their merger history (e.g., a quadratic relation for ellipticals driven by dry mergers; \citet{2023arXiv231203999G}). Furthermore, utilizing morphologically-aware whole-galaxy mass relations ($M_{\rm BH}- M_{\star}$) may provide more robust calibrations for gravitational wave event predictions where bulge/disc decompositions are unavailable.

Up to $z \sim 2$, we observe a general agreement amongst the majority of the models. In the stellar mass range $\rm M_{*} = 10^9$--$10^{11.5}\msun$, the model predictions agree within $< 1 \,\rm dex$, whereas in the lower mass end, $M_{\rm *} < 10^9\msun$, a larger dispersion, of the order of a factor of $\sim$30, is driven by the different seeding prescriptions assumed. This agreement at high mass and low redshift is mostly by construction, as simulations are typically optimized to produce the same scaling relations in massive galaxies where hierarchical assembly and accretion tend to smooth out the initial differences. A clear outlier is \newHorizon, wherein most of the MBHs are undermassive with respect to the relation traced by observations for $M_{\star}\leqslant 10^{10}\msun$ \citep{2023MNRAS.523.5610B}. This is because their growth is strongly regulated by SN feedback until their galaxies are massive enough to overcome this effect (i.e., $M_{\star}\geqslant 5\times 10^{9}\msun$). In addition, the MBH seed mass is similar to the star particle mass, causing the MBHs to struggle to sink and stay in galaxy centres. The \newHorizon simulation also used an AGN feedback efficiency calibrated in lower-resolution simulations, which may have made it more difficult for them to grow.

The same SN-regulated growth in low-mass galaxies is found in \obelisk, which employs a similar SN feedback model, as well as in \barausse, \eagle, and \flares. These simulations share a similar shape of the relation, but show different normalizations, depending on other choices, such as repositioning of the MBHs in galaxy centres and implementation of SN feedback (\eagle and \flares), higher ratio of MBH-to-particle mass and/or efficiency of dynamical friction (\obelisk). Unlike \newHorizon and \obelisk, which employ a more efficient mechanical feedback from SNae, \horizonAGN employs a weaker kinetic feedback, whose effect on MBH growth is milder. Amongst the other large-scale simulations, \romulus, \illustris, \massiveblack, \ketju, and \astrid also employ a modelling of SN feedback that has a relatively weak effect on MBH growth, though the seed mass also plays a role in determining this. Simulations with higher seed masses are less likely to see this effect. Simulations such as {\sc TNG} have an efficiency of SN feedback which depends more strongly on redshift (more efficient at high redshift), and shape differently the $M_{\rm BH}$--$M_{\star}$ relation at different redshifts. In the vast majority of cosmological simulations, even in models with quite massive initial seed masses (e.g., \romulus{}), MBHs do not grow substantially through accretion within low-mass galaxies, due to feedback from SNae, MBHs, or a combination thereof. Instead, galaxies and MBHs follow a roughly constant relationship between accretion rate and star formation rate \citep{2019MNRAS.489..802R}. This results in a nearly flat relation at low masses, with MBH masses similar to the seed mass and scatter due to variations in MBH merger histories. Only MBHs more massive than $>10^7\msun$ have growth histories dominated by accretion. However, simulations like \romulus that have many high-redshift seeds will see more mergers take place, potentially contributing to the relation being less flat at low masses than other simulations.

Moving on to semi-analytical models, \shark, \lgalaxies, and \barausse show a close agreement at $z \sim 0$ above $10^{10}\msun$. In the low-mass end instead, \shark shows a trend which is intermediate between the \lgalaxies and \barausse predictions, likely due to different seeding, SN efficiency, and dynamics. In particular, since \barausse limits MBH growth in low-mass galaxies by mimicking the effects of SN feedback, this leads to typically smaller MBH masses compared to the \lgalaxies and \shark models in the low-mass end. While an efficient growth of BHs at $z>2$ in \lgalaxies produces the rapid build-up of the $M_{\rm BH}$--$M_{\rm\star}$ relation, which appears to be in place already at $z\sim2$, at later times, the increase of the host galaxies stellar mass brings \lgalaxies  closer to all other models. 

At $z\geq5$, we see a stronger scatter among all simulations, due to differing seeding prescriptions. All models also broadly fail to produce a median $M_{\rm BH}$--$M_{\star}$ relation that falls on the peak of the observed distribution in galaxies with mass $<10^{10} \msun$. However, flux-limited observations are prone to select overmassive MBHs \citep{2007ApJ...670..249L}, and even the assumptions on the scalings to obtain MBH masses may differ for highly accreting MBHs \citep{2024A&A...689A.128L}. Stellar mass measurements at high redshift have also equally large uncertainties. Some of the models are specifically dedicated to this $z \gtrsim5$ redshift range.  For instance, \delphi has been recently updated to match the properties of JWST-detected AGN, and this leads to very massive MBHs ($>10^{10}\msun$) already at $z \sim 5$ (see similar result in the most updated version of \lgalaxies{}, \citealt{2025arXiv250912325B}). 
Along with \shark and, in part, \cat, these are the only models that reach the masses of the $z\sim 6$ quasars shown as a cloud in the panels at stellar masses $>10^{10}\msun$ at $z \sim 5$. The tentative detection of AGN at $z>10$ poses a challenge to all theoretical models. 

To summarise, the differences in the galaxy low-mass regime are driven by a combination of several factors: the seed mass employed in the various models, their modelling of gas accretion, their modelling of SN and AGN feedback, the efficiency of MBH mergers, and, generally speaking, the calibration of free parameters adopted in the models. Not surprisingly, models with small seed mass and low accretion rates/efficient SN feedback are able to reproduce the lower part of the cloud of data at $M_{\star}\sim 10^{10} \msun$ in the observations in the local Universe, but struggle to produce the most massive BHs at fixed $M_{\star}$. Models that allow for efficient MBH accretion, however, better reproduce the massive BHs while missing this lower part of the cloud. For the galaxy massive end, most models produce a median. $M_{\rm BH}$--$M_{\star}$ relation that converge at $z \sim 0$--2 to the empirical scaling relations that have been derived for massive galaxies, mostly by construction of the models, which are generally optimized to produce such results.

\begin{figure}
\centering
\includegraphics[width=1.\columnwidth]{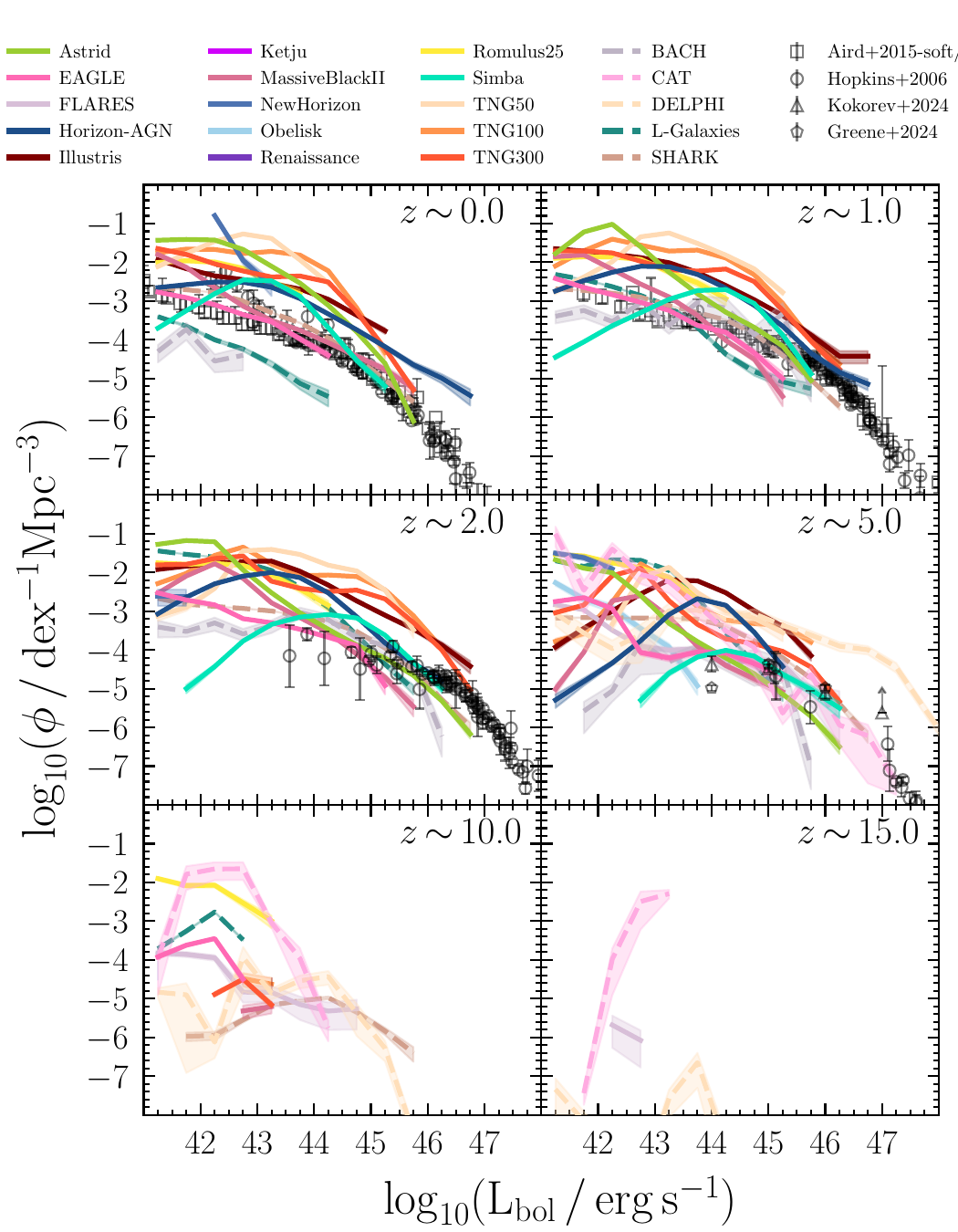}
\caption{AGN luminosity function from the semi-analytical models and cosmological simulations. 
All MBHs produced by the different models are included, although in some simulations (\tng, \illustris, \massiveblack, \simba, \eagle) we retain only the most massive MBHs per galaxy, a choice that primarily affects the most massive galaxies, which can host a population of non-evolved wandering MBHs. This selection has no effect on the luminosity function. For each model, we show only bins with a minimum of 5 counts. 
Shaded areas indicate the 1$\sigma$ region accounting for Poisson errors. For example, the AGN of \ketju do not appear at $z\sim 0$ because there are less than 5 per bin. }
\label{fig:FullMBHpopulation_AGN_LF}
\end{figure}

\subsubsection{AGN luminosity function}\label{AGNLF}

In Fig.~\ref{fig:FullMBHpopulation_AGN_LF}, we show the AGN luminosity function produced by all the models. For simplicity, we assume that all AGN are radiatively efficient and therefore compute the bolometric luminosity as $L_{\rm bol}=\epsilon_{\rm r}\dot{M}_{\rm BH} c^{2}$, with $c$ being the speed of light in vacuum and $\epsilon_{\rm r}$ the radiative efficiency provided by each model. We include all the simulated AGN and do not add any corrections for the fraction of those that could be obscured. 

The modelling of most large-scale cosmological simulations (e.g., MBH seeding, accretion, and  feedback processes) is calibrated on the massive end of one of the local observed $M_{\rm BH}$--$M_{\star}$ relations, and is not calibrated to reproduce any observational constraints on the AGN luminosity function. Conversely, some semi-analytical models employ a calibration on high-redshift properties. For example, CAT and DELPHI are calibrated to reproduce the observed properties (mass and bolometric luminosity) of $z \geqslant 5$--6 MBHs.

Simulations and semi-analytical models produce luminosity functions that can differ by more than two orders of magnitude in normalization. For most redshifts, the largest discrepancies arise at the AGN luminosity function faint-end, i.e., $L_{\rm bol}< 10^{45}\, {\rm erg~s}^{-1}$. This is mainly driven by the differences in seed masses, accretion modelling, and feedback efficiency in preventing MBH growth, as discussed for the $M_{\rm BH}$--$M_{\star}$ relations (see Fig.~\ref{fig:MBH_Mstellar_Comparison_Observations} and Section~\ref{sec:MBH-stellar relation}). In this regime, the agreement with observational constraints is particularly affected by uncertainties in the number of obscured AGN, a correction that is not applied to the theoretical models here \citep[][for the impact in cosmological simulations]{2022MNRAS.509.3015H}.
 
There is a reasonable agreement between models at $z<2$ for luminous AGN with $L_{\rm bol}\geqslant10^{45}\, {\rm erg~s}^{-1}$, which also broadly agree with current observational constraints. For the specific case of \lgalaxies{} at $z\!\sim\!0$, the disagreement with observational data is only due to the limited volume and high mass-resolution of the \texttt{Millennium-II} merger-trees. Indeed, as noted in \cite{2023MNRAS.518.4672S} these factors concur to produce a lack of MBHs shining as faint AGN (i.e. $L_{\rm bol}\leqslant10^{45}\, {\rm erg~s}^{-1}$) at $z<1$. Recent works employing \lgalaxies{} on the larger \texttt{Millennium} simulation show good agreement of the AGN luminosity function down to $z=0$ \citep[e.g.,][]{2020MNRAS.495.4681I,2025arXiv250912325B}.


\begin{center}
\begin{figure*}
\includegraphics[width=2.\columnwidth]{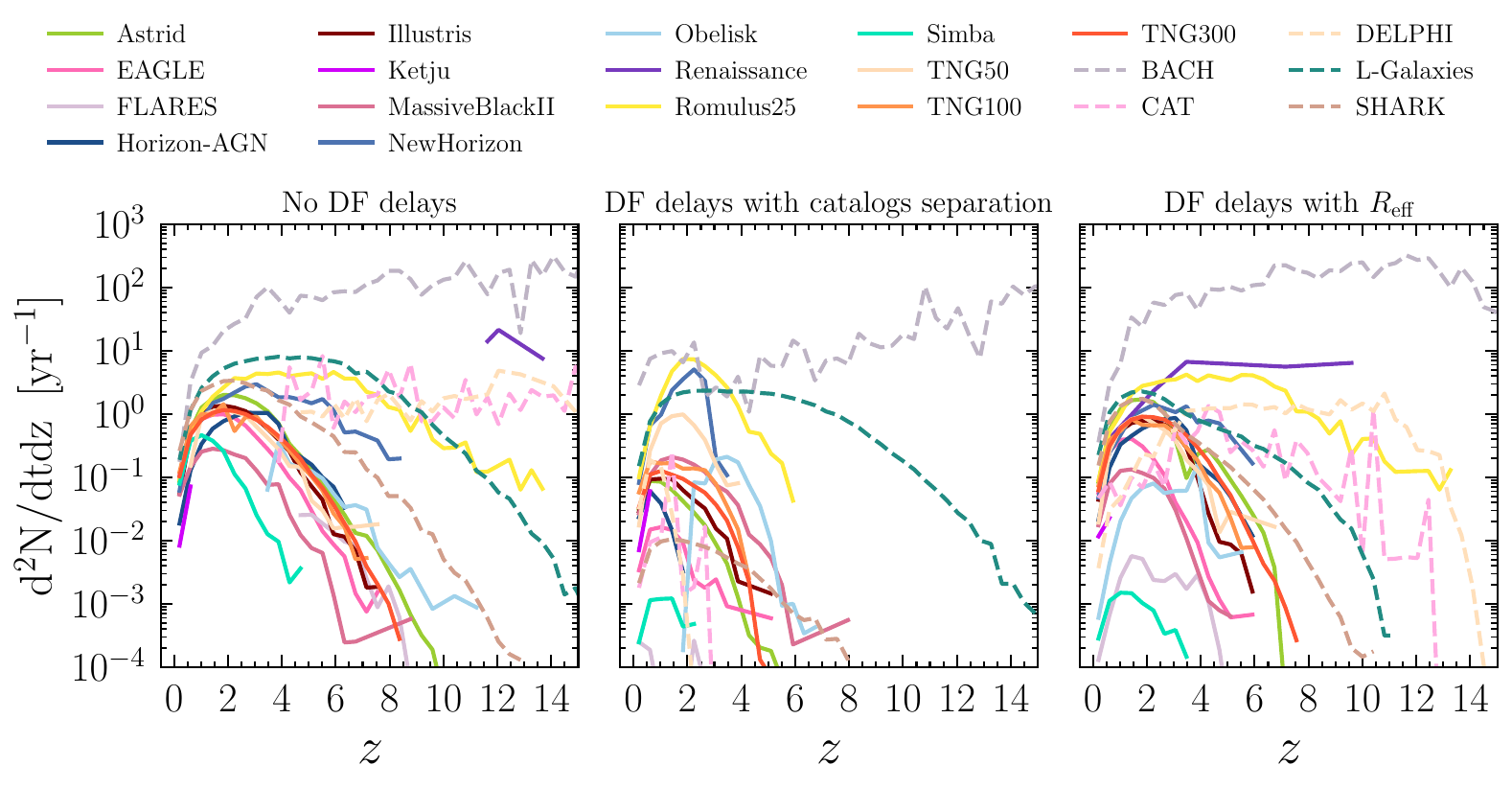}
\caption{Merger rate of MBHBs as a function of redshift predicted by the different models, for three different assumptions about applied delays. No cut in the mass of the MBHBs is used, nor in the mass ratio. Redshift bins have a fixed width of $\sim 0.4$. {\it Left-hand panel}: Instantaneous mergers with no post-processing delays. Models that incorporate delays (i.e., \barausse, \lgalaxies, \newHorizon, \obelisk, and  \horizonAGN) have these delays disabled. {\it Middle panel}: mergers with delays computed based on dynamical friction using the separation between the two MBHs given by the catalogues. {\it Right-hand panel}: mergers with dynamical friction delays, computed using the effective radius for early-type galaxies (independently of the true nature of galaxies, Eq.~\ref{eq:Reff_scale}) and a velocity dispersion $\sigma$ based on that effective radius (Eq.~\ref{eq:sigma_2}). 
}
\label{fig:BinaryMBHpopulation_MergerRate}
\end{figure*}
\end{center}

At $z \sim 2$, we observe a tendency for models to overproduce the faint end of the luminosity function. Discrepancies between models diverge as redshift increases and observational constraints decline: even with JWST, constraints suffer from large uncertainties and incompleteness. At $z \sim 10$ and higher, the results are completely dominated by choices on seeding and accretion. Hydrodynamical simulations and semi-analytical models based on N-body simulations are also limited by volume and resolution. Only a few models capture a volume, or halo masses in the case of extended Press-Schechter-based models, large enough to produce AGN as luminous as those observed. We recall that observations are, on the one hand, limited by sensitivity and therefore struggle to have a complete census at the faint end, and on the other hand, the currently observed volumes where AGN are spectroscopically characterized by JWST are severely limited. 

As for the previous two sections, this discussion is not intended to rate models, but to help interpret the comparison of merger rates and properties of merging MBHs. Models that overpredict the mass function and/or luminosity function are more likely to predict a high merger rate, fixing all other parameters (e.g., volume, resolution, minimum and maximum halo mass). Furthermore, the regime wherein models differ the most is the most relevant for LISA:  $M_{\rm BH}\sim 10^{4-7}\,\rm M_{\odot}$. A final important point to keep in mind, specifically when focusing on mergers for LISA, is that models that have low resolution/high minimum halo mass will underpredict the merger rate because low-mass galaxies (and therefore the MBHs they host) are not present in the model \citep{2020MNRAS.498.2219V}. This can be assessed in particular from Fig.~\ref{fig:FullMBHpopulation_BHmassfunction}, where one can see where the MBH mass function cuts off. It is more complicated to judge based on the luminosity function because luminosity is a combination of MBH mass and Eddington ratio.

\subsection{Massive black hole merger rates}\label{sec:merger_rates}
After characterising and comparing the global population of MBHs across all models, we now explore MBH mergers. To this end, we compute the intrinsic differential merger rate at a given redshift as:
\begin{eqnarray}
\frac{\mathrm{d}N}{\mathrm{d}t \,\mathrm{d}z} = \frac{\mathrm{d}n}{\mathrm{d}z}\times 4\pi c \, \left(\frac{d_{\rm L}}{1+z}\right)^2~,
\end{eqnarray}
\noindent where $\mathrm{d}n/{\mathrm{d}z}$ is the differential number of MBH mergers per redshift interval and comoving volume, and $d_L(z)$ is the luminosity distance at the same redshift.

In Fig.~\ref{fig:BinaryMBHpopulation_MergerRate}, we show the MBH merger rates as a function of redshift predicted by all models for different assumptions regarding the applied delays (described in Section~\ref{ModellingDelays}). MBHBs of any mass and any mass ratio are considered. In the left-hand panel, the rates are presented with no delays (i.e., even in the models where such delays can be computed, they are switched off). As can be seen, there is a very large spread in the different predictions, especially at high redshift, where models are less firmly anchored to observational data and different seeding prescriptions are adopted. The spread is smaller at low redshift, where all models are calibrated to a plethora of local observables. Most models do not predict mergers at redshift $z>10$, but some (\romulus, \delphi, \lgalaxies, \barausse, \cat, \renaissance) predict large numbers. This is due to the different assumptions about seeding and different parameters such as resolution. 

In the middle panel of Fig.~\ref{fig:BinaryMBHpopulation_MergerRate}, the rates are computed applying dynamical friction delays from the initial MBH pair separation provided by each model, as described in Section~\ref{ModellingDelays}. This results in fewer mergers at high redshift for several models, but does not yield a clear reduction in the observed scatter between the different models. 
Adding delays introduces discrepancies between the merger rates of \flares and \eagle, despite their similar modeling and MBH populations (BH mass function, AGN luminosity function, and MBH-galaxy stellar mass relation). This difference arises because \flares only has numerical mergers down to $z\sim 5$; when DF delays are applied, these mergers are redistributed over the range $0<z<5$. In contrast, the higher merger rate in \eagle is driven by the additional contribution of numerical mergers occurring at $z<5$.

A similarly large spread of the models' merger rate is observed even when the initial pair's separation is set based on fits to observations by adopting the effective radius from Eq.~\eqref{eq:Reff_scale} for early-type galaxies, as shown in the right-hand panel of Fig.~\ref{fig:BinaryMBHpopulation_MergerRate}. In this case, the merger rate is less suppressed at high redshift. We attribute this behaviour to the scaling of $R_{\rm eff}$ with redshift. Due to the mass-size relation of galaxies, the binary separation at which delays are applied decreases with increasing redshift, implying shorter delays and, as a consequence, a larger number of mergers. In the case of separations provided by each model (middle panel), very often these are commensurable with the resolution of simulations, which does not change significantly with redshift, except in cases wherein spatial resolution is expressed in comoving units -- then the physical resolution scales as $(1+z)^{-1}$ (see Section~\ref{sec:methods_models} for details on each model).

This comparison suggests that, while there is a large spread in predicted merger rates, several models consistently predict at least a few mergers per year, even if delays are included in the modelling. We stress that predicted merger rates are also affected by resolution: lower-resolution models such as \horizonAGN, \illustris, \tngm, \tngh, \simba, and \eagle do not resolve the assembly of dwarf galaxies, which are hosts to many potential LISA sources. Therefore, these models miss a potentially large number of mergers at such low galaxy masses \citep{2020MNRAS.498.2219V}. 

We can obtain more insight on the scatter in merger rates and the impact of considering delays by exploring the distribution of merger rates in different MBHB mass ($M_{\rm bin}$) intervals. In Table~\ref{tab:MergerRate}, we show the merger rate from different models binned in source-frame binary mass, with and without any delays. We find that some models predict much higher rates at binary masses below $10^6\msun$, due to the choice of seeding prescription. For example, \barausse, which adopts a very efficient heavy-seed scenario in order to match PTA results, is characterised by very large merger rates at the low mass end ($M_{\rm bin} \in\, 10^{4}$--$10^{6}\, \rm M_{\odot}$). However, as expected, the differences between models become more nuanced at high masses, where models are more firmly anchored to observations and the memory of the initial seeding mechanism is washed out.

\setlength{\tabcolsep}{4pt} 
\begin{sidewaystable*}
\begin{tabular}{llcccccccccc}
 & Redshifts &
  \multicolumn{2}{c}{Merger rate {[}$\rm yr^{-1}${]}} &
  \multicolumn{2}{c}{Merger rate {[}$\rm yr^{-1}${]}} &
  \multicolumn{2}{c}{Merger rate {[}$\rm yr^{-1}${]}} &
  \multicolumn{2}{c}{Merger rate {[}$\rm yr^{-1}${]}} &
  \multicolumn{2}{c}{Merger rate {[}$\rm yr^{-1}${]}} \\
 & covered &
  \multicolumn{2}{c}{$\rm M_{bin}\in 10^4 - 10^5 M_{\odot}$} &
  \multicolumn{2}{c}{$\rm M_{bin}\in  10^5 - 10^6 M_{\odot}$} &
  \multicolumn{2}{c}{$\rm M_{bin}\in 10^6 - 10^7 M_{\odot}$} &
  \multicolumn{2}{c}{$\rm M_{bin}\in  10^7 - 10^8 M_{\odot}$} &
  \multicolumn{2}{c}{$\rm M_{bin}\in  10^8 - 10^{10} M_{\odot}$} \\\cline{3-12}

                            & by model &    Intrinsic & Delayed &  Intrinsic & Delayed & Intrinsic & Delayed &  Intrinsic & Delayed &  Intrinsic & Delayed \\ 
\\
\hline \hline
\cat                        & 4-20   &      -                    & -                    & 12.93                & 0.12                 & 1.20                  & $8.21\times 10^{-6}$  & $9.49\times 10^{-2}$ & 0.00                 & $3.88\times 10^{-3}$ & $3.44\times 10^{-6}$ \\     
\delphi                     & 0-20   &      3.46                 & $1.06\times 10^{-2}$ & 10.43                & 0.14                 & 0.86                  & $1.75\times 10^{-5}$  & 0.24                 & $1.45\times 10^{-7}$ & 0.11                 & $2.24\times 10^{-4}$ \\
\lgalaxies                  & 0-20   &      13.01                & 5.73                 & 7.23                 & 1.48                 & 6.16                  & 0.85                  & 2.84                 & 0.25                 & 0.87                 & $5.83\times 10^{-2}$ \\
\shark                      &        &      2.48                 & $6.93\times 10^{-3}$ & 3.16                 & $3.95\times 10^{-3}$ & 2.86                  & $1.11\times 10^{-2}$  & 1.82                 & $8.78\times 10^{-3}$ & 0.94                 & $2.77\times 10^{-3}$ \\
\barausse                   & 0-20   &      1271.63              & 299.54               & 961.32               & 235.23               & 2.28                  & 0.53                  & 0.31                 & $3.94\times 10^{-3}$ & 0.31                 & $2.57\times 10^{-2}$ \\
\astrid                     & 1-5    &      $1.38\times 10^{-3}$ & $9.06 \times 10^{-6}$& 3.44                 & $2.63\times 10^{-2}$ & 1.82                  & $6.48\times 10^{-2}$  & 0.21                 & $3.64\times 10^{-2}$ & $3.64\times 10^{-2}$ & $1.03\times 10^{-2}$ \\
\massiveblack               & 0-12   &      -                    & -                    & -                    & -                    & 0.37                  & 0.25                  & 0.21                 & 0.16                 & 0.12                 & $7.43\times 10^{-2}$ \\
\eagle                      & 0-10   &      -                    & -                    & 1.31                 & $4.45\times 10^{-5}$ & 0.23                  & $1.40\times 10^{-4}$  & 0.65                 & $1.10\times 10^{-2}$ & 0.23                 & $1.46\times 10^{-2}$ \\
\horizonAGN                 & 0-5    &      -                    & -                    & $3.96\times 10^{-2}$ & 0.00                 & 0.63                  & 0.00                  & 1.76                 & $8.35\times 10^{-3}$ & 0.60                 & $3.93\times 10^{-2}$ \\
\newHorizon                 & 0-10   &      10.05                & 6.16                 & -                    & -                    & -                     & -                     & -                    & -                    & -                    & -                    \\
\illustris                  & 0-10   &      -                    & -                    & $7.60\times 10^{-2}$ & $1.22\times 10^{-4}$ & 1.17                  & $1.20\times 10^{-2}$  & 1.38                 & $7.18\times 10^{-2}$ & 0.96                 & $9.88\times 10^{-2}$ \\
\tngm               & 0-10   &      -                    & -                    & -                    & -                    & 1.09                  & 0.13                  & 0.95                 & 0.19                 & 1.23                 & 0.11                 \\
\tngh                      & 0-10   &      -                    & -                    & -                    & -                    & 1.84                  & $4.50\times 10^{-2}$  & 0.77                 & 0.13                 & 0.75                 & $8.21\times 10^{-2}$ \\
\simba                      & 0-5    &      $2.83\times 10^{-2}$ & 0.00                 & $8.29\times 10^{-3}$ & 0.00                 & $1.50\times 10^{-2}$  & $4.91\times 10^{-5}$  & 0.22                 & $5.16\times 10^{-4}$ & 0.42                 & $1.26\times 10^{-3}$ \\
\romulus                    & 0-15   &      -                    & -                    & -                    & -                    & 21.44                 & 11.55                 & 4.72                 & 3.40                 & 0.30                 & 0.26                 \\
\obelisk                    & 3.5-15.5 &    0.28                 & 0.21                 & 0.17                 & 0.14                 & $1.03\times 10^{-2}$  & $8.71\times 10^{-3}$  & $1.07\times 10^{-3}$ & $8.93\times 10^{-4}$ & $1.93\times 10^{-4}$ & $1.21\times 10^{-4}$ \\
\tngl                       & 0-10   &      -                    & -                    & -                    & -                    & 0.57                  & 0.29                  & 1.32                 & 0.98                 & 1.04                 & 0.47                 \\
\renaissance                & 10-16  &      8.72                 & 2.90                 & -                    & -                    & -                     & -                     & -                    & -                    & -                    & -                    \\
\flares                     & 5-15   &      -                    & -                    & $1.60\times 10^{-2}$ & $4.81\times 10^{-5}$ & $8.69\times 10^{-3}$  & $1.20\times 10^{-4}$  & $1.23\times 10^{-2}$ & $6.11\times 10^{-5}$ & $6.70\times 10^{-3}$ & $6.82\times 10^{-5}$ \\
\ketju                      & $<1$   &      -                    & -                    & -                    &    -                 & -                     & -                     & -                    & -                    &$2.92\times 10^{-2}$ & $2.45\times 10^{-2}$ \\
\hline \hline
\end{tabular}
\caption{MBH merger rates for the different semi-analytical models and simulations, integrated over all redshifts. Results on both the intrinsic merger rates from the models and rates corrected for DF delays (as adopted in middle panel of Fig.~\ref{fig:BinaryMBHpopulation_MergerRate} and Fig.~\ref{fig:BinaryMBHpopulation_MergerRate_specificMassBands}) are presented in MBHB mass bins. The second column shows the redshift range over which each model contains at least one MBH. These ranges do not account for the application of our delay models. Predictions of merger rates differ significantly in the MBH low-mass regime, where different assumptions regarding seeding have a direct effect on merger rates (see e.g., BACH where the use of heavy seeds results in a larger than average rate)}. 
\label{tab:MergerRate}
\end{sidewaystable*}
\setlength{\tabcolsep}{6pt}  

\begin{figure*}
\centering
\includegraphics[width=2.\columnwidth]{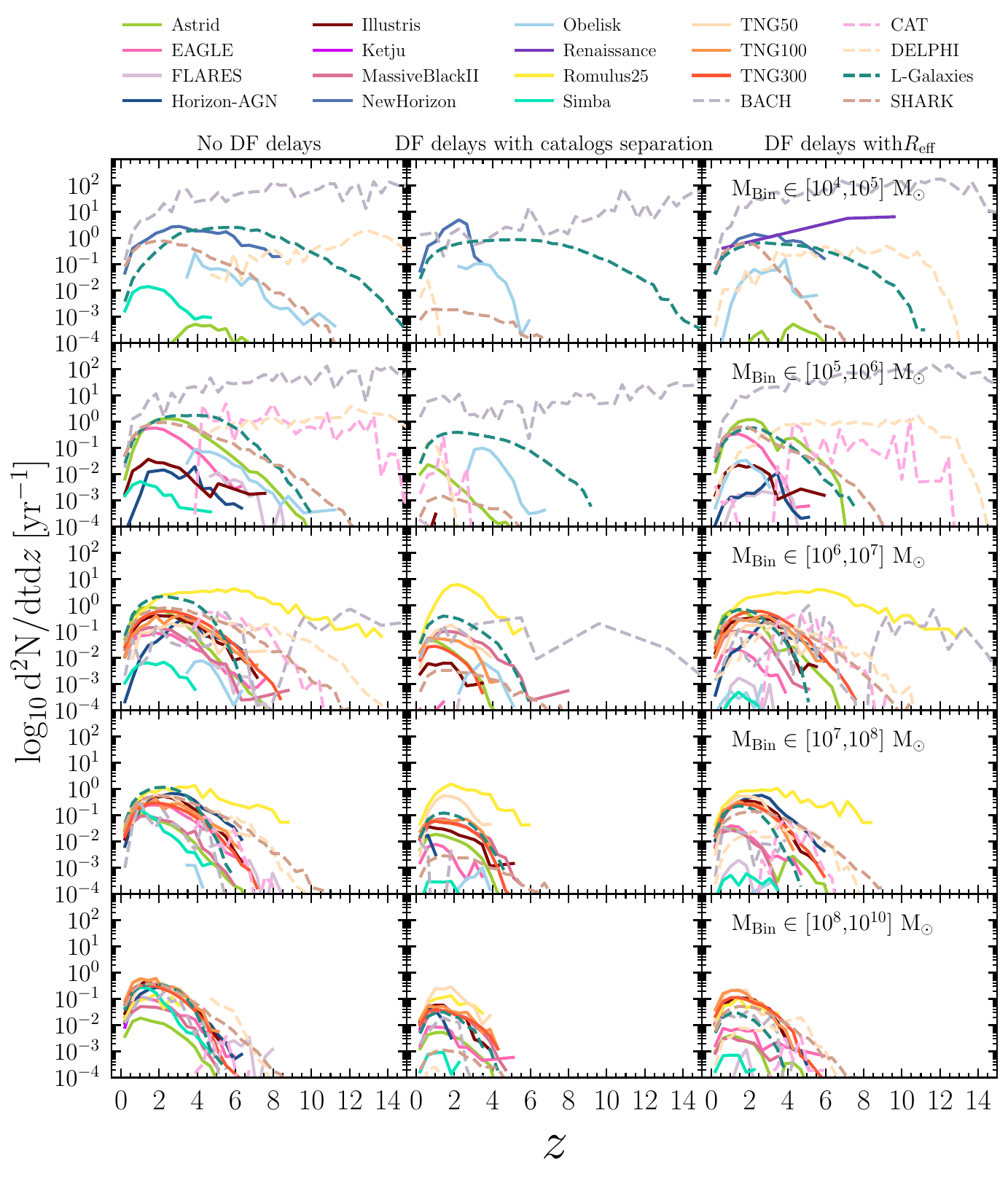}

\caption{Merger rate of MBHBs as a function of redshift predicted by the different models, for five different MBHB mass bins ($M_{\rm Bin} = M_{\rm BH,1} + M_{\rm BH,2}$) and three assumptions about delays: no post-processing (left-hand panels), with models that incorporate delays (i.e., \barausse, \lgalaxies, \newHorizon, \obelisk, \horizonAGN) having these delays disabled, and delays computed based on dynamical friction, using either the separation between the two MBHs given by the catalogs (middle panels) or using the effective radius for early-type galaxies (right-hand panels). Redshift bins have a fixed width of $\sim$0.4.}
\label{fig:BinaryMBHpopulation_MergerRate_specificMassBands}
\end{figure*}


\begin{figure*}
\centering
\includegraphics[width=1.8\columnwidth]{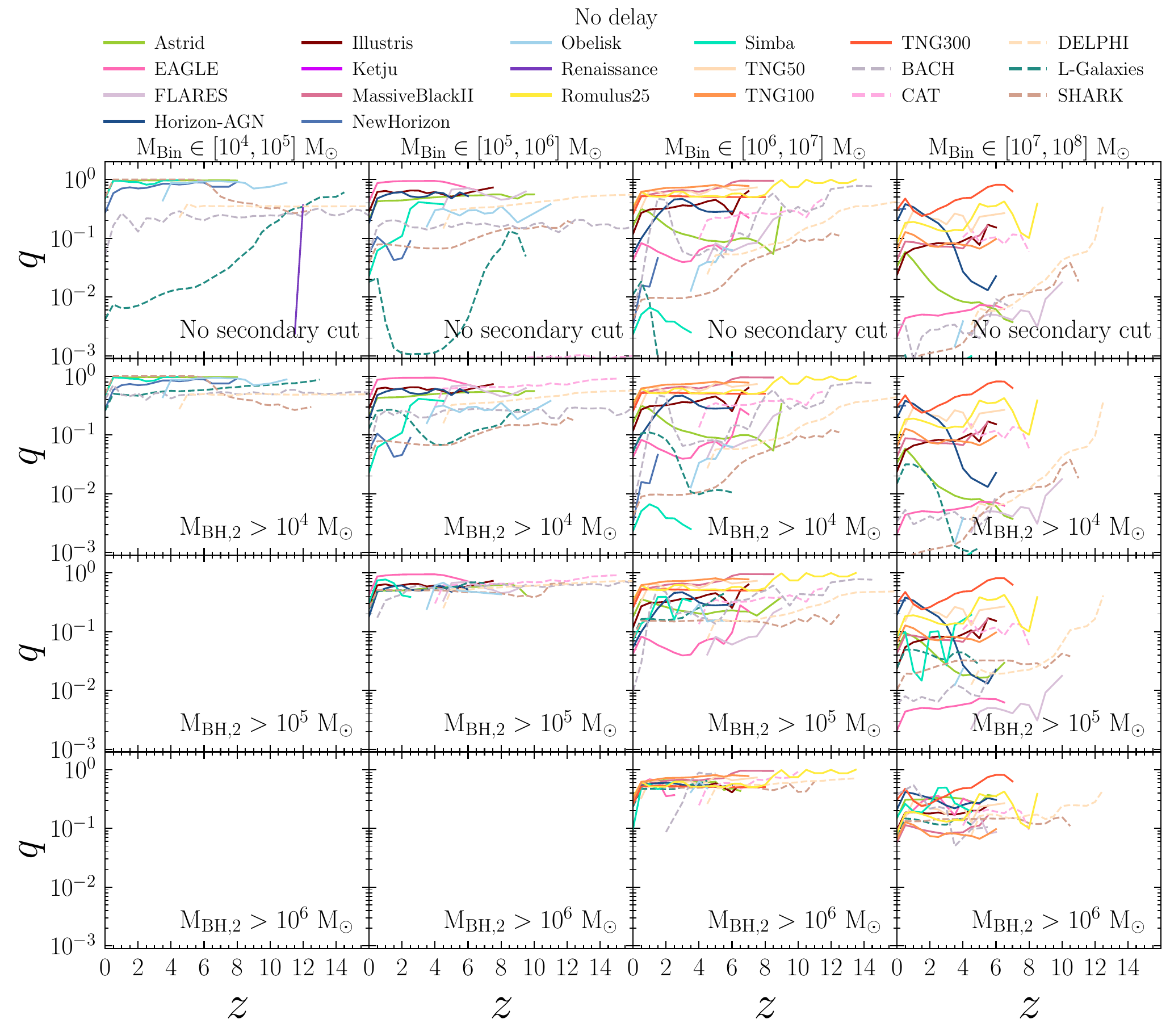}
\caption{Mass ratio $q=M_{\rm BH, \,2}/M_{\rm BH, \,1}$ (secondary / primary) of the binaries for the MBH mergers of the models without any  delay. {\it First row:} mass ratios are binned by binary mass (i.e., $M_{\rm BH, \,1}+M_{\rm BH, \,2}$). In the left-hand panel, the large differences in the mass ratios are driven by the range of seed masses used in the models. For example, the small ratios found for \lgalaxies are a consequence of the inclusion of light seeds in the model: low-mass BHs of $10^{4-5}\, \rm M_{\odot}$ can merge with seed-mass BHs of $\sim 100\, \rm M_{\odot}$, for example. {\it Remaining rows:} To limit the impact of the different seed masses amongst the models, we add thresholds for the mass of the secondary BH: $M_{\rm BH, \,2}\geqslant \, 10^{4}, 10^5, 10^6\msun$.}
\label{fig:BinaryMBHpopulation_MergerRate_massratio}
\end{figure*}


Similar conclusions can be drawn from Fig.~\ref{fig:BinaryMBHpopulation_MergerRate_specificMassBands}, which shows the merger rates as a function of redshift calculated with the same delay options as in Fig.~\ref{fig:BinaryMBHpopulation_MergerRate}, but for different intervals of the total MBHB mass ($M_{\rm bin} = M_{\rm BH,1} + M_{\rm BH,2}$, being $M_{\rm BH,1}$ and $M_{\rm BH,2}$ the mass of the heavier and lighter MBH involved in the merger). A large scatter is seen in the lower mass bins for any choice of delays, while better agreement between the models is found in the larger mass bins, and in particular for $M_{\rm bin} > 10^7\msun$, due to the lack of mergers at high redshift. LISA's sensitivity  at the lower mass end will therefore provide a powerful tool to learn about MBH seeds and early MBH evolution.


\begin{figure*}
\centering
\includegraphics[width=1.8\columnwidth]{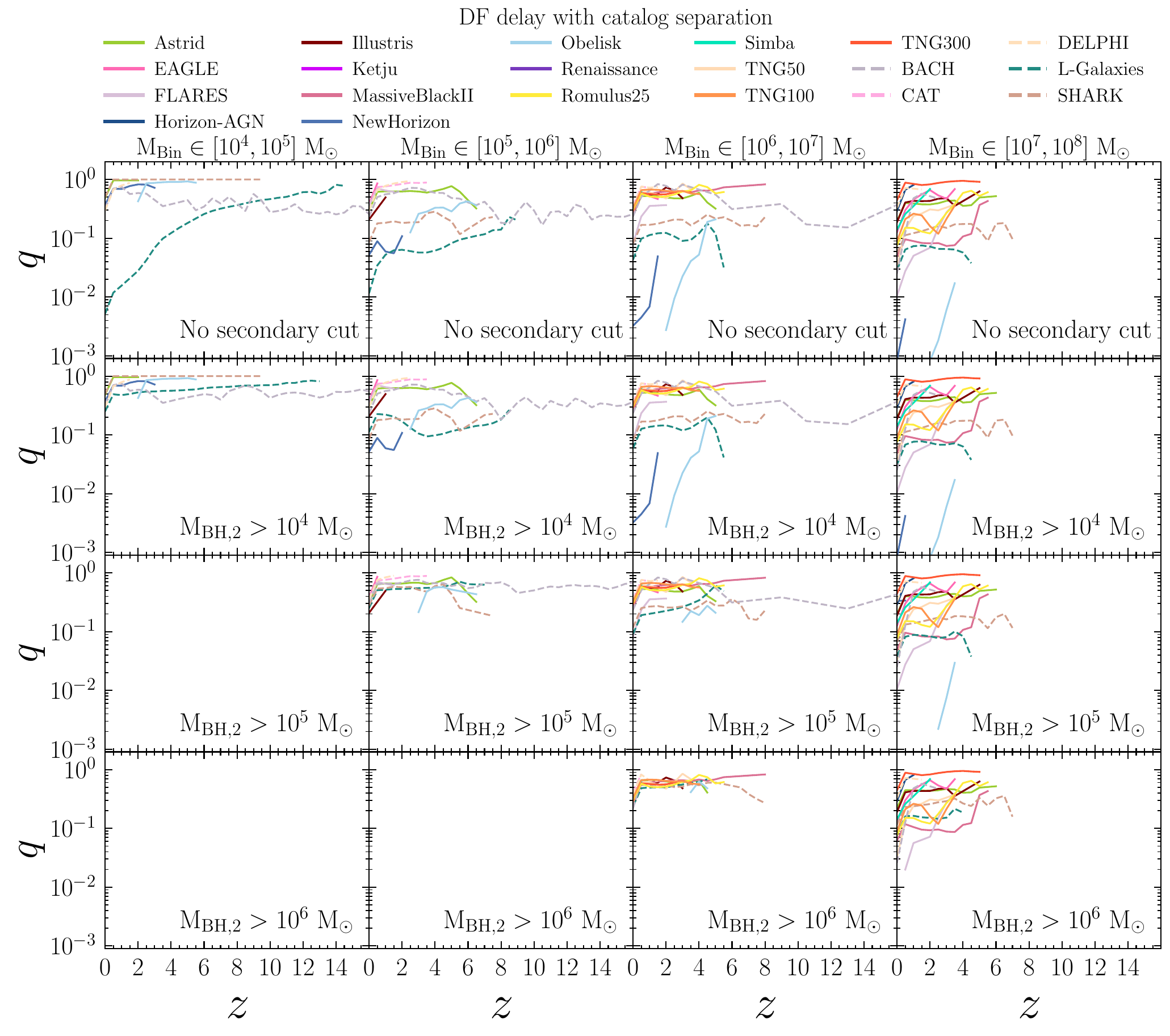}
\caption{
Similar to Fig.~\ref{fig:BinaryMBHpopulation_MergerRate_massratio}, but with dynamical friction timescales calculated using the separation of MBHs obtained from the respective models.}
\label{fig:BinaryMBHpopulation_MergerRate_massratio_catsep}
\end{figure*}



\begin{figure*}
\centering
\includegraphics[width=1.8\columnwidth]{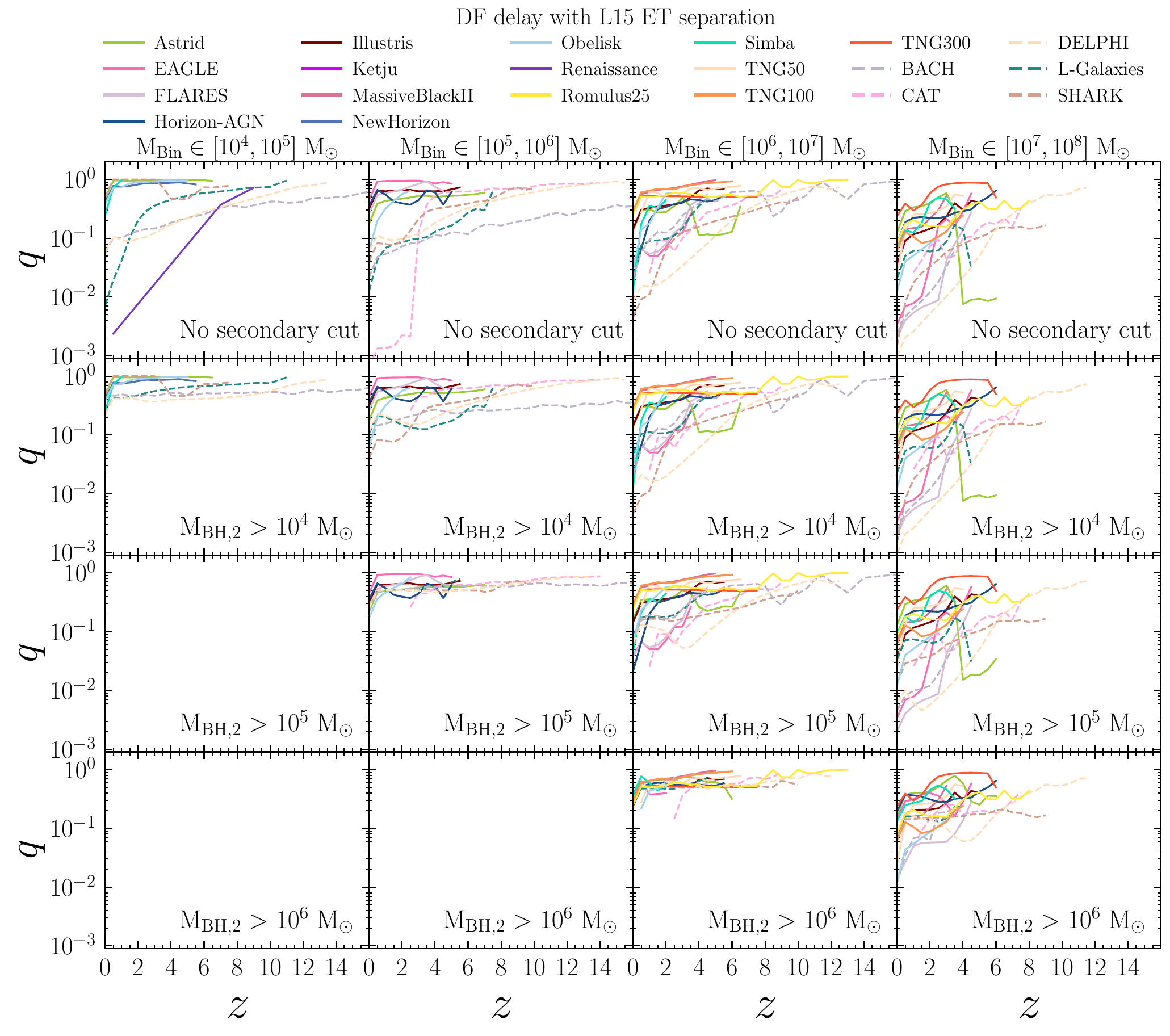}
\caption{
Similar to Fig.~\ref{fig:BinaryMBHpopulation_MergerRate_massratio}, but with dynamical friction timescales computed assuming $R_{\rm eff}$ for early-type galaxies \citep{2015MNRAS.447.2603L} following Eq.~\eqref{eq:Reff_scale} and a redshift dependence $R_{\text{eff}} \propto (1 + z)^{\delta}$  \citep{2014ApJ...788...28V}.}
\label{fig:BinaryMBHpopulation_MergerRate_massratio_ET}
\end{figure*}


\subsection{Massive black hole binary population} 

Fig.~\ref{fig:BinaryMBHpopulation_MergerRate_massratio} shows the median binary mass ratio $q\equiv M_{\rm BH,2}/M_{\rm BH,1}$ as a function of redshift within various mass bins for all of the models, and with no additional delays included. As the binary mass increases going from left to right, more models show mergers with very unequal mass ratios, wherein one of the two MBHs is close to its initial mass. Considering high-mass secondary MBHs, this naturally converges, as the mass ratio for each binary mass bin gets more constrained. With no or minor cuts to the secondary mass, the effect of different seed-mass prescriptions becomes evident (top right panels). In particular, we see a clear separation of mass ratios for high-mass binaries ($10^7<M_{\rm bin}/\msun<10^8$) between simulations that include low-mass seeds (typically semi-analytical models, lower ratios) and those that only include massive seeds (typically cosmological simulations, higher ratios). This effect is particularly extreme for the \lgalaxies model, which includes a large population of very low-mass seeds (from $10\msun$), producing very low-mass ratio mergers, going down as low as $q=10^{-5}$. For simulations with higher seed masses, such as \romulus and \tngm{} ($\sim$$10^{6}\msun$), the mass ratios are typically very high at all redshifts (typically at or above $10^{-1}$), but this is more a consequence of the assumed seed mass than a prediction of the simulation.

In Fig.~\ref{fig:BinaryMBHpopulation_MergerRate_massratio_catsep} and  Fig.~\ref{fig:BinaryMBHpopulation_MergerRate_massratio_ET}, we show that the effect of adding delays to mergers is to remove high-redshift, low-mass ratio mergers, either delaying them to lower redshift or removing them entirely from the merger sample. The overall effect is to decrease the differences between models. It is worth noticing that when delays are calculated using the galaxy effective radius, as shown in Fig.~\ref{fig:BinaryMBHpopulation_MergerRate_massratio_ET}, the effect is much stronger and the models differ even less. This is because MBHs start from a uniform separation, regardless of the resolution/assumptions. 

We note that Fig.~\ref{fig:BinaryMBHpopulation_MergerRate_massratio}, Fig.~\ref{fig:BinaryMBHpopulation_MergerRate_massratio_catsep}, and  Fig.~\ref{fig:BinaryMBHpopulation_MergerRate_massratio_ET} only show the median mass ratios for the sake of clarity. However, there is often a wide range of binary ratios, even when delays are accounted for. For some models, even including merger delays still results in mass ratios as low as $10^{-3}$, albeit occurring at a lower rate. Waveform modelling is complex for mass ratios $1/20$--$1/100$ \citep{2023arXiv231101300L} and, while these become significantly rarer once merger delays are introduced, most simulations still predict that such mergers will be a non-negligible component of the LISA data stream. The models that predict no such mergers are typically only those with high seed masses.

\subsection{The hosts of massive black hole binaries} 


\begin{figure*}
\centering
\includegraphics[width=1.8\columnwidth]{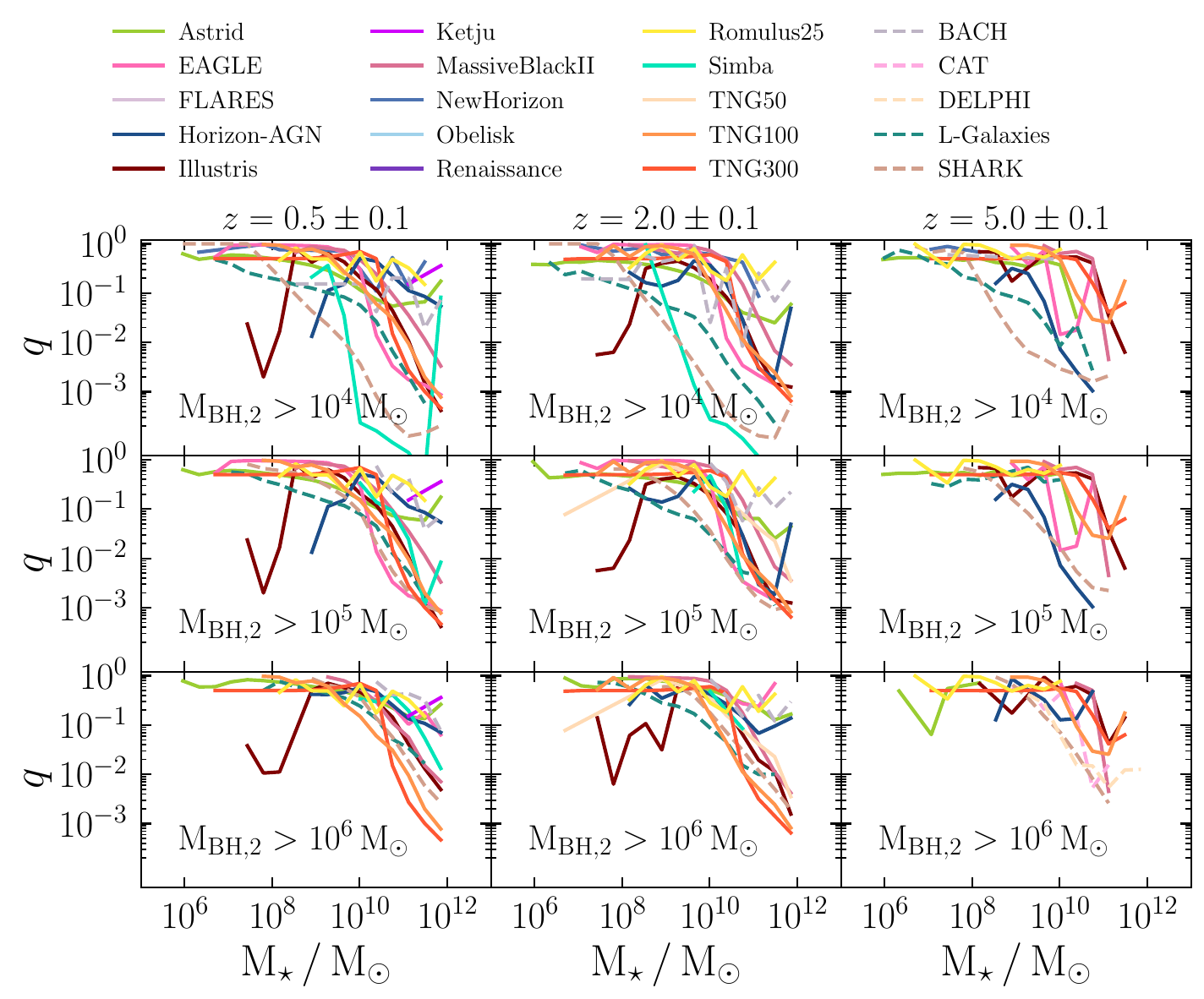} 
\caption{
Median mass ratios of the merging MBHBs, defined as $q=M_{\rm BH,2}/M_{\rm BH,1}$ (secondary/primary), as a function of the stellar mass of the remnant host galaxies. Here we consider the numerical mergers of the models, and do not apply our post-processing DF delays. We show different redshifts (columns) and cuts in secondary MBH mass (rows).}
\label{fig:mass_ratios_withMstar}
\end{figure*}


Now, we turn our focus to investigating the properties of the host galaxies in which mergers occur. Fig.~\ref{fig:mass_ratios_withMstar} shows the MBHB mass ratios as a function of the host stellar mass for different redshifts and mass cuts for the secondary MBH. A typical trend is the decrease of $q$ as $M_\star$ increases, starting from $M_{\star}\sim10^{10}\msun$, regardless of redshift. The decrease is driven by seed-like MBHs merging with the central MBHs of massive galaxies. \shark and \lgalaxies exhibit a more continuous decrease with stellar mass, owing mostly to the high rate of binaries forming from low-mass seeds. Placing more strict limits on the secondary mass removes this difference. The significant drop of $q$ at low galaxy mass for \illustris{} arises for a small number of mergers in which the secondary galaxy is briefly identified as the remnant galaxy. This can occur when a low-mass secondary MBH is repositioned onto a more massive host before the galaxy merger is complete. Mass ratios do not have a strong $M_\star$ dependence only for models that start from already heavy seeds, e.g., \romulus. 

\begin{figure*}
\centering
\includegraphics[width=2\columnwidth]{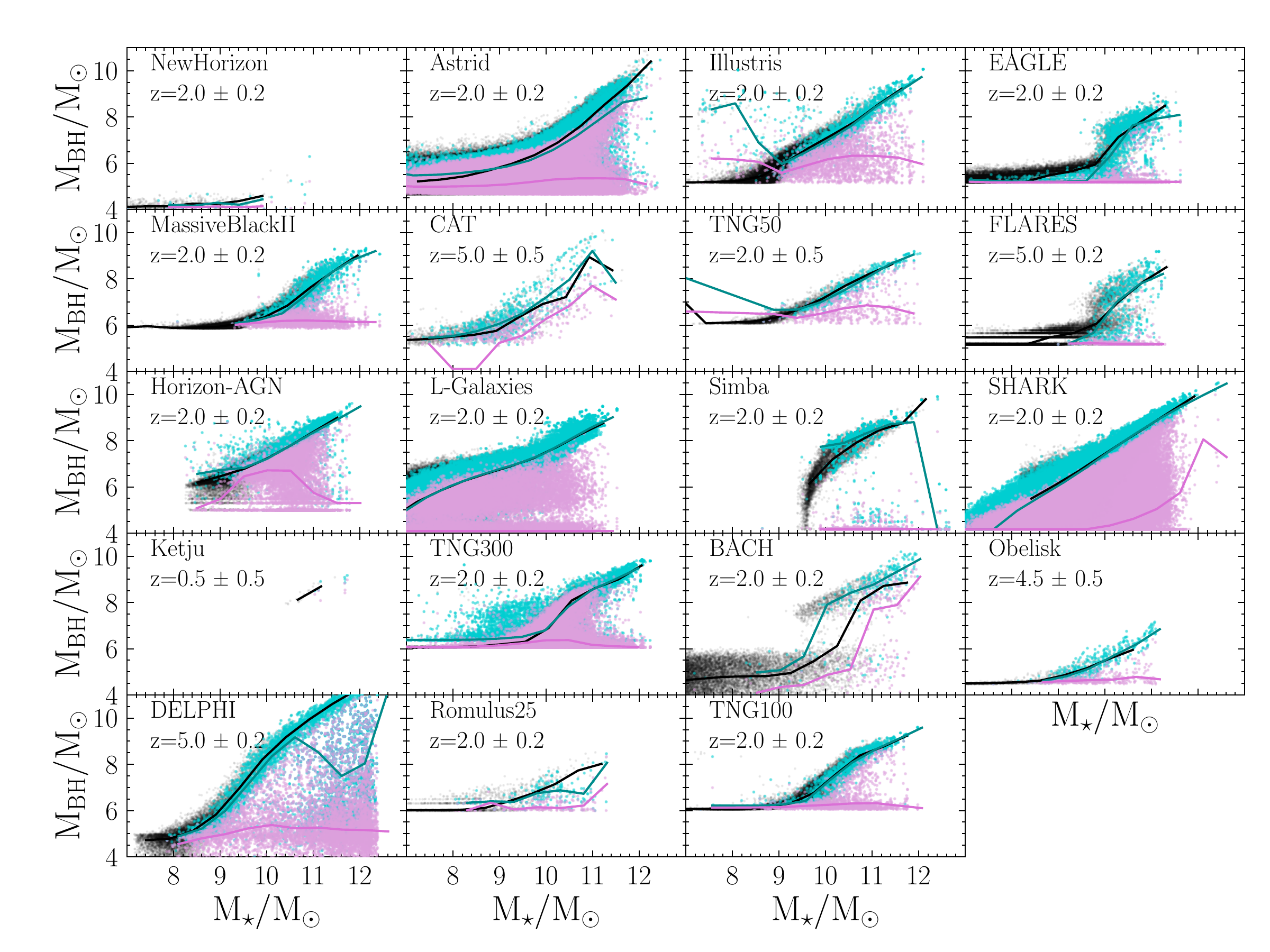}
\caption{Mass comparison between primary (green) and secondary (pink) MBHs of merging systems and the full MBH population (black) produced by the models. Solid lines indicate the median of the populations. The redshift that is chosen for the comparison is $z=2\pm 0.2$ and, if there are not enough mergers at that time, we select the closest redshift available in the catalogs. The selected redshift is indicated in each panel. 
}
\label{fig:Hostproperties}
\end{figure*}

Fig.~\ref{fig:Hostproperties} 
compares more directly the population of MBHBs with the typical MBHs hosted in galaxies at $z\sim 2$ (when that redshift is available in the catalogs) within each simulation. We omit \renaissance, since its mergers are limited to very high redshifts. The median masses of the primary and secondary MBHs are shown in green and pink, respectively, whereas the median for the total MBH population produced by each model is shown in black. As in Fig.~\ref{fig:FullMBHpopulation_BHmassfunction}, Fig.~\ref{fig:MBH_Mstellar_Comparison_Observations}, and Fig.~\ref{fig:FullMBHpopulation_AGN_LF}, we use restrict the median for the full population (black lines) to the most massive MBHs per galaxy in massive galaxies with $M_{\star}\geqslant 10^{11}\, \rm M_{\odot}$ for \tng, \illustris, \massiveblack, \simba, and \eagle. As mentioned before, some low-mass galaxies (at the resolution limit) in \illustris and \tng
with $M_{\star}\leqslant 10^{9}\, \rm M_{\odot}$ are temporary matched to the incorrect MBHs. It leads to artificially high primary and secondary MBH masses with $10^8 \msun$ in galaxies with $10^8 \msun$ $M_\star$ in \illustris{} or with $10^7 \msun$ $M_\star$ in TNG50, in Fig.~\ref{fig:Hostproperties}.


For the most part, MBHBs have total masses that are consistent with typical MBHs for their given host galaxy. In addition, \delphi, \astrid and \romulus produce a sizable population of mergers in galaxies with $M_\star >10^{10} \msun$ where even the primary is 
undermassive ($\leqslant 10^{7-8}\, \rm M_{\odot}$ for \astrid and $\leqslant 10^{8-9}\, \rm M_{\odot}$ for \delphi, $\leqslant 10^{6.5-7}\, \rm M_{\odot}$ for \romulus) with respect to the median of the full MBH population). 
In \delphi this is due to an increase in the gas mass and the associated star formation rate, and SN feedback in merging systems that limits the growth of the central black holes. In \astrid and \romulus, this is due to mergers between wandering MBHs (or secondary MBHs) in massive galaxies. Other models, such as \horizonAGN, \obelisk{}, and \newHorizon{} consider such mergers spurious and remove them from catalogues, since in reality it is not very plausible that two wandering MBHs ``find'' each other: the cross section of MBHs in simulations is of the order of the resolution (10s-100s of pcs at best), while for real MBHs it would be $<$ mpc. 

In conclusion, MBH mergers involve the typical population of MBHs in galaxies, but in all models we identify at least a few mergers with very low mass ratio $q < 1/20$. In models wherein seeds form down to low masses and then merge efficiently (because delays are not included), this intermediate mass ratio is well populated. However, when dynamical friction delays are included, a larger fraction of binaries merge with a mass ratio $\sim 0.1$, even when low-mass seeds ($<10^6 \msun$) exist in the models, e.g., \lgalaxies, \barausse, and  \delphi. The very low mass ratios are now limited to lower redshift, involving MBHs that have not grown much. Dynamical delays decrease the size of this population, but they do not remove it completely. It is important for LISA's scientific development to keep this population into consideration, and to develop waveforms and data analysis techniques capable to identify these mergers. 

\subsection{Models with intrinsic delays}\label{ModelsWithDelays}

\begin{figure*}
\centering
\includegraphics[width=1.3\columnwidth]{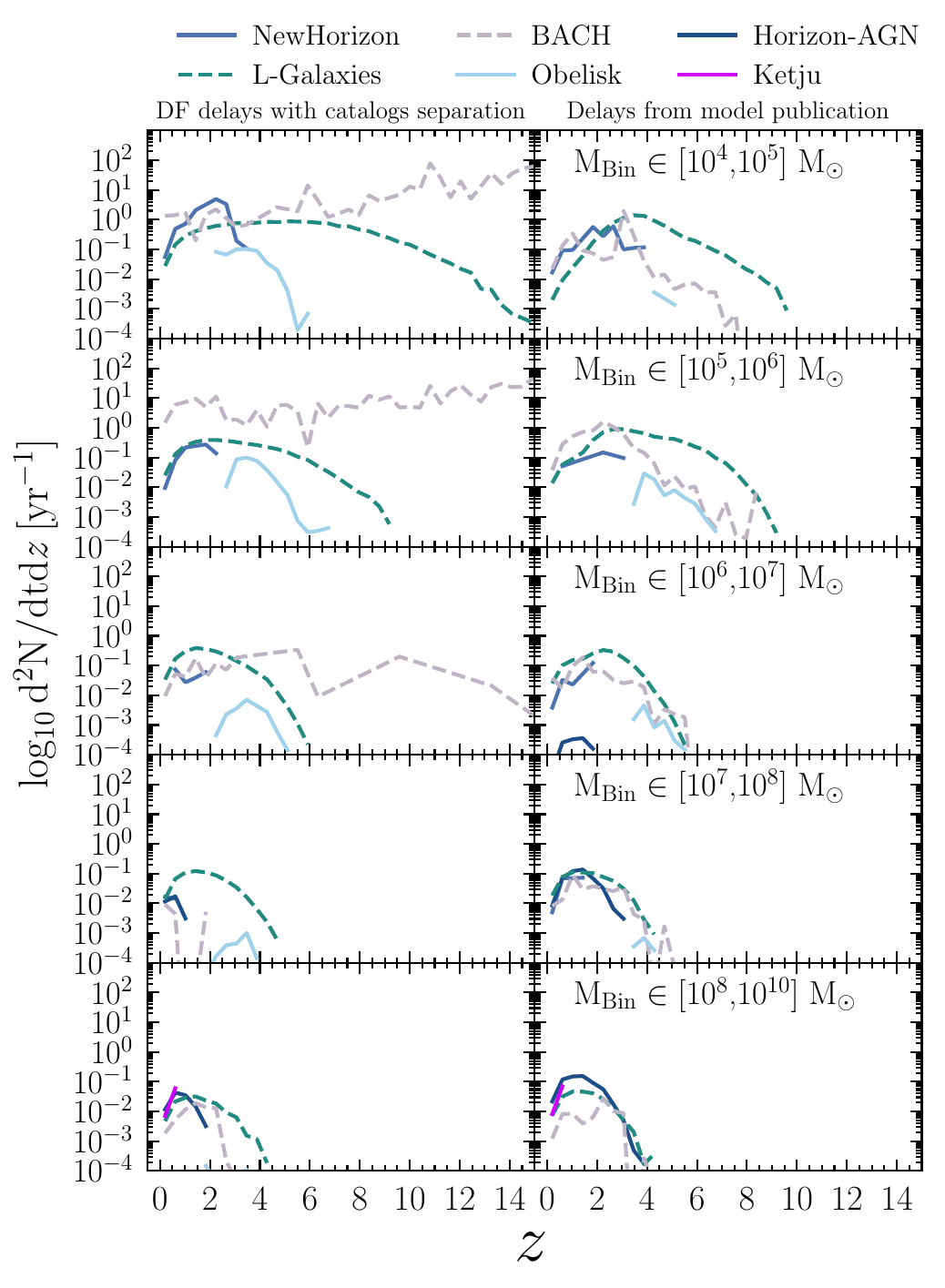}
\caption{Merger rate of MBHBs as a function of redshift, for five different MBHB mass bins ($M_{\rm Bin} = M_{\rm BH,1} + M_{\rm BH,2}$). Redshift bins have a fixed width of $\sim 0.4$. The figure illustrates the differences between the DF delays computed in this paper and the delays computed in previous studies for \barausse, \lgalaxies, \ketju, \newHorizon, \lgalaxies, \obelisk, and  \horizonAGN.}
\label{fig:difference_DFfromthispaper_previousstudies}
\end{figure*}

In Section~\ref{sec:merger_rates}, we compared binary merger rate predictions across different models, for all of which we used a common approach to estimate the dynamical friction timescale needed to shrink the binary from its initial separation to the regime where stellar and gas hardening drive further evolution. Some of the simulations included in this study, however, already account for binary evolution below their resolution limits. In this section, we compare the merger time delays predicted by these models with the post-processed delays calculated through Eq.~\eqref{Eq_df_old}, to assess any discrepancies and quantify the potential errors in merger rate estimates. The results are shown in Fig.~\ref{fig:difference_DFfromthispaper_previousstudies}.

\lgalaxies tracks the evolution of MBHBs right after the merger of two galaxies. Specifically, it assumes that their evolution is driven by three different phases: dynamical friction, hardening, and GW emission. The dynamical friction phase of MBHs occurs right after the galaxy merger and lasts for a time given by Eq.~\eqref{Eq_df_old}, where the initial separation between the two MBHs corresponds to the radius at which the tidal forces removed 80\% of the satellite's stellar mass. When this phase ends, the two MBHs form a gravitationally bound binary and enter the hardening and GW inspiral phase. The separation and eccentricity of the binary are numerically evolved self consistently by taking into account the environment in which the system resides. In gas-rich environments, the binary evolves by interacting with the circumbinary discs and the GW emission. In gas-poor environments, the interaction with the stellar component of the galactic bulge (distributed according to a Sérsic distribution) and the GW emission harden the binary down to the final coalescence. Finally, triple interactions are also taken into account and a preferential accretion is assumed to occur during the MBHB growth.

\barausse{} models the dynamical friction between two merging haloes, and later between galaxies, accounting also for tidal effects (disruption and evaporation) that occur in the process. At smaller separations, MBHs sink toward the galactic cent\soutPC{e}re under the effect of dynamical friction from the surrounding matter. \barausse describes this phase resorting to results from cosmological simulations \citep[][]{2018MNRAS.475.4967T}. At even smaller separations, when the MBHs finally form a bound binary, dynamical friction becomes inefficient. Stellar hardening, gas-driven migration (in gas-rich systems) and interactions with other black holes (triple systems) are included at the bound binary stage, as well as at the GW emission stage. At this final step, \barausse{} follows the merger remnant's mass, spin and kick velocity imparted by GW emission~\citep{2010ApJ...719.1427V,2012ApJ...758...63B,2016ApJ...825L..19H}.


In \ketju, self consistent evolution of MBHBs in the dynamical friction and hardening phases is automatically implemented, as the code employs zero softening for stellar encounters with MBHs. Additionally, it contains PN terms up to 3.5 order, allowing the binaries to evolve also in the GW-dominated regime.  

In \newHorizon, \horizonAGN, and \obelisk, MBHs evolve under the effect of dynamical friction implemented directly in the simulation until two MBHs reach a separation corresponding to 4 resolution elements. At that point, we have what we defined as a ``numerical merger''. Additional evolution is calculated in post-processing. The full model is described and analyzed in \citet{2020MNRAS.498.2219V}. They first calculate the timescale for dynamical friction sourced by the stellar distribution up until the MBHs form a binary. Contrary to what was assumed in Section~\ref{ModellingDelays}, dynamical friction does not start from the MBH relative separation, but from the distance of each of the two MBHs from the centre of the host galaxy prior to the MBH merger. They also add a correction factor of 0.3 to account for cosmological orbits being eccentric \citep{2006A&A...445..403K}. After the dynamical friction time has elapsed, they select the galaxy hosting the virtual binary -- technically, the galaxy hosting the already merged MBH -- at the nearest galaxy snapshot and calculate the time until coalescence via stellar hardening and  
GW emission, and migration in a circumbinary disc 
and GW emission based on the stellar structure and the accretion rate. They consider as time to merger the shortest of these two timescales. 

The merger rate calculated self-consistently in the models is in good agreement when comparing directly the mass and redshift ranges where models overlap. For instance, \lgalaxies, \barausse, \obelisk{} and \newHorizon{} for masses $<10^7 \msun$; while \ketju{} and \horizonAGN, which focus on massive galaxies and MBHs, become comparable with \lgalaxies{} and \barausse{}for masses $>10^8 \msun$. This suggests that the self-consitent approaches taken in the various models is in good agreement. We highlight in particular that \ketju{} performs a full integration of the MBHB orbits and therefore represents an important validation for all models.  
We note, however, that adding delays based on the separations reported in the catalogs have significant differences up to binaries with mass $10^8 \msun$: this implies that a ``one-size-fits-all'' approach is less reliable than a customized one.

\subsection{LISA detection rates}

\begin{figure*}
\centering
\includegraphics[width=2\columnwidth]{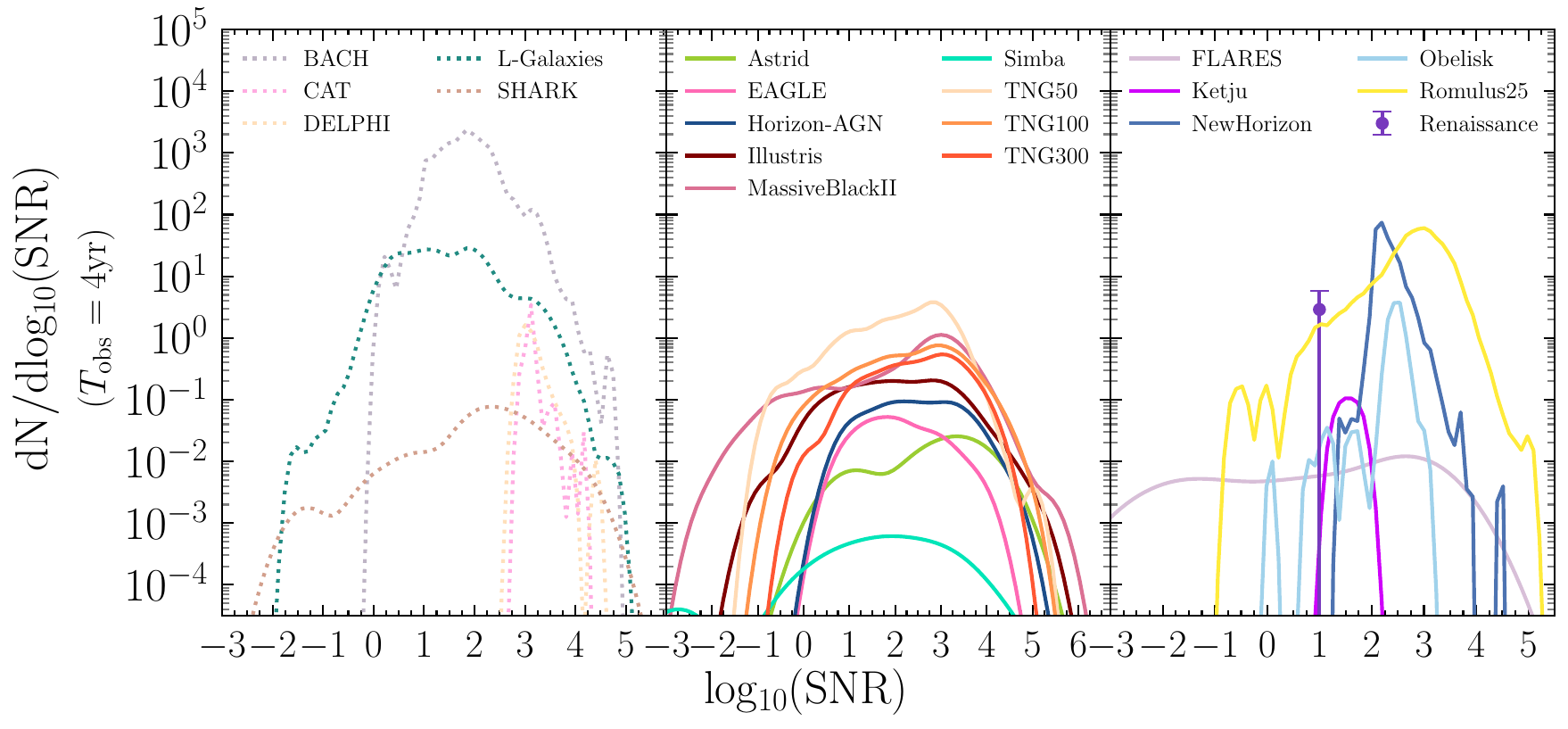}
\caption{SNR distribution for the models explored in this work, assuming four years of LISA observation. To increase the readability, we divided the results in three different panels. In the left-hand panel we grouped the SNR distribution from semi-analytical models, the middle panel shows the large-scale cosmological simulations of $\geqslant 100^{3}\, \rm cMpc^{3}$, and  the right-hand panel shows the results for the highest-resolution simulations. A KDE estimator was used to smooth the curves. The \renaissance simulation yields few mergers, although it has a high rate, 
and only one of them is left after applying our delay prescription. In order to acknowledge this low statistics, we show this single data point together with its Poisson error.}
\label{fig:SNR_distribution}
\end{figure*}

To efficiently evaluate the signal-to-noise ratio (SNR) of MBHBs from the different models with the dynamical friction delays computed using the catalogue separations (middle panels of Fig.~\ref{fig:BinaryMBHpopulation_MergerRate} and Fig.~\ref{fig:BinaryMBHpopulation_MergerRate_specificMassBands}), we proceed as follows:

\begin{itemize}

\item[$\bullet$] for each model, we numerically assemble the expected number of mergers in a specific time frame $T_{\rm obs}$ as:

\begin{equation}
\begin{split}
\frac{\mathrm{d}N}{\mathrm{d}z \,\mathrm{d}M_{\rm BH,1} \,\mathrm{d}M_{\rm BH,2}} =
\frac{\mathrm{d}n}{\mathrm{d}z \,\mathrm{d}M_{\rm BH,1} \,\mathrm{d}M_{\rm BH,2}} \\
\times 4\pi c \left(\frac{d_{\rm L}}{1+z}\right)^2 T_{\rm obs}~;
\end{split}
\label{eq:rate}
\end{equation}

\item[$\bullet$] we split the parameter space in 80 equally spaced bins between 0 and 20 in $z$, and 70 logarithmically spaced bins between $10~\msun$ and $10^9\msun$ for the masses. In every bin, we sample $N_{\rm bin}$ binaries assuming that the number of binaries follows a Poisson distribution with mean equal to $\frac{\mathrm{d}N}{\,\mathrm{d}z \,\mathrm{d}M_{\rm BH,1} \,\mathrm{d}M_{\rm BH,2} }$;

\item[$\bullet$] for every $N_{\rm bin}$ binary, we generate its properties, i.e., $(z_{\rm samp},m_{1,\rm samp},m_{2,\rm samp})$, from flat distributions within the corresponding bin edges of redshift, primary and secondary masses. As the spin information is not provided in all catalogues, we conservatively set the spin of each MBH to zero. As discussed below, we have verified the impact of this assumption;

\item[$\bullet$] for each binary, we draw 500 sets of inclination angle (with cosine uniformly distributed between $-1$ and 1), polarization, phase (uniformly distributed between $-\pi$ and $\pi$), and sky location (uniformly distributed on the sphere), and compute the averaged SNR over these 500 realisations;

\item[$\bullet$] we repeat the above procedure $N_{\rm realisation}$ times in order to increase statistics.
    
\end{itemize}

For the mission duration, we take $T_{\rm obs} = 4$~yr. This procedure generates catalogues of merging binaries during the mission duration for each model. Two of the models, \flares and \simba, yield at most a few observations during $T_{\rm obs}$, because of the low predicted rates. These simulations do output many events, but with very small weights. For these models, we use directly the catalogs outputted by the simulations. For each merger event in the catalog, we compute its averaged SNR as described above and use the weights given by Eq.~\eqref{eq:rate} to compute the distribution of SNRs for these models. Conversely, in \renaissance, including DF delays reduces the original catalog to a single merger event, but this event carries a large weight, leading to a high predicted rate. In order to account for this low statistics, we consider this single point alone and compute the associated statistical (Poisson) error, without going through the binning step. 

The SNR is calculated with the following expression:

\begin{equation}
    \mathrm{SNR}^2 = \sum_{\alpha=A, \ E, \ T}4 \int_{f_{\rm min}}^{f_{\rm max}}\frac{\tilde{h}_\alpha^2(f)}{S_{n,\alpha}(f)} {\rm d}f~,
    \label{eq:SNR_def_old}
\end{equation}

\noindent where the $\tilde{h}_{\alpha}(f)$ are the GW signals in each of the LISA time-delay-interferometry channels~\citep[A,E,T][]{2021LRR....24....1T}, which constitute three noise-independent measurements, and $S_{n,\alpha}(f)$ are the power spectral densities in these channels.

We compute SNRs with the \texttt{lisabeta} package \citep{2021PhRvD.103h3011M}, using the PhenomXHM waveform model for MBHBs \citep{2020PhRvD.102f4002G} and the \texttt{SciRDv1} LISA noise curve~\citep{2021arXiv210801167B}, with a white dwarf background estimated for 4 years of observations. The integral in Eq.~\eqref{eq:SNR_def_old} is evaluated numerically between $f_{\rm min}$ and  $f_{\rm max}$, representing, respectively, the minimum and maximum frequency spanned by the source during its evolution. Specifically, $f_{\rm min}$ is the maximum between $10^{-5} \ {\rm Hz}$ and the initial frequency of the binary with a corresponding merging time drawn from a uniform distribution between 0 and the nominal LISA mission duration ($T_{\rm obs} = 4$~yr). The value of $10^{-5} \ {\rm Hz}$ corresponds to the lower limit of the LISA sensitivity region. We use $f_{\rm max}=0.5 \ {\rm Hz}$, the higher limit of the LISA sensitivity region, though, in practice, most signals have already vanished by then. 

The SNR distributions for the merging MBHBs of the different models is shown in Fig.~\ref{fig:SNR_distribution}. The results from semi-analytical models are displayed in the left-hand panel, whereas the findings from hydrodynamical simulations are split between the central and right-hand panels for clarity.

We can see that, for the vast majority of the models, the SNR distribution firmly lies above $\rm SNR \gtrsim 10$, meaning that a large fraction of the predicted mergers can be detected by LISA. 
Notable exceptions are \lgalaxies, \shark, the cosmological simulations with volume boxes $\geqslant 100 \, \rm cMpc$ shown in the middle panel (for example, \massiveblack{}), and the higher-resolution simulation \romulus{}. There are two different reasons for these exceptions. For \lgalaxies, the MBH mass function extends to quite low masses because of the low-mass seeding prescriptions, and such MBHs are merging outside the LISA sensitivity curve, that can capture only part of the inspiral, resulting in low SNRs. For the large-volume simulations and \shark,  the reason is the opposite: the presence of very massive MBHs ($>10^8\msun$) merging at relatively high redshift. Such MBHs are also not ideal LISA sources, because in this case the inspiral phase is at too low frequency and only the merger and ringdown are in the detector's sensitivity band. This is further exacerbated when the mass ratio is very different from unity, which is often the case for very massive primary MBHs. In summary, for MBHBs below $10^4\msun$ and above $10^8\msun$ beyond $z=5$, the sensitivity of LISA worsens significantly, and therefore such coalescences cannot be properly detected, producing a long tail of events with low SNR. 
We stress that, despite these low SNR tails, in all models more than 50\% of the MBHBs merging throughout the universe will be confidently detected by LISA.
For \renaissance, the single datapoint is consistent with other high-resolution simulations.

To assess the impact of assuming that all MBHs have a spin of zero in the previous approach, we recomputed the SNRs averaged over the angles and the time to coalescence assuming a spin of 0.5 (with both spins aligned with the orbital angular momentum). We found that the SNR distributions shown in Fig.~\ref{fig:SNR_distribution} are slightly shifted towards higher values, without altering the overall trend in the figure. Additionally, the number of systems with an SNR $>$ 10 increases by at most a few per cent across all catalogues.

Semi-analytical models with high resolution and run to $z=0$ (\lgalaxies, \barausse) predict the largest number of detections. This is inherently connected with the ability to capture, though semi-analytically, the evolution of low-mass galaxies, in which many merger events occur throughout cosmic time. This is not possible in large cosmological simulations, which lack the necessary spatial and mass resolution to capture the evolution of low-mass systems. For this reason, their predicted merger rate, and consequentially the number of detections, remains smaller with respect to that of semi-analytical models. High resolution simulations run to $z=0$ (\romulus, \newHorizon, \tngl) start reaching the parameter space probed by semi-analytical models.

Focusing on the high end of the SNR distributions, the models with the highest SNRs are those able to simulate the cosmological evolution down to low redshift, while models that provide results only down to a certain redshift, like \renaissance or \obelisk, produce only moderately high SNRs. This is due to the combined effect of larger luminosity distances and lower intrinsic MBH masses at high redshifts, both conspiring to reduce the strength of the signal.

The main conclusion that we can draw from Fig.~\ref{fig:SNR_distribution} is the very large scatter amongst models in the number of detectable MBHBs. This is intrinsically connected to the large scatter we found for the merger rates (see Fig.~\ref{fig:BinaryMBHpopulation_MergerRate}), making the heterogeneous number of detections deeply connected with the astrophysical uncertainties affecting our modelisation of MBHB-galaxy co-evolution. Further to the unknowns concerning the physics driving the formation and evolution of MBHs and MBHBs, the efforts of this work also highlight that the limited resolution determined by our current computational capabilities plays an important role in shaping the results we find, thus a comprehensive prediction about the MBHB detection rate needs to account for, and incorporate the uncertainties caused by, the longstanding dilemma between high resolution and large volumes.

\section{Conclusions}
\label{sec:conclusions}

In this collaborative effort of the LISA Astrophysics Working Group, we compared various theoretical predictions of MBH merger rates to quantify the spread, and evaluate the global astrophysical uncertainties of the LISA event rates. The 20 models analyzed in this study differ in their underlying techniques, modelling of galaxy and MBH physics,  assumptions for parameters, and how different groups define different quantities (Section~\ref{sec:methods_models}, Fig.~\ref{fig:SeedMass}, Fig.~\ref{fig:Separation}). Consequently, the MBH and MBHB catalogues do not share identical properties. These intrinsic differences have to be accounted for when comparing models to assess the uncertainties in the MBH merger rate: the uncertainties are a combination of astrophysical assumptions and technical differences. 
To assess the role of these differences, our work represents the first comprehensive comparison of MBHB merger rates and LISA detectability across 20 state-of-the-art models. Below, we summarize our main findings.

The evolution of the MBH mass function, the $M_{\rm BH}$--$M_{\star}$ relation, and the AGN luminosity function across the different models (Fig.~\ref{fig:FullMBHpopulation_BHmassfunction}, \ref{fig:MBH_Mstellar_Comparison_Observations}, \ref{fig:FullMBHpopulation_AGN_LF}) reveal both areas of convergence and substantial divergence. While broad agreement is found at lower redshifts ($z\leqslant 1$) due to model calibration on available observational constraints, significant discrepancies appear at higher redshifts ($z \geqslant 4$--5) and in the low-mass regime at any redshift ($M_{\rm BH}\leqslant 10^{7}\, \rm M_{\odot}$ and/or $M_{\star}\leqslant 10^{10}\, \rm M_{\odot}$), which is particularly relevant for LISA. 

The predicted rates of MBH mergers (Fig.~\ref{fig:BinaryMBHpopulation_MergerRate}, Fig.~\ref{fig:BinaryMBHpopulation_MergerRate_specificMassBands}) and their detectability by LISA (Fig.~\ref{fig:SNR_distribution}) vary significantly across models from $\lesssim 1$ up to thousands for a LISA mission lifetime of 4 years, primarily driven by differences in seeding prescriptions, simulation resolution, and the inclusion (or omission) of MBH dynamical modelling. Cosmological simulations with large volumes but relatively low resolution (e.g., \horizonAGN, \illustris, \tngm, \tngh) often fail to capture mergers in dwarf galaxies. In contrast, high-resolution simulations and semi-analytical models, which are capable of resolving the assembly of these low-mass galaxies, tend to predict higher merger rates. Low-mass galaxies could be key contributors to the LISA detection space and therefore warrant increased attention from the modelling community. 

The seed mass and the contribution of low-mass MBHs to the total merger rate are important. Semi-analytical models that implement physically motivated MBH formation channels -- with both light and heavy seeds forming at redshifts $z\sim 10$ -- typically predict higher merger rates than cosmological simulations. This is largely because many simulations adopt a fixed seed mass of $\sim 10^{6}\, \rm M_{\odot}$, near the upper limit of LISA’s sensitive mass range, thereby missing a significant population of low-mass mergers.

The inclusion of dynamical friction delays generally reduces the predicted merger rates, with the strongest suppression occurring at high redshifts and for mergers involving low-mass MBHs. Despite these differences, several models consistently predict a few detectable MBHB mergers per year. Most models also yield a diverse binary population, with mass ratios spanning several orders of magnitude (Fig.~\ref{fig:BinaryMBHpopulation_MergerRate_massratio}, Fig.~\ref{fig:BinaryMBHpopulation_MergerRate_massratio_catsep}, Fig.~\ref{fig:BinaryMBHpopulation_MergerRate_massratio_ET}, Fig.~\ref{fig:mass_ratios_withMstar}). Our results suggest that the development of GW waveforms in the notoriously challenging intermediate mass ratio regime $10^{-3}<q<10^{-1}$ is needed for the analysis of LISA data.

\section*{Acknowledgments}
{\small DIV acknowledges the support of the ``la Caixa'' Foundation fellowship (ID 100010434). The project that gave rise to these results received the support of a fellowship from ``la Caixa'' Foundation (ID 100010434). The fellowship code is LCF/BQ/PI25/12100024. MH acknowledges support from the Swiss SNSF Starting Grant (grant no. 218032). MB acknowledges support from the Italian Ministry for Universities and Research (MUR) program “Dipartimenti di Eccellenza 2023-2027”, within the framework of the activities of the Centro Bicocca di Cosmologia Quantitativa (BiCoQ). SB acknowledges support from the Spanish Ministerio de Ciencia e Innovación through project PID2021-124243NB-C21 AG acknowledges support from the STFC grant ST/Y002385/1 MV acknowledges funding from the French National Research Agency (grant ANR-21-CE31-0026, project MBH\_waves) and from the Centre National d’Etudes Spatiales. EB acknowledges support from the European Union’s Horizon ERC Synergy Grant ``Making Sense of the Unexpected in the Gravitational-Wave Sky'' (Grant No. GWSky-101167314). This work has been supported by the Agenzia Spaziale Italiana (ASI), Project n. 2024-36-HH.0, ``Attività per la fase B2/C della missione LISA''. AKB acknowledges support from NSF-AST 2510738 and NSF-AST 2346977 LB acknowledges support from National Science Foundation grants AST-2307171 and AST-2509457. ABR is supported by the Funda\c{c}\~{a}o Carlos Chagas Filho de Amparo \`{a} Pesquisa do Estado do Rio de Janeiro (FAPERJ), Grant No E-26/200.149/2025 \'{e} 200.150/2025 (304809) EB acknowledges support from the European Union’s Horizon Europe programme under the Marie SkłodowskaCurie grant agreement No 101105915 (TESIFA). PRC acknowledges support from the Swiss National Science Foundation under the Sinergia Grant CRSII5\_213497 (GW-Learn). MC acknowledges support from the Excellence Department BiCoQ P. Dayal warmly acknowledges support from an NSERC discovery grant (RGPIN-2025-06182). RPD and CP acknowledge funding from the South African Radio Astronomy Observatory (SARAO), which is a facility of the National Research Foundation (NRF), an agency of the Department of Science, Technology and Innovation (DSTI). RPD acknowledges funding by the South African Research Chairs Initiative of the DSTI/NRF (Grant ID: 77948). CD acknowledges support for HPC use from the CIMUSE consortium. D.H. acknowledges funding from the NSERC Arthur B. McDonald Fellowship and Discovery Grant programs and the Canada Research Chairs (CRC) program. PHJ acknowledges the support by the European Research Council via ERC Consolidator grant KETJU (no 818930) and the Research Council of Finland grant 339127. AK acknowledges the support by the European Research Council via ERC Consolidator grant KETJU (no. 818930) and the Vilho, Yrjö and Kalle Väisälä Foundation. This material is based on work supported by Tamkeen under the NYU Abu Dhabi Research Institute grant CASS. SL acknowledges support from the National Natural Science Foundation of China (NSFC) grant (no. 12473015). AM acknowledges support from the postdoctoral fellowships of IN2P3 (CNRS). This project has received funding from the European Union’s Horizon 2020 research and innovation program under the Marie Skłodowska-Curie grant agreement No. 101066346 (MASSIVEBAYES). FP, AC and LC acknowledge support from the Romanian Ministry of Research, Innovation and Digitalization under the Romanian National Core Program LAPLAS VII - contract no. 30N/2023 and from the ESA PRODEX project RoLISASpace. AR acknowledges support from the University of Helsinki Research Foundation and the European Research Council via ERC Consolidator grant KETJU (no 818930). JR acknowledges support from the Royal Society and Research Ireland under grant number URF\textbackslash R1\textbackslash 191132  and support from the Research Ireland Laureate programme under grant number IRCLA/2022/1165. BR acknowledges funding through ANID (CONICYT-PFCHA/Doctorado acuerdo bilateral DAAD/62180013), DAAD (funding program number 57451854), the International Max Planck Research School for Astronomy and Cosmic Physics at the University of Heidelberg (IMPRS-HD), and the support by the European Research Council via ERC Consolidator grant KETJU (no. 818930). JSR acknowledges financial support from the National Science Foundation (NSF) EMIT Program (NSF-2125764). MR acknowledges support from the Generalitat Valenciana Grant CIDEGENT/2021/046,  by the Spanish Agencia Estatal de Investigación (grant PID2024-159689NB-C21) funded by MICIU/AEI/10.13039/501100011033 and by FEDER / EU. A.S. acknowledges support by the European Union’s H2020 ERC Advanced Grant ``PINGU'' (Grant Agreement: 101142079). G.M.S acknowledges the financial support provided under the European Union’s H2020 ERC Consolidator Grant B Massive (Grant Agreement: 818691) and Advanced Grant PINGU (Grant Agreement: 101142097). JS acknowledges financial contribution from the Bando Ricerca Fondamentale INAF 2022 Large Grant, 'Dual and binary supermassive black holes in the multi-messenger era: from galaxy mergers to gravitational waves’ and the Bando Ricerca Fondamentale INAF 2024 Large Grant, 'The Quest for dual and binary massive black holes in the gravitational wave era'. D.S. acknowledges the financial support provided under the European Union’s H2020 ERC Advanced Grant ``PINGU'' (Grant Agreement: 101142079). A.T. is supported by MUR Young Researchers Grant No. SOE2024-0000125, ERC Starting Grant No.~945155--GWmining, Cariplo Foundation Grant No.~2021-0555, MUR PRIN Grant No.~2022-Z9X4XS, Italian-French University (UIF/UFI) Grant No.~2025-C3-386, MUR Grant ``Progetto Dipartimenti di Eccellenza 2023-2027'' (BiCoQ), and the ICSC National Research Centre funded by NextGenerationEU. AT acknowledges financial support from the Bando Ricerca Fondamentale INAF 2023, Mini-grant ``Cosmic Archaeology with the first black hole seeds" (RSN1 1.05.23.04.01) and from the PRIN MUR “2022935STW" funded by European Union-Next Generation EU. RV acknowledges support from PRIN MUR "2022935STW" funded by European Union-Next Generation EU, Missione 4 Componente 2 CUP C53D23000950006, from the Bando Ricerca Fondamentale INAF 2023, Theory Grant "Theoretical models for Black Holes Archaeology", and from the ASI-INAF agreement n. 2024-36-HH.1-2025" LG acknowledges support from the Amaldi Research Center funded by the MIUR program “Dipartimento di Eccellenza” (CUP:B81I18001170001). RS acknowledges support from the PRIN 2022 MUR
project 2022CB3PJ3—First Light And Galaxy aSsembly (FLAGS) funded by the European Union—Next Generation EU, and from the INFN TEONGRAV initiative JHW acknowledges support from NSF grant AST-2510197.}

\bibliographystyle{aasjournal}
\bibliography{paper_OJA} 

\begin{appendix}
\section{Uncertainties in estimating dynamical friction delays}\label{sec:uncertainties}

\begin{figure*}
\centering
\includegraphics[width=0.95\columnwidth]{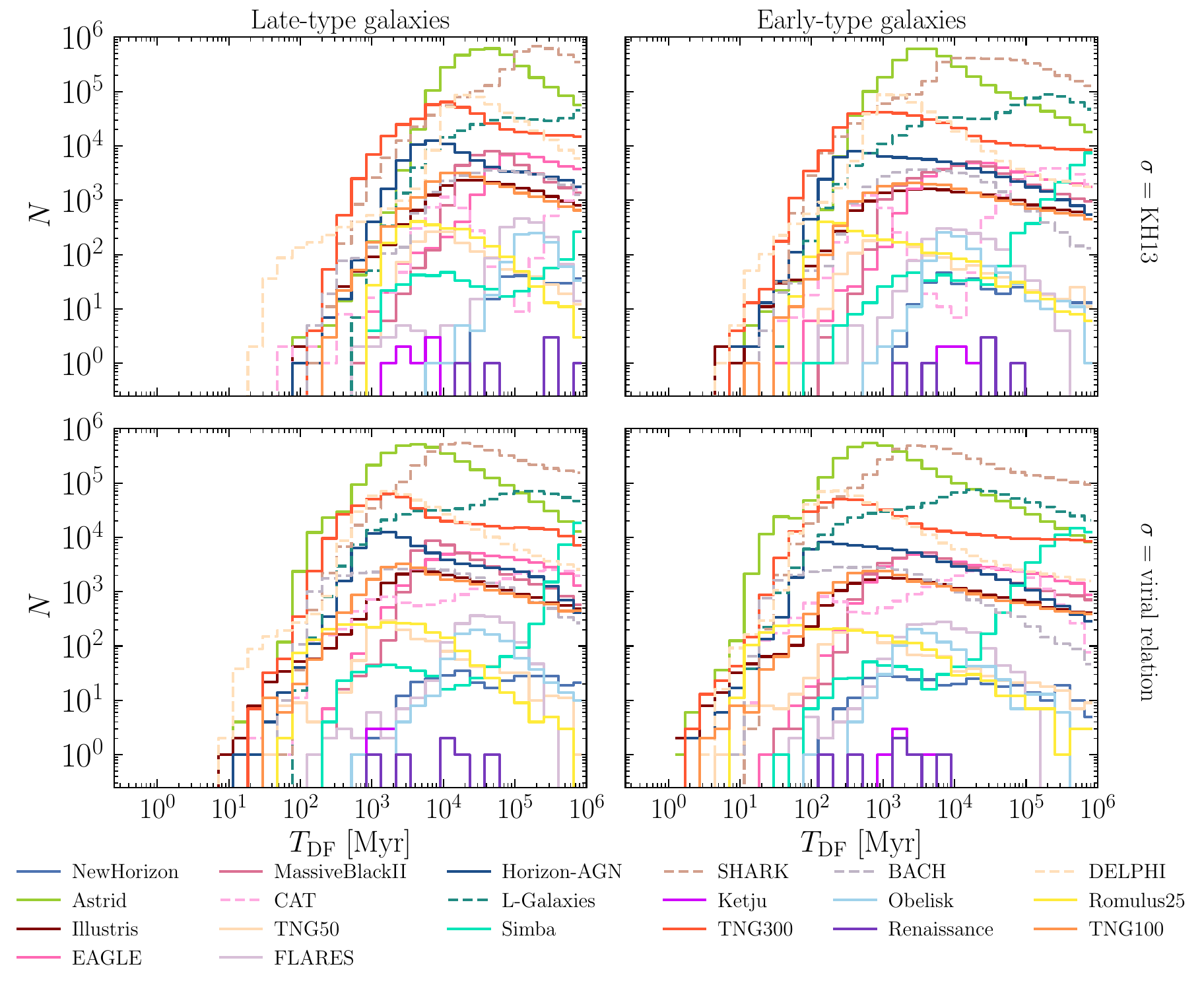}
\caption{Dynamical friction delays obtained from Eq.~\eqref{Eq_df_old} assuming the galaxy effective radius as the initial binary separation. The results are shown for different scaling relations for $\rm R_{eff}$ and $\sigma$ in different panels. For $\rm R_{eff}$, we considered the relation for late-type \citep[][left-hand panels]{2003MNRAS.343..978S} and early-type \citep[][right-hand panels]{2015MNRAS.447.2603L} galaxies, including the redshift dependence from \citet{2014ApJ...788...28V}. The galaxy velocity dispersion has been derived both through the \citet{2013ARA&A..51..511K} scaling (upper panels) and virial estimations (lower panels, see text in Section~\ref{ModellingDelays}). 
}
\label{fig:DFtimescales_Reff}
\end{figure*}

Fig.~\ref{fig:DFtimescales_Reff} shows the dynamical friction timescales obtained according to the model described in Section~\ref{ModellingDelays} assuming an initial MBHB separation given by the effective radius of the galaxy. The results obtained for different assumed scaling relations (early- and late-type galaxies), the addition of redshift evolution for $\rm R_{eff}$, and different choices of the stellar velocity dispersion $\sigma$ are shown in different panels. We find that the resulting delays span a large range of timescales, from a few hundred Myr to longer than a Hubble time, with a broad peak around 1--10~Gyr. The differences can mainly be attributed to differences in the models, with resolution playing a significant role. For example, the \ketju simulations, which model binaries down to and below pc-scale separations, shows a narrower distribution of delays peaked at around 1~Gyr, with a tail to tens of Gyr.


We note again that these delays can be considered as upper limits to the dynamical friction timescale experienced by the binaries, as they consider only the mass of the secondary MBH rather than the total mass of the binary and what remains of its surrounding galaxy. 

The modelling of dynamical friction according to Chandrasekhar's formalism has limitations, and there is scope for improvement in the future. For example, \citet{2015ApJ...806..220A} report that it fails in the inner regions of galaxies and provide an improved expression that is additionally a function of the orbital eccentricity and the negative logarithmic slope of the density profile. \citet{2006AJ....132.2701G} provide an equation for the negative logarithmic slope of the density model from \citet{1997A&A...321..111P}, which is based on S\'ersic model parameters. 

We note that virial mass estimators ($M \sim \sigma^2R_{\rm eff}/G$) also suffer from significant limitations, especially for disc-dominated merger remnants resulting from a galactic major merger, if the galaxies are treated as single-component systems \citep{2023MNRAS.522.3588G}.

Finally, $M_{\rm BH}$--$\sigma$ scaling relations are also a source of uncertainty. While we have adopted generally accepted forms, as described in Section~\ref{ModellingDelays}, modifications may be considered in the near future based on observational evidence. Data from \citet{2013ApJ...764..151G, 2018ApJ...852..131B, 2019ApJ...887...10S} reveal a steepening of the $M_{\rm BH}$--$\sigma$ distribution, in agreement with expectations from \citet{2023MNRAS.518.6293G}. In particular, brightest cluster galaxies seem to follow a steep $M_{\rm BH}$-$\sigma^{8\pm1}$ trend, while spiral galaxies, thought not to have experienced a major merger event, show a considerable scatter and a shallower trend.

\clearpage
\section{Cosmology and model resolution}

\end{appendix}

\end{document}